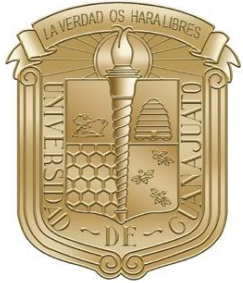

# Gestión y Medición del Conocimiento en Organizaciones Públicas

## TESIS

Que para obtener el grado de
Maestría en Administración

**PRESENTA:**

Héctor Pérez López Portillo

**DIRECTORES DE TESIS:**

Dr. Hugo Arnoldo Mitre Hernández

**CODIRECTOR DE TESIS:**

Mtro. Jorge Alberto Romero Hidalgo

**SINODAL DE TESIS:**

Dra. Claudia Susana Gómez López

Guanajuato, Viernes, 1 de Julio de 2016.

# Agradecimientos

# Gestión y Medición del Conocimiento en Organizaciones Públicas

**Héctor Pérez López Portillo**

Universidad de Guanajuato

**Director de tesis:** Dr. Hugo Arnoldo Mitre Hernández,

**Co-director de tesis:** Mtro. Jorge Alberto Romero Hidalgo

## Nota de autenticidad

El autor declara que el presente trabajo es de su autoría y que en él ha depositado esfuerzo intelectual y académico para la consecución del objetivo. Para lograr lo anterior, este trabajo de tesis ha sido *parcialmente* presentado en distintos foros académicos, y publicado *parcialmente,* en un capítulo de libro, como se detalla a continuación:

- **Actas de conferencia**: 16th European Conference on Knowledge Management (Udine, Italy. 2015), Mitre-Hernández, H. A., Mora-Soto, A., López-Portillo, H. P., & Lara-Alvarez, C. (2015). *Strategies for fostering Knowledge Management Programs in Public Organizations.* In A. Garlatti & M. Massaro (Eds.), 16th European Conference on Knowledge Management (pp. 539–547). Reading: Academic Conferences and Publishing International Limited. ISSN: 2048-8971. Disponible en: http://arxiv.org/abs/1506.03828

- **Capítulo de libro publicado**: Pérez López-Portillo, H., Romero Hidalgo, J. A., & Mora Martínez, E. O. (2016). Factores Previos para la Gestión del Conocimiento en la Administración Publica Costarricense. En *Administrar lo Público 3* (pp. 102–129). San José, Costa Rica: CICAP, Universidad de Costa Rica, ISBN: 978-9968-932-22-6. Disponible en: https://goo.gl/fT5DYU

- **Ponencia:** International Research Society for Public Management, 20th Annual Conference, Hong Kong 2016, Pérez López-Portillo, H., Vázquez-González, E. R. & Romero-Hidalgo, J. R. (2016). *Knowledge management metrics for Public Organizations: A literature review-based proposal.* (In Press).

# RESUMEN

En la actualidad, la gestión del conocimiento es componente estratégico que permite el desarrollo, crecimiento y mejora continua de las organizaciones del sector público. En ese contexto específico se circunscribe la presente tesis, al estudiar de manera crítica y exhaustiva los factores que caracterizan a la gestión del conocimiento, las estrategias que favorecen su desarrollo y, finalmente, proponer métricas que permitan su evaluación, de cara a la mejora continua y sistemática de este tipo de prácticas en las organizaciones públicas.

Principalmente se atienden problemas relacionados con la falta de literatura académica que explique los elementos propios de la gestión del conocimiento en el contexto determinado. Al mismo tiempo se atiende a la falta de un análisis profundo sobre cuáles son los factores que hacen posible la gestión del conocimiento y cuáles favorecen su éxito en las organizaciones públicas. Además, se identifica como problema la ausencia de criterios para medir y evaluar la gestión del conocimiento en las organizaciones públicas, desde una perspectiva distinta a la perspectiva empresarial.

La contribución de la presente tesis es que aporta elementos valiosos para un debate académico sobre los factores previos, las estrategias para favorecer y las métricas con las cuales se puede evaluar las distintas iniciativas de gestión del conocimiento en las instituciones públicas.

Para lograr el objetivo de investigación de esta tesis se ha realizado una exhaustiva revisión sistemática de la literatura para conocer cuáles son los principales factores críticos de éxito que se han estudiado hasta el momento, asimismo se integran algunas métricas, a manera de propuesta, para evaluar el desempeño y éxito de la gestión del conocimiento en las organizaciones públicas. Finalmente, se ha realizado un estudio a profundidad entre distintas organizaciones del sector público para conocer cuáles son los factores críticos de éxito que deben considerar previamente en la implementación de iniciativas de gestión del conocimiento.

Finalmente, con base en esta propuesta metodológica y como resultado de este trabajo de investigación se pudieron conocer los factores críticos de éxito, se identificaron las estrategias para incentivar el éxito de la gestión del conocimiento y se integraron algunas métricas propuestas atendiendo distintos enfoques para la gestión del conocimiento en las organizaciones del sector público.

**Palabras Clave:** Gestión del conocimiento, Organizaciones del Sector Público, Factores Críticos de Éxito, Estrategias y Medición

# ABSTRACT

Nowadays, Knowledge Management is a strategic component that enables development, growth and continuous improvement of Public Sector Organizations. This thesis is bounded to this specific context. Indeed, we critically and comprehensively study the factors that characterize Knowledge Management strategies and those that foster its development and success in Public Sector Organizations; then finally we propose metrics to measure and evaluate, in order to continuous and systematically improve Knowledge Management practices in Public Organizations.

Main problems are related to the lack of academic literature that explains the elements that Knowledge Management address in the given context. At the same time, we look to address the lack of a thorough analysis on what are the factors that enable Knowledge Management and those who foster its success in Public Organizations. In addition, it is identified as a problem the lack of criteria for measuring and evaluating Knowledge Management in Public Organizations, from a different viewpoint than business perspective.

The contribution of this thesis is that it provides valuable elements for an academic debate on the previous factors, strategies and metrics to promote Knowledge Management initiatives in Public Institutions.

To achieve the research objective of this thesis we performed a comprehensive systematic literature review in order to discover what the Knowledge Management critical success factors are. Also we integrated and proposed some metrics to evaluate and assess the performance and success of Knowledge Management in Public Organizations Subsequently we conducted an in-depth study among different Public Sector Organizations in order to learn what are the critical success factors that must be first considered before implement Knowledge Management initiatives. based on this.

Finally, based on this methodological proposal and as a result of this research, we were able to analyze and explain the critical success factors, we identified some strategies to encourage the success of Knowledge Management and, then finally we integrated a proposal of metrics, from different approaches, for Knowledge Management in Public Sector Organizations.

# TABLA DE CONTENIDO

# Lista de ilustraciones

# Lista de Tablas

# Lista de Acrónimos

**AP:** Administración Pública

**BM**: Banco Mundial

**FCE**: Factores Críticos de Éxito

**GC:** Gestión del Conocimiento

**MPS**: Mejora de Procesos de Software

**OCDE**: Organización para la Cooperación y el Desarrollo Económicos

**ONU**: Organización de las Naciones Unidas

**OP:** Organizaciones Públicas

**OSP:** Organizaciones del Sector Público

**PGC:** Proceso de Gestión del Conocimiento

**RSL:** Revisión Sistemática de la Literatura

**SGC:** Sistema de Gestión del Conocimiento

**TIC:** Tecnologías de la Comunicación y la Información

# CAPÍTULO I

# 1. INTRODUCCIÓN

*«El conocimiento une a cada uno consigo mismo y a todos con todos»,*

José Saramago.

Desde inicios del siglo pasado nuestro mundo ha experimentado distintas y muy diversas reconfiguraciones sociales, que han tenido lugar desde los comienzos de la industrialización, el desarrollo de nuevas tecnologías, la producción intensa de información y la proliferación de las comunicaciones. Lo que ha propiciado, como nunca antes en la historia, distintas transformaciones en todas las dimensiones de la vida en sociedad. Tan sólo la población mundial aumentó en más de 4 billones de habitantes en los últimos 55 años, como se da cuenta en la ilustración 1.1.

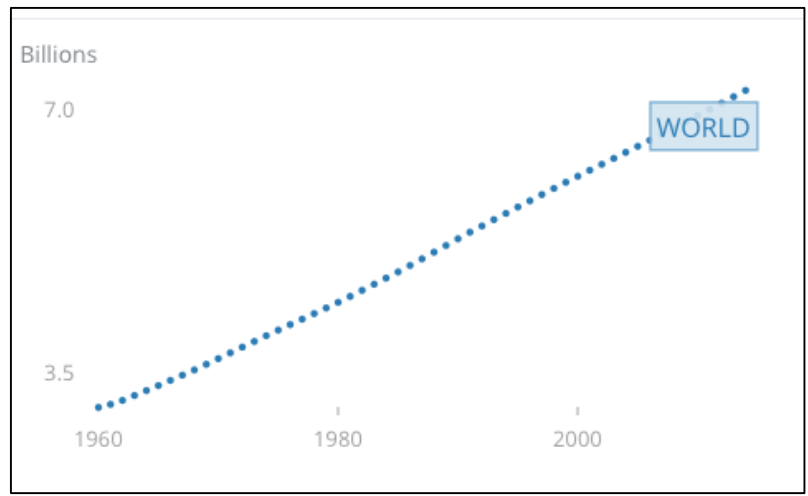

**Ilustración 1-1. Comportamiento en el crecimiento de la población mundial**
Fuente: Banco Mundial (2016)

La denominada y multicitada "Era de la Información", ha incubado distintas mudanzas sociales. Desde comienzos del siglo XX, la producción del conocimiento se ha incrementado sustancialmente en distintos sectores de la sociedad (OECD, 2000). Este incremento ha sido tan sustancial, que desde finales de ese siglo se comenzó a hablar de una Economía del Conocimiento (Powell & Snellman, 2004), que como su nombre lo indica, se caracteriza por el uso del conocimiento como su principal activo e insumo para la producción de más conocimiento.

Así, la Economía del Conocimiento ha ido configurando lo que distintos autores, denominan Sociedades del Conocimiento (Bindé & Matsuura, 2005; Minati, 2012; Pellegrino, 2012; Stehr, Adolf, & Mast, 2013; UCR, 2013; VV AA, 2009), en las que el conocimiento es el recursos más importante, de mayor valor, y que generalmente se caracterizan por ser economías industrialmente avanzadas (Hislop, 2005), porque se auto-actualizan y se renuevan sistemáticamente.

Una sociedad del conocimiento, es una sociedad que se nutre de sus diversidades y capacidades (UNESCO & Bindé, 2005); este tipo de sociedades y economías necesitan organizaciones basadas en el conocimiento (David Rooney, Hearn, & Ninan, 2005). Las economías del conocimiento suponen nuevos paradigmas para la innovación y para el avance en el conocimiento con relación al desarrollo económico (OECD, 2004).

El tema que se pretende estudiar es relevante y pertinente a la luz de su poder transformador en la sociedad actual (OECD, 2004, 2015c). Nuestra época ha sido escenario de tantas transformaciones y cambios radicales considerables que aun cuando se seguía hablando de una tercera revolución industrial (UNESCO & Bindé, 2005, p. 25), encaramos una cuarta revolución industrial, caracterizada por sistemas ciber-físicos, que desafían lo conocido hasta ahora, la forma en que hacemos lo que hacemos y nuestros paradigmas para relacionarnos y para habitar la tierra. En ese sentido, nos encontramos en una era denominada de la híper-conexión (World Economic Forum, 2016), caracterizada por la complejidad, la velocidad y la interdependencia.

Ante tales desafíos, las Organizaciones del Sector Público (OSP) no están ajenas; al contrario, desde los años 80's distintas transformaciones han ocurrido en su interior. La nueva gestión pública (*New Public Management*) (Hood, 1991), trajo consigo una serie de preceptos que transformaron a las organizaciones públicas convencionales, como por ejemplo: la medición, la eficiencia, la efectividad y la mejora continua de estas organizaciones (Bezes et al., 2011; Brown, 2005; Osborne, 2006). En la sociedad del conocimiento las organizaciones públicas o los gobiernos son uno de los mayores consumidores y productores de conocimiento (P. Jain, 2009). Ello, a raíz de una fuerte motivación por la actualización y modernización de las organizaciones públicas.

Indudablemente, el conocimiento es un aliado estratégico de las organizaciones públicas de la actualidad, que les permite convertirse en gobiernos que actúan de forma más eficiente, transparente, sensible a las necesidades de los ciudadanos y eficaz en el logro de sus objetivos (De Angelis, 2013), además de fortalecer la gobernabilidad pública (Puron-Cid, 2014) y, consecuentemente, el desarrollo de la sociedad (Ragab & Arisha, 2013, p. 877), en su conjunto.

Como resultado, en la presente tesis, exploramos el conocimiento como una fuente de ventaja competitiva para las organizaciones públicas de la actualidad. En entornos caracterizados por la velocidad a la cual ocurren las transformaciones y las operaciones, así como la gran cantidad de información y de conocimiento que se produce, se almacena y se utiliza para producir nuevo conocimiento. Por ello, es indudable que se necesitan realizar propuestas que permitan una mejor gestión de este activo intangible, que repercute significativamente en la competitividad, la efectividad y la eficiencia de las organizaciones.

La contemporaneidad, representa un desafío abierto en términos de gestión para posibilitar un mejor desarrollo de las organizaciones en el presente y una base sustentable hacia el futuro. Son tiempos en los que la velocidad, la interconectividad, la interdependencia se anuncian como imperativos naturales de nuestras actividades cotidianas. Es por ello, que la presente tesis quiere contribuir al debate académico, serio y crítico, sobre el conocimiento y la gestión del conocimiento en las organizaciones públicas, pasando por una revisión de los factores críticos de éxito que hacen posible la gestión del conocimiento y formulando propuestas de estrategias para favorecerla y métricas para medir su desempeño.

## 1.1.    Sentencia del problema

A continuación, se exponen detalladamente los principales problemas y dificultades que se presentaron para la realización de la presente tesis.

1. La aún incipiente literatura académica, casos o prácticas documentadas, relacionadas con la gestión del conocimiento en las organizaciones públicas (Dumay, Guthrie, & Puntillo, 2015; Massaro, Dumay, & Garlatti, 2015);
2. La falta de consenso metodológico para implementar, evaluar y analizar el impacto de la gestión del conocimiento (C. S. Lee & Wong, 2015), y la falta de consciencia sobre los beneficios que tiene ésta en las organizaciones públicas (Chawla & Joshi, 2010);
3. Existen aún múltiples riesgos asociados con la gestión del conocimiento en las organizaciones públicas, sobre todo en temas relativos a seguridad y privacidad de los funcionarios, confidencialidad de la información (Yang & Maxwell, 2011), imprecisiones, inconsistencias o estado incompleto de la información, incompatibilidad o complejidad tecnológica, presupuestos anuales y relaciones intergubernamentales, tal como señala Weerakkody et al. (2013).

## 1.2. Motivación

La existencia de una literatura académica más robusta y empíricamente validada en relación con la gestión del conocimiento en las organizaciones públicas podría permitir analizar y proponer distintos modelos de implementación y evaluación de estas iniciativas en el sector público. Así, podrían verse beneficiadas las organizaciones públicas que pretendan implementar estrategias e iniciativas de gestión del conocimiento en sus actividades cotidianas, porque contarían con marcos o modelos de gestión probados y validados en otras organizaciones.

Por esta razón, estudiar la gestión del conocimiento en las organizaciones públicas se vuelve interesante y necesario en los tiempos actuales. Para llevar a cabo el presente trabajo de tesis, se ha estudiado la literatura más relevante con relación a la gestión del conocimiento en organizaciones públicas, además se ha analizado, mediante un instrumento de evaluación, la presencia de factores críticos para la gestión del conocimiento en las instituciones públicas. Al finalizar este trabajo, se proponen algunas estrategias para favorecer y fortalecer la gestión del conocimiento, así como algunas métricas para evaluar el desempeño del conocimiento en las instituciones públicas.

Finalmente, también contar con mayor periodo de tiempo y con las condiciones necesarias que permitan comparar geográficamente el estudio realizado con otros países podría dar cuenta de variables o elementos que no han sido observados hasta el momento, pero que también tienen relevancia en el estudio de la gestión del conocimiento en las organizaciones públicas. Variables que, por su naturaleza, obedecen al entorno externo de las organizaciones.

## 1.3.    Objetivos y preguntas de investigación

*Objetivos de investigación*

- [O1] Analizar y describir los factores previos que favorecen las prácticas de gestión del conocimiento en las organizaciones públicas mediante un estudio en distintas instituciones.
- [O2] Definir un modelo de evaluación de factores culturales, estratégicos y de infraestructura para obtener un diagnóstico previo a la implementación de iniciativas de gestión del conocimiento en las organizaciones públicas.
- [O3] Ejecutar y discutir la implementación del modelo de evaluación (previamente mencionado) para conocer la contribución (influencia) de los constructos de cultura, infraestructura y estrategia sobre la gestión del conocimiento.

*Preguntas de investigación*

- P1 ¿Cuáles son los factores que favorecen la gestión del conocimiento en las organizaciones públicas?
- P2: ¿Qué estrategias podrían favorecer el desarrollo de las iniciativas de gestión del conocimiento en las organizaciones públicas?
- P3: ¿Cuáles podrían ser las métricas para evaluar el desempeño de la gestión del conocimiento en las organizaciones públicas?
- P4: ¿Cuál es la influencia de los factores culturales, estratégicos y de infraestructura sobre la gestión del conocimiento en las organizaciones públicas?

## 1.4.    Resumen de la propuesta

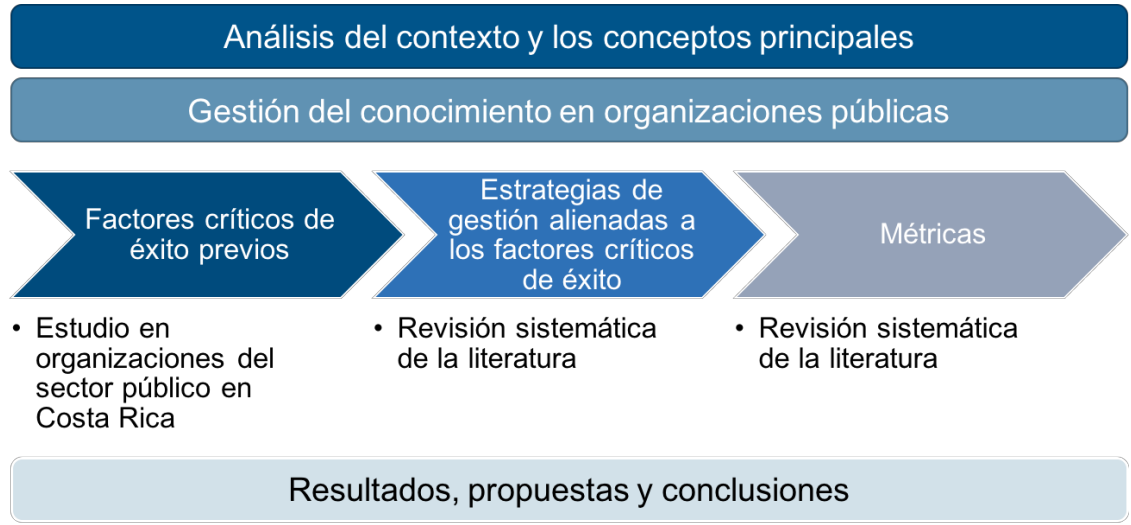

**Ilustración 1-2. Descripción gráfica de la tesis**
Elaboración propia (2016)

A continuación, en la ilustración 1.2., se presenta de manera gráfica la propuesta conceptual a través de la cual se desarrolla la presente tesis. En un primer momento se realiza un análisis del contexto y los principales conceptos relacionados con el objeto de estudio, siendo estos la Gestión del Conocimiento (GC) y las Organizaciones Públicas (OP). Enseguida, se analiza particularmente el fenómeno de la GC en las OP, a través de experiencias previas descritas en trabajos académicos o en experiencias documentadas. Posteriormente, derivado de los resultados de una encuesta realizada entre Organizaciones del Sector Público (OSP), se analizan cuáles son los Factores Críticos de Éxito (FCE) previos a la implementación de iniciativas de GC en OSP; asimismo, derivado la Revisión Sistemática de la Literatura (RSL), se plantean algunas estrategias que podrían favorecer el éxito de estas iniciativas alineadas con los factores previos descritos en el estudio realizado en primera instancia. Finalmente, se proponen algunas métricas para la GC, basadas en la RSL.

En la última sección de la tesis, se ofrecen al lector los principales resultados de los métodos empleados y, de igual manera, se realizan algunas propuestas que pueden contribuir al debate académico sobre el tema y las conclusiones derivadas del trabajo realizado.

## 1.5.    Estructura de tesis

En el Capítulo I. En este capítulo se realiza un resumen general del contenido de la tesis, asimismo se integran los elementos esenciales que configuran este trabajo de investigación, se plantea el objetivo general y los particulares, se describe la propuesta de análisis conceptual y se proyecta el contexto en el cual se circunscribe este trabajo exponiendo para ello las motivaciones para su realización. Asimismo, se explica la estructura general y las preguntas de investigación.

En el Capítulo II. Se presenta la construcción del marco teórico de este trabajo de investigación a partir de una exhaustiva revisión sistemática de la literatura académica, que da sustento metodológico al objeto de estudio en el contexto específico. Se analiza en esta sección los distintos conceptos de gestión del conocimiento (GC), sus elementos asociados, los factores críticos de éxito (FCE), los sistemas de gestión del conocimiento (SGC), el proceso de gestión del conocimiento (PGC), así como las barreras, los riesgos y se da cuenta de la importancia que ha ido ganando el conocimiento dentro de la economía del conocimiento y la sociedad actual.

En el Capítulo III, se presente la propuesta metodológica para la realización de este estudio. Se describen los elementos y la justificación para realizarlo de esa manera. Asimismo, se describen los pasos seguidos para lograr el objetivo de la presente tesis y responder a las preguntas de investigación formuladas. Finalmente, derivado de los resultados obtenidos y a partir de la propuesta metodológica, se han construido algunas propuestas de estrategias y se proponen métricas para la GC en las OP. En esta sección, se presentan los métodos utilizados para analizar y comprender el objeto de estudio.

En el Capítulo IV, se integran las discusiones producto de los resultados obtenidos mediante la propuesta metodológica. En esta sección se presentan algunos elementos que podrían contribuir al debate académico. Finalmente, se ofrecen al lector las conclusiones, limitaciones del estudio y futuras líneas de investigación de cara a la continuidad en el análisis del objeto de estudio.

# CAPÍTULO I

# 2. MARCO TEORICO

Para el desarrollo de esta tesis tomaremos algunos conceptos que son claves para aterrizar esta propuesta de investigación, tal como se da cuenta en la ilustración 2.1. Por una parte, se analiza el conocimiento, quizá el activo intangible más importante de este siglo (Davenport & Prusak, 1998a; Polanyi, 1962; Spender, 1996), y sus múltiples manifestaciones y definiciones (Scarso & Bolisani, 2015); sus Factores Críticos de Éxito (FCE) (Kuan Yew Wong, 2005); y las estrategias para incentivar la GC (Mitre-Hernández, Mora-Soto, López-Portillo, & Lara-Alvarez, 2015; Suwannathat, Decharin, & Somboonsavatdee, 2015); y por otra parte se proponen métricas para evaluar el desempeño de la GC y su contribución organizacional (Kuah & Wong, 2011; Ragab & Arisha, 2013).

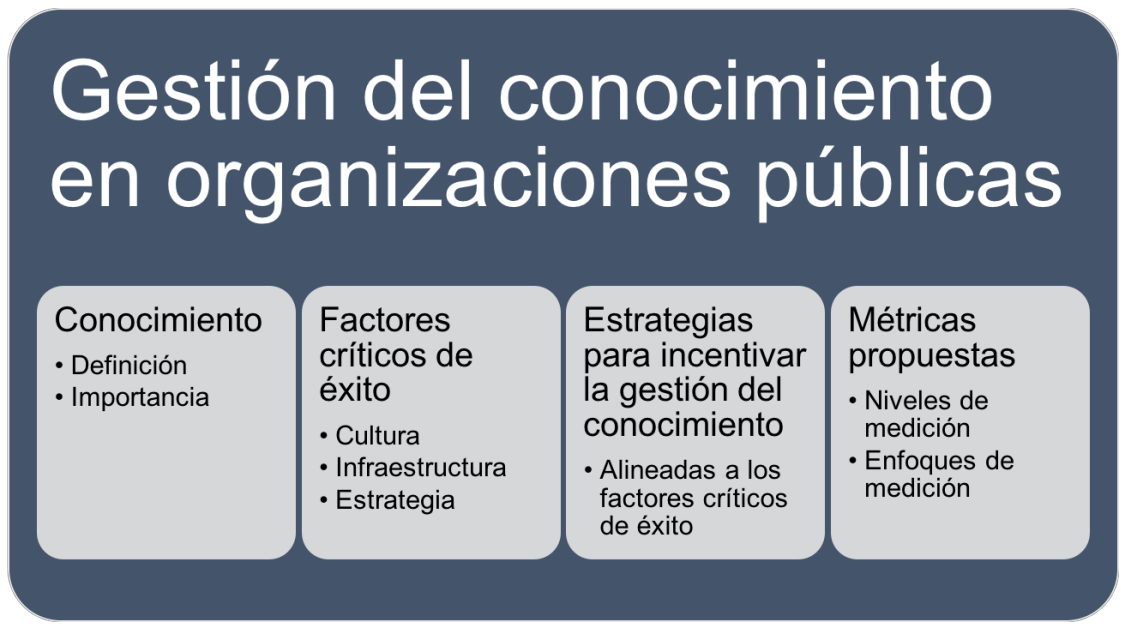

**Ilustración 2-1. Estructura del marco teórico**
Elaboración propia (2016)

## 2.1. Tendencias de desarrollo del conocimiento y la información en el mundo

El mundo en el que vivimos está en constante evolución y cambio. La innovación, hoy en día es una fuente crucial de competitividad, de desarrollo económico y de transformación de la sociedad moderna (OECD, 2004, p. 7), imperativo para el desarrollo del mundo actual (Cornell University, INSEAD, & WIPO, 2015; OECD, 2015c), elemento clave del crecimiento económico y del desarrollo, que permite a las naciones alcanzar estándares de vida más altos (OECD, 2015c).

La producción de nuevo conocimiento, las innovaciones en todos los sectores y los enérgicos cambios en las tendencias de crecimiento en el mundo, hacen de éste un mundo complejo y desafiante. Vivimos, cada vez más, en una sociedad red (Castells & Cardoso, 2005), cuya característica esencial es la interconexión y la interdependencia.

Hoy en día, se estima que el 46.4% de la población mundial utiliza internet (Miniwatts Marketing Group, 2016), esto equivale a más de 3.3 billones de usuarios y a una producción de casi 1.5 millones de Gigabytes por minuto (BGR Media, 2016). De hecho, la empresa IBM habla de que tan sólo en los últimos dos años se ha producido el 90% de los datos en el mundo (IBM, 2016). Para el caso de México, se estima que son 60 millones de usuarios de internet, lo que equivale a un 73.2% de la población. Innegablemente nos encontramos en la era de la híper-conectividad, en nuestro país existen 82.2 líneas de teléfono celular por cada 100 habitantes, tal como se aprecia en la ilustración 2.2. Los usuarios de internet mexicanos, refieren que lo usan principalmente para obtener información y para consultar sus redes sociales (INEGI, 2015).

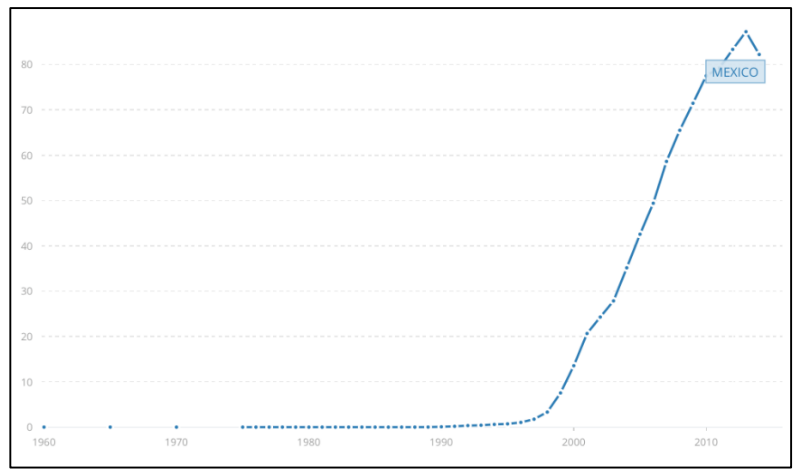

**Ilustración 2-2. Líneas celulares por cada 100 habitantes**
Fuente: Banco Mundial (2016)

En la actualidad, hablar de ciencia, tecnología e innovación resulta común y hasta ordinario. Nos hemos, al menos en los últimos años, acostumbrado a estos conceptos de alta plasticidad semántica, de definición multívoca, de la cual incluso se ha abusado en sus referencias, pervirtiendo los conceptos hasta volverles traslaticios. La producción científica, por su parte, ha crecido considerablemente, la ilustración 2.3, muestra el crecimiento en el número de artículos científicos publicados en las áreas de física, biología, química, matemática, medicina clínica, investigación biomédica, ingeniería y tecnología, y ciencias de la tierra y el espacio, desde el año 1985 al año 2015.

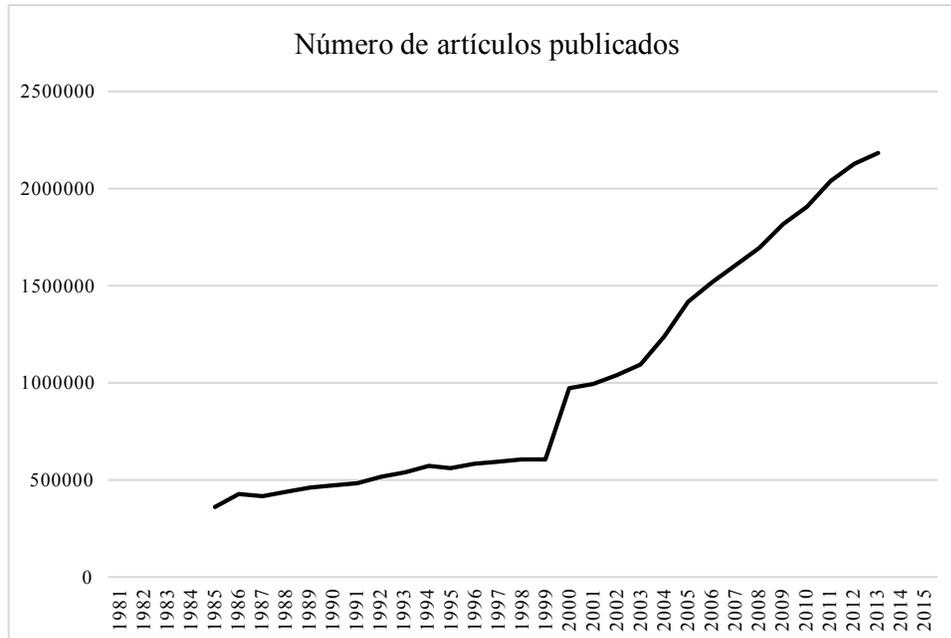

**Ilustración 2-3. Número de artículos científicos publicados en ciencias.**
Fuente: Banco Mundial (2016)

De la misma manera, el número de investigadores de tiempo completo ha ido en aumento conforme el paso de los años. En algunos países, unos con un mayor auge que en otros, pero con una pronunciada tendencia al alza, tal como se puede observar en la ilustración 2.4.

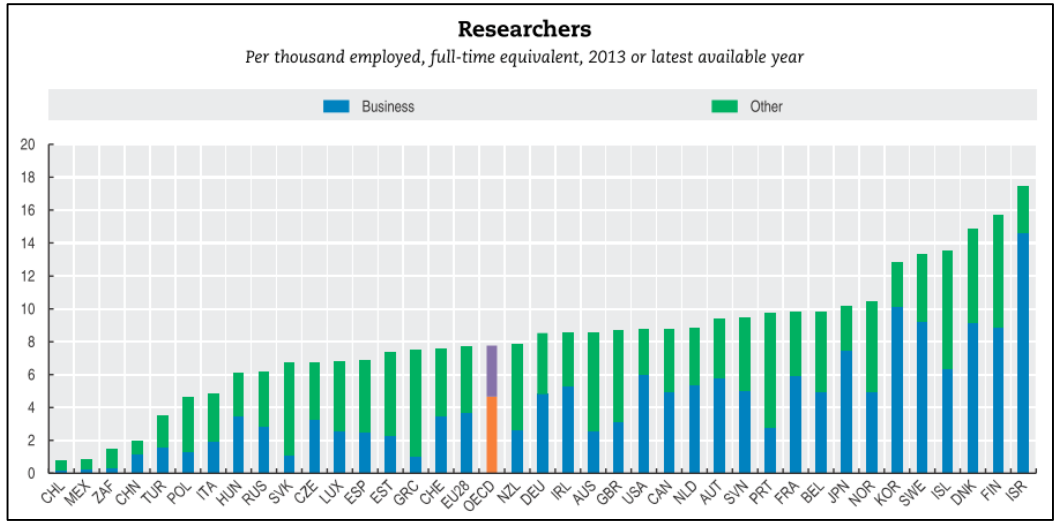

**Ilustración 2-4. Número de investigadores de tiempo completo**
Fuente: OCDE Factbook (OECD, 2016)

Sin embargo, aún hoy, en nuestro país nos encontramos con grandes retos, como la inversión en investigación y desarrollo con respecto al Producto Interno Bruto (OECD, 2016), tal como se puede ver en la ilustración 2.5. Para el caso particular de México, la segunda mayor economía de América Latina, éste ha crecido, de acuerdo con cifras del Banco Mundial (2016) 3.5% en términos reales durante los últimos cuatro años.

Pero, por otra parte, también enfrentamos grandes desafíos en cuanto a la modernización de las instituciones, el establecimiento de políticas que favorezcan la innovación para afrontar los grandes retos sociales, incrementar la vinculación entre la ciencia y la industria, fortalecer la capacidad e infraestructura en Investigación y Desarrollo, e incrementar las capacidades de los recursos humanos para elevar el desarrollo tecnológico y científico (OCDE, 2014, pp. 52–55), así como incrementar el gasto en educación superior, que de acuerdo con la ilustración 2.6, es uno de los más bajos en comparación con otros países de la Organización para la Cooperación y Desarrollo Económico (OCDE).

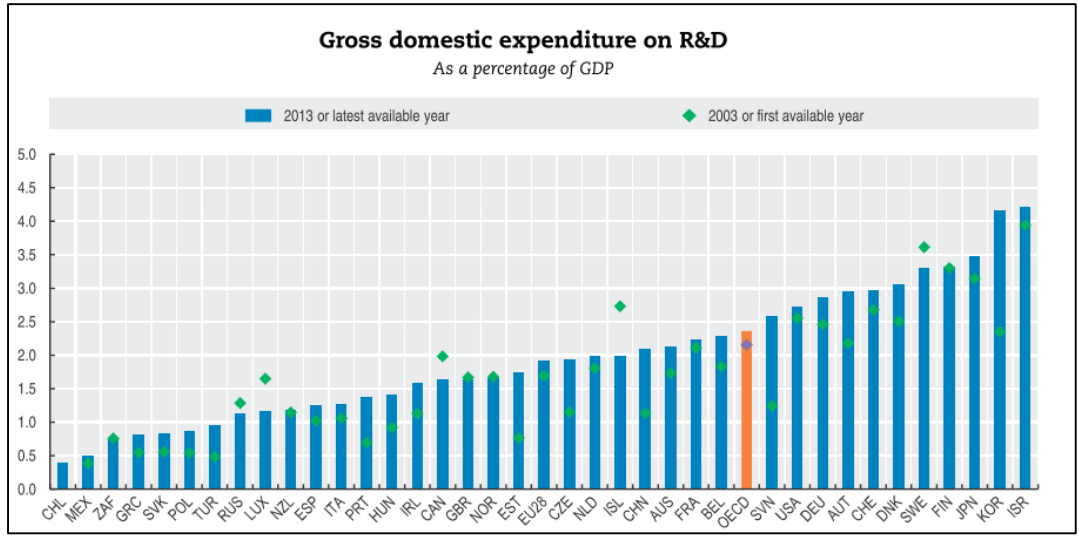

**Ilustración 2-5. Gasto en Investigación y Desarrollo con respecto al Producto Interno Bruto**
Fuente: OCDE Factbook (OECD, 2016)

Desde mediados de la década de los 80, el gasto medio de los países que integran a la OCDE en investigación básica ha aumentado más rápido que el gasto en investigación aplicada y desarrollo experimental, reflejo del énfasis que muchos gobiernos dan al financiamiento de la investigación científica. La investigación básica sigue muy concentrada en las universidades y en los centros de investigación públicos. Un porcentaje considerable de la I+D que se realiza en esas instituciones se dedica al desarrollo en Corea (35%) y en China (43%). En términos generales, en 2013 China invirtió relativamente poco (4%) en investigación básica en comparación con la mayoría de las economías de la OCDE (17%) y su gasto en I+D sigue muy orientado al desarrollo de infraestructura para ciencia y tecnología; es decir, inmuebles y equipos (OECD, 2015b).

Una nueva generación de Tecnologías de la Información y la Comunicación (TIC), como las relacionadas con el Internet de las Cosas, los datos masivos y la computación cuántica, más una ola de invenciones en salud y materiales avanzados están sentando las bases para transformaciones profundas en la forma en que vamos a trabajar y vivir en el futuro. Entre los años 2010 a 2012, Estados Unidos, Japón y Corea lideraron la invención en esos campos (representando en conjunto más del 65% de las familias de patentes solicitadas en Europa y Estados Unidos) seguidos por Alemania, Francia y China (OECD, 2015b).

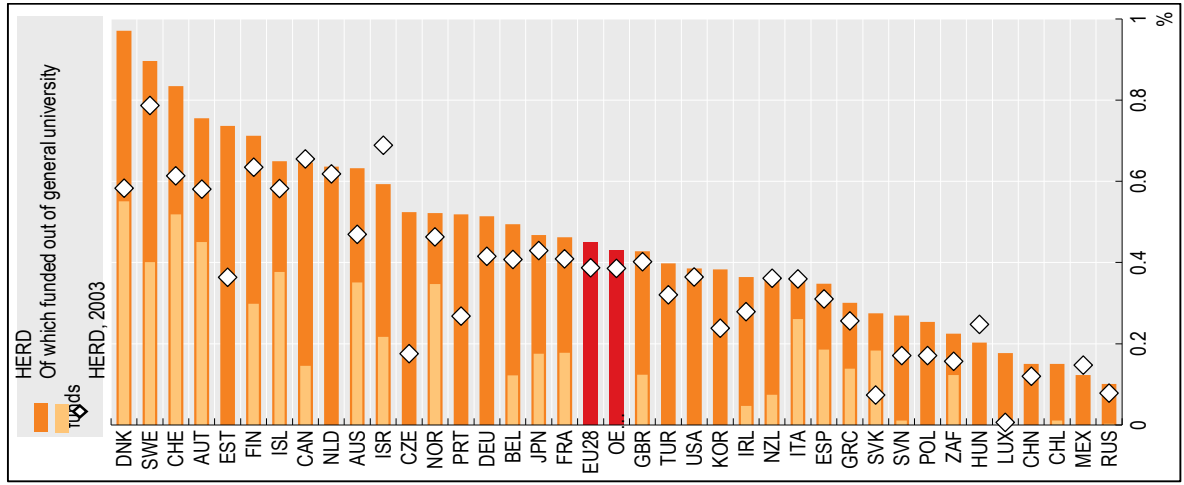

**Ilustración 2-6. Gasto en Educación Superior**

Fuente: OCDE Factbook (OECD, 2016)

## 2.2. La sociedad, el conocimiento y las instituciones públicas

### 2.2.1. El conocimiento

Desde las postrimerías del siglo XX, la producción del conocimiento ha incrementado considerablemente. El conocimiento un recurso fundamental. Es por excelencia el activo más importante de la denominada, por antonomasia, economía del conocimiento (OECD, 2004; Powell & Snellman, 2004; David Rooney et al., 2005). Incluso, antes en el siglo XVI, con la, ya por excelencia y multicitada, frase "*el conocimiento es poder*", atribuida al célebre filósofo inglés del Trinity College, Sir Francis Bacon, podemos observar como este recurso esencial se ha ido consumando como un elemento transformador del entorno.

No tan lejos de aquellos años, en nuestro tiempo, el conocimiento es un recurso fundamental y estratégico de las organizaciones (Wilcox King & Zeithaml, 2003), generalmente relacionado con otro recurso de igual importancia: el tiempo (Ragab & Arisha, 2013). Por tal motivo, es quizá el activo más importante del siglo XXI (Tianyong Zhang, 2010, p. 572). El conocimiento engloba las ideas guardadas en la mente, realidades, conceptos, datos y técnicas de la memoria humana (Nonaka & Takeuchi, 1999; Nonaka, 1991). Su fuente es la mente humana, y se basa en la información que se obtiene a través de la experiencia, las creencias y los valores personales. Su transformación ocurre cuando se asocia con decisiones o acciones (M. Allameh, Zamani, & Davoodi, 2011, p. 1227).

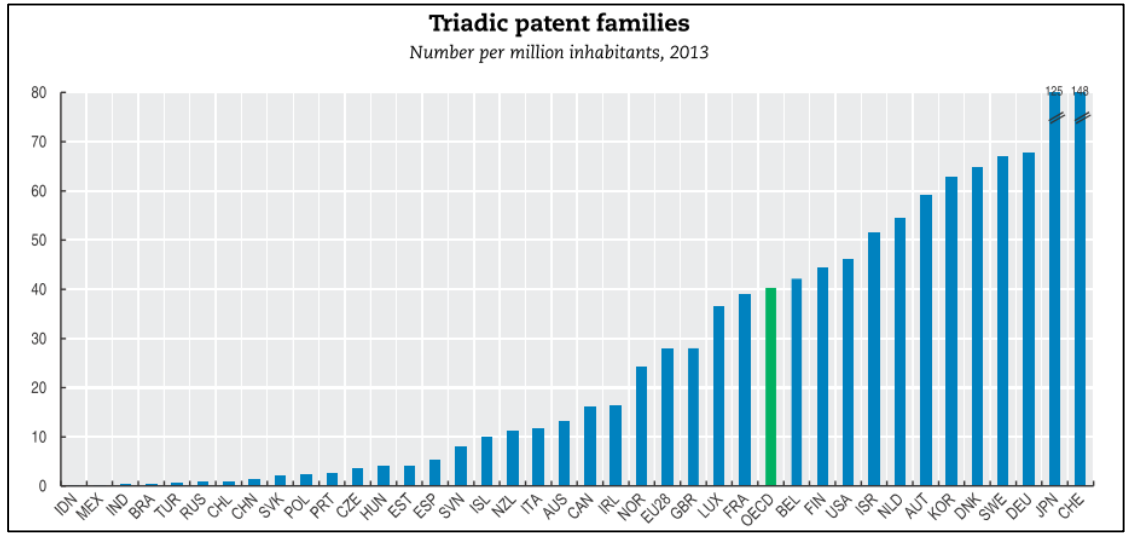

**Ilustración 2-7. Patentes registradas en las tres principales oficinas de registro de patentes: Europa, Japón y Estados Unidos**
Fuente: OCDE Factbook (OECD, 2016)

Definir el conocimiento en ocasiones resulta complicado, debido a su naturaleza intangible, la subjetividad y, algunas veces, por el eclecticismo que envuelve el campo de la administración (Alsadhan, Zairi, & Keoy, 2008). Entonces, diremos que el conocimiento es "un proceso humano dinámico de justificación de la creencia personal en busca de la verdad" (Nonaka & Takeuchi, 1999, p. 63), un recurso crucial para el funcionamiento, la innovación, el desempeño y la competitividad de las organizaciones (C.W. Holsapple & Joshi, 2001). La creación de conocimiento organizacional es un proceso interminable que se actualiza a sí mismo continuamente (Nonaka & Takeuchi, 1999) y convertirse en una organización basada en el conocimiento es un imperativo para el éxito de las organizaciones hoy en día (Bose & Ranjit, 2004). El conocimiento es la moneda en circulación de la economía actual (Ragab & Arisha, 2013, p. 873).

Dentro de la economía del conocimiento (Powell & Snellman, 2004), este activo intangible tiene un papel importante dentro de las economías nacionales para mantener el crecimiento económico sostenido y para crear, ganar y mantener una ventaja competitiva (Choy Chong, Salleh, Noh Syed Ahmad, & Syed Omar Sharifuddin, 2011, p. 497; Ragab & Arisha, 2013, p. 873).

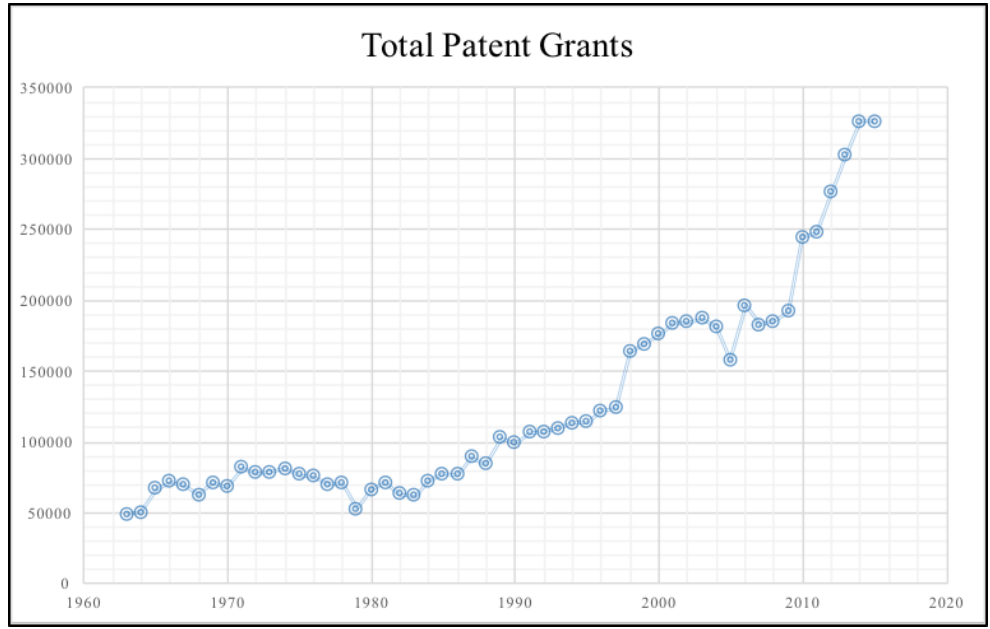

**Ilustración 2-8. Número de patentes concedidas por la USPTO**
Fuente: (U.S. PATENT AND TRADEMARK OFFICE, 2016).

En consecuencia, gestionar el conocimiento se ha vuelto cada vez más necesario en las organizaciones, para que éstas "conozcan lo que saben" (Davenport & Prusak, 1998b; Kogut & Zander, 1992) y como una fuente de ventaja competitiva en entornos altamente cambiantes (Coff, 2003). Por esa razón, aparece la gestión del conocimiento (GC), que puede ser definida, como el arte de crear valor con los activos intangibles de una organización (Sarvary, 1999), y observada como un proceso que busca optimizar la aplicación efectiva del capital intelectual para alcanzar los objetivos organizacionales (Tocan, 2012), a través de la integración de todas las unidades de la organización para identificar y compartir los conocimientos que se producen y se acumulan (Mojibi, Khojasteh, & Khojasteh-Ghamari, 2015), con el objetivo de mejorar el manejo sistemático del conocimiento dentro de la organización (Heisig, 2009).

Como se ha referido, la producción mundial de conocimiento ha incrementado con gran impulso en los últimos años. El número de patentes concedidas, por la United States Patents Office (USPTO), desde los años 60's a la fecha, ha ido creciendo exponencialmente. Tal como se puede observar en la ilustración 2.8.

## 2.3.    Gestión del Conocimiento

La Gestión del Conocimiento (GC) es un fenómeno relativamente nuevo que aparece en la esfera de las Organizaciones del Sector Público (OSP) trayendo a la vez, nuevos paradigmas de gestión organizacional, retos, riesgos y oportunidades para su implementación, desarrollo y evaluación en las instituciones públicas. Este enfoque de gestión se ha desarrollado paralelamente con la evolución y el uso de las tecnologías de la información y la comunicación (TIC). En esta tesis se analiza a la GC con el objetivo de conocer los factores y sus componentes asociados que favorecen su implementación, y los beneficios percibidos con respecto a ésta. Este estudio busca contribuir al debate teórico sobre los factores previos para implementar iniciativas de GC entre las OSP.

Por su parte, en la literatura académica se enuncian distintos objetivos de la GC, como nivelar el conocimiento en la organización, para crear una ventaja competitiva que permite identificar recursos críticos y áreas con conocimiento relevante con el objetivo de *saber lo que una organización sabe* y qué están haciendo bien y por qué (Garlatti, Massaro, Dumay, & Zanin, 2014, p. 176). Particularmente, para el caso de las OSP la GC puede mejorar la efectividad y viabilidad del dominio del conocimiento en las Organizaciones Públicas (OP) (Mbhalati, 2014), para mejorar sistemáticamente el manejo del conocimiento real y potencial dentro de las organizaciones (Heisig, 2009, p. 5)".

De acuerdo con Tsui et al. (2009, p. 986), el propósito de la GC es "*proveer acceso en línea y en tiempo real al conocimiento de la organización a la organización y a sus clientes, que actúan como facilitadores y catalizadores para la aplicación innovadora de las agencias gubernamentales* (H. Tsui et al., 2009, p. 988)". Por tanto, la GC es vista como una solución efectiva que puede apoyar las actividades administrativas y la modernización de las agencias gubernamentales.

Es ampliamente reconocido que la GC integra a las personas, la tecnología, los procesos y la estructura de una organización (Tianyong Zhang, 2010), consiste en "llevar el conocimiento correcto a las personas adecuadas en el momento que lo necesitan con el fin de tomar una acción concertada" (Murray E Jennex & Smolnik, 2011, p. 75). Por tanto, puede ser vista como "un esfuerzo sistemático y deliberado para coordinar a las personas, la tecnología y la estructura de una organización y su ambiente a través de la reutilización del conocimiento y la innovación" (Tianyong Zhang, 2010, p. 572). Su éxito se materializa cuando el conocimiento es reutilizado para mejorar el desempeño de la organización (M.E. Jennex & Olfman, 2004).

Además, la GC puede ser considerada como "*un conjunto de prácticas destinadas a la interacción entre el conocimiento tácito y explícito para adquirir y crear nuevas competencias (conocimientos + habilidades + actitudes), que permitan a una organización actuar inteligentemente en diferentes entornos*" (De Angelis, 2013, p. 808) y como una competencia esencial (Chong Siong Choy, Yew, & Lin, 2006; de Sordi & Carvalho Azevedo, 2008; Hafeez & Abdelmeguid, 2003) que deben desarrollar las organizaciones de la actualidad. De este modo, la GC es vital no sólo para el éxito de las organizaciones, sino también para desarrollo de la sociedad en general (Ragab & Arisha, 2013, p. 877).

En los últimos años, las investigaciones y prácticas relacionadas con la GC han explorado distintos y diversos aspectos relacionados con este tema, desde el Proceso de Gestión Del Conocimiento (PGC) (S. M. Allameh, Zare, & Davoodi, 2011; Chang Lee, Lee, & Kang, 2005; Ding, Liang, Tang, & van Vliet, 2014; Ho, Hsieh, & Hung, 2014; Kuah, Wong, & Wong, 2012); los Sistemas de Gestión del Conocimiento (SGC) (Chang Lee et al., 2005; Gourova & Toteva, 2014; Liao, 2003; Ragab & Arisha, 2013; Savvas & Bassiliades, 2009); y los factores que favorecen la GC (S. M. Allameh et al., 2011; Alsadhan et al., 2008; C. S. Choy, 2006; C.W. Holsapple & Joshi, 2001; Liberona & Ruiz, 2013; Mas-Machuca & Martínez Costa, 2012; Loo Geok Pee & Kankanhalli, 2008; Pinho, Rego, & Cunha, 2012; Kuan Yew Wong & Aspinwall, 2005).

Adicionalmente, se han analizado los beneficios (Alsqour & Owoc, 2015; Edvardsson & Durst, 2013; P. R. Massingham & Massingham, 2014; Weerakkody et al., 2013; Yahyapour, Shamizanjani, & Mosakhani, 2015); los riesgos (Gil-García & Pardo, 2005; Weerakkody et al., 2013; Yang & Maxwell, 2011); las barreras (Akhavan, Reza Zahedi, & Hosein Hosein, 2014; C. Lin, Wu, & Yen, 2012; Marouf & Khalil, 2015; Singh & Kant, 2007); los resultados (Chong Siong Choy et al., 2006; H.-F. Lin, 2015; P. R. Massingham & Massingham, 2014; Migdadi, 2009); y las tendencias futuros para la GC (Asrar-ul-Haq, Anwar, & Nisar, 2016; Tsai, 2013; E. Tsui, 2005; K. Y. Wong, Tan, Lee, & Wong, 2015).

Finalmente, es importante puntualizar que la GC es diferente del gobierno electrónico o *e-government*. Así, mientras que el gobierno electrónico se centra en el Estado y su gente (usuarios) y en el uso de la tecnología, la GC se centra más en la organización (pública y privada) y sus empleados (trabajadores de conocimiento) y no se trata solamente acerca de la tecnología, sino que integra, como se ha dicho, procesos, personas y otros factores dentro de una organización (Concha, Astudillo, Porrúa, & Pimenta, 2012; Mbhalati, 2014). Por lo tanto, podemos decir que bajo esta perspectiva el gobierno electrónico está circunscrito dentro de la GC.

### 2.3.1. Contexto internacional sobre la Gestión del Conocimiento

A nivel internacional el acceso a las Tecnologías de la Información y la Comunicación (TIC) aún supone grandes retos. La denominada "brecha digital" todavía existe y algunas personas aún no tienen acceso a las redes de comunicación móviles, mientras que la adopción de la banda ancha alcanza su nivel de madurez en los países desarrollados, todavía no se ha materializado en los países menos adelantados. Así, aproximadamente 4,300 millones de personas en todo el mundo aún no utilizan Internet y el 90% de ellos viven en países en desarrollo (UIT, 2014).

La Unión Internacional de Telecomunicaciones (UIT), observa como actualmente las entidades públicas y los gobiernos son los principales usuarios de las TIC y como están utilizando cada vez más Internet para prestar servicios a los ciudadanos (UIT, 2014, p. 7). Contrariamente a lo que se podría suponer, el ingreso nacional de los países ciertamente no constituye ni garantiza el desarrollo y los avances en materia de administración pública electrónica (United Nations, 2014b, p. 19). Esta tendencia se va revirtiendo poco a poco.

La Organización Mundial de las Naciones Unidas (ONU) observa como principal factor para el éxito de la administración pública electrónica al desarrollo de marcos de gobernanza efectivos que apoyan y gestionan modelos de prestación de servicios centrados en los ciudadanos, políticas nacionales de TIC y una estrategia clara sobre la administración pública electrónica, así como el fortalecimiento de las instituciones y el desarrollo de capacidades en los servidores públicos (United Nations, 2014b).

Por otra parte, Mbhalati (2014), observa como los países con ingresos más altos están mejor posicionados en los rankings de gobierno electrónico a nivel mundial. Esto, el autor, lo relaciona con los requerimientos necesarios de inversión en infraestructura de telecomunicaciones, educación de funcionarios y servicios en línea. Aunque, como se ha señalado esta tendencia parece que se ha ido revirtiendo en los últimos años, dada la proliferación y el uso de TIC más eficientes y de menor costo (United Nations, 2014b).

En este contexto, se propician algunas condiciones (acceso a información, realización de transacciones y transparencia en la gestión) que son necesarias para iniciar con el proceso de GC. Al respecto, Peluffo y Catalán Contreras (2002), indican que el antecedente de la GC es la "Economía basada en el Conocimiento y el Aprendizaje (EBCA)", que sustenta las bases de una nueva economía basada en la información, que como se ha venido indicado, resulta en un activo intangible que permite generar ventajas competitivas.

Por lo tanto, desde un punto de vista evolutivo, la GC pasa también por un proceso de madurez en las organizaciones, públicas y privadas, que se visualiza en la ilustración 2.9, sobre las etapas de maduración de la GC, propuestas por Peluffo y Catalán Contreras (2002, pp. 34–35):

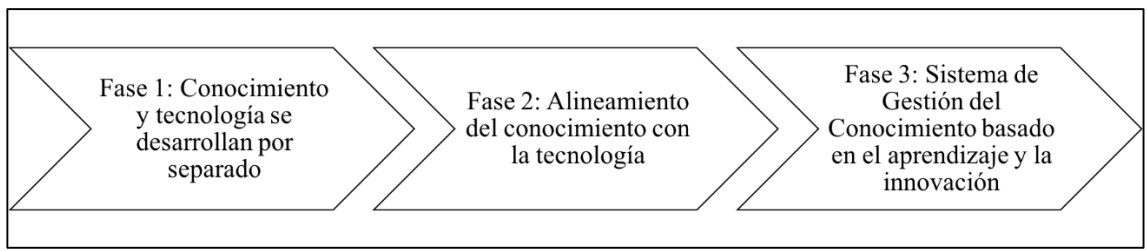

**Ilustración 2-9. Etapas de maduración de la Gestión del Conocimiento**
Elaboración propia, basado en Peluffo y Catalán Contreras (2002).

Para el caso de América Latina, Criado y Gil-García (2013), reflexionan sobre el por qué es importante el conocimiento para las OSP, teniendo en cuenta las potencialidades para mejorar la gestión gubernamental y las políticas públicas, además reconocen que aún hay algunos retos de cara hacia el futuro como la inclusión digital, las redes sociales y el gobierno 2.0, el gobierno abierto y la transparencia administrativa, el intercambio de información, la interoperabilidad y la computación en la nube. Así como, los datos masivos (*big data*) y el modelado de políticas como resultado de la participación virtual de los ciudadanos. Además de las ciudades inteligentes, los gobiernos electrónicos locales y el gobierno móvil.

### 2.3.2. Sistemas de Gestión del Conocimiento

Del mismo modo, para llevar a cabo las actividades relacionadas con la GC, comúnmente las organizaciones, se apoyan en Sistemas de Gestión del Conocimiento (SGC) (Gourova & Toteva, 2014), que son "sistemas de información especializados, que usan tecnologías modernas (por ejemplo, internet, intranets, navegadores, almacenes de datos y agentes de software) para sistematizar, facilitar y agilizar la GC en toda la organización" (Chang Lee et al., 2005, p. 471). Los SGC usan las TIC para integrar a los recursos gerenciales, técnicos y organizativos (Ragab & Arisha, 2013) de las instituciones.

### 2.3.3. Proceso de Gestión del Conocimiento

Estos sistemas se alinean al Proceso de Gestión del Conocimiento (PGC) (Chang Lee et al., 2005), que se refiere a la transformación del conocimiento implícito, fragmentado y privado de los individuos o grupos, tanto dentro como fuera de la organización en los activos intelectuales valiosos para la organización (Ho et al., 2014, p. 736). Este proceso generalmente se integra por actividades como la (i) creación, (ii) almacenamiento, (iii) intercambio, (iv) utilización e (v) internalización del conocimiento.

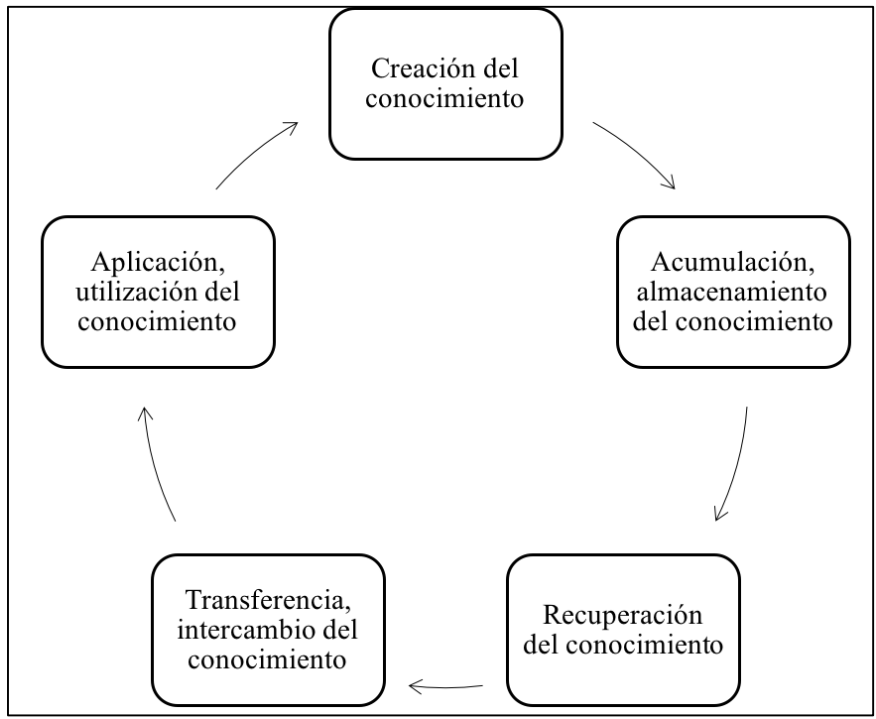

**Ilustración 2-10. Proceso de Gestión del Conocimiento**
Elaboración propia, a partir de Ding et al., (2014) y Lee et al., (2005)

De esta manera la GC se realiza, y puede ser observada en las organizaciones cuando: (1) el conocimiento que se produce es reutilizado constantemente para producir nuevo conocimiento (Ding et al., 2014); (2) cuando el conocimiento es utilizado racionalmente para tomar decisiones (; Revere et al., 2007); y (3) cuando el conocimiento que se produce es útil para otras OP (Ding et al., 2014; Sarantis, Charalabidis, & Askounis, 2011). Consecuentemente, en la ilustración 2.10, se muestra el PGC propuesto por Ding et al., (2014).

### 2.3.4. Barreras a la Gestión del Conocimiento

Por otra parte, existen barreras, "obstáculos individuales, socio-organizaciones, o tecnológicos que pueden impedir la GC en las organizaciones" (Pinho et al., 2012, p. 215). Estos obstáculos limitan la implementación, el desarrollo y el éxito de las iniciativas de la GC. Para comodidad del lector, en la tabla 2.1, se han agrupado las barreras observadas en la literatura académica, de acuerdo al factor de análisis (cultura, infraestructura y estrategia) con el que comúnmente se asocian.

| Cultura | Infraestructura | Estrategia |
|---|---|---|
| • Colaboración deficiente y confianza débil para compartir el conocimiento (Ho et al., 2014), <br> • No existe una cultura que fomente la GC en la organización (Singh & Kant, 2007), <br> • Resistencias de los individuos hacia los cambios que conlleva la GC (Liberona & Ruiz, 2013) | • Poca preparación de las instituciones públicas para el cambio tecnológico (Ferro, Helbig, & Gil-Garcia, 2011), <br> • La denominada brecha digital (OECD, 2003b), <br> • La falta de tecnologías apropiadas (S. M. Allameh et al., 2011), <br> • Bajos niveles de alfabetización en general (Ferro et al., 2011), <br> • La falta de niveles adecuados de infraestructura tecnológica, experiencia, compromiso y recursos, especialmente en países en desarrollo (United Nations, 2007, 2014b) | • Barreras legislativas y reglamentarias, marcos presupuestarios que restringen estas iniciativas (OECD, 2003b), <br> • La falta de liderazgo e involucramiento de la alta dirección con las iniciativas de GC (Anantatmula & Kanungo, 2007; Singh & Kant, 2007), <br> • Ausencia de voluntad política (Mbhalati, 2014) para iniciar los cambios, <br> • No hay claridad respecto de qué es la GC, falta de tiempo y apoyo de los directivos (Liberona & Ruiz, 2013) |

**Tabla 2.1. Principales Barreras para la Gestión del Conocimiento**
Elaboración propia (2016)

### 2.3.5. Factores críticos de éxito para la Gestión del Conocimiento

Los Factores Críticos de Éxito (FCE) para la GC son "elementos que facilitan la adquisición, creación, intercambio y transferencia del conocimiento dentro de y entre las organizaciones" (Pinho et al., 2012, p. 217), y que deben ser abordados con el fin de asegurar la implementación exitosa de las iniciativas de GC (S. M. Allameh et al., 2011; Alsadhan et al., 2008). Generalmente, se atienden aquellos factores del ambiente interno de la organización que pueden controlarse (C. S. Choy, 2006).

Con base en una extensa revisión de la literatura académica, para la presente investigación se han analizan los factores: (i) cultura, (ii) infraestructura y (iii) estrategia, como elementos que acompañan, y hacen posible, la implementación de la GC en las OP. Estos factores se analizan debido a que han sido profusamente relacionados con el éxito de las iniciativas de GC (S. M. Allameh et al., 2011; Alsadhan et al., 2008; C.W. Holsapple & Joshi, 2001; Mas-Machuca & Martínez Costa, 2012; Loo Geok Pee & Kankanhalli, 2008; Kuan Yew Wong & Aspinwall, 2005; Kuan Yew Wong, 2005).

| Factores | Componentes |
|---|---|
| Cultura | (COL) Colaboración<br>(CON) Confianza<br>(EXP) Experiencia<br>(INN) Innovación<br>(COM) Comunicación |
| Infraestructura | (INF) Infraestructura tecnológica<br>(FOR) Formación y aprendizaje<br>(SIS) Sistemas<br>(PRO) Procesos<br>(EVA) Evaluación |
| Estrategia | (EST) Estrategia y visión<br>(LID) Liderazgo<br>(NOR) Normatividad<br>(PRE) Presupuesto |

**Tabla 2.2. Factores y sus componentes, asociados a la Gestión del Conocimiento**
Elaboración propia (2016).

Adicionalmente, en la literatura académica es posible identificar los siguientes beneficios comúnmente asociados con la GC:

- (B1) Ayuda a reducir la corrupción
- (B2) Ayuda a generar mejores políticas públicas
- (B3) Permite trabajar de manera más eficiente
- (B4) Fomenta la transparencia en la organización
- (B5) Mejora la percepción ciudadana y la confianza en la AP
- (B6) Mejora la toma de decisiones
- (B7) Favorece la profesionalización de los funcionarios
- (B8) Impulsa el logro de los objetivos estratégicos

### 2.3.6. Descripción de los factores críticos de éxito y sus elementos

*Cultura*

La cultura organizacional es considerado uno de los FCE más importantes en las iniciativas de GC (M. Allameh et al., 2011; Ho et al., 2014), tiene influencia esencial en la decisión acerca de cuándo, dónde y con quién debe ser intercambiado el conocimiento (Syed-Ikhsan & Rowland, 2004). La cultura puede ser definida como "los valores de una organización, principios, normas, reglas no escritas y procedimientos. Este recurso, se compone de suposiciones y creencias básicas que rigen las actividades de los participantes" (C.W. Holsapple & Joshi, 2001, p. 46). El factor cultura se relaciona positiva y significativamente con el PCC, y por tanto con el éxito de las iniciativas de GC (Ho et al., 2014). Integra componentes interdependientes, como la colaboración, la confianza, la experiencia, la innovación y la comunicación.

La **colaboración** es el grado de voluntad que los individuos exhiben para apoyarse entre ellos (H. Lee & Choi, 2003). La colaboración efectiva entre los organismos públicos en todos los niveles de gobierno es esencial, como lo es con los actores no gubernamentales, para garantizar el buen gobierno y los buenos resultados de desarrollo (United Nations, 2014a, 2014b). La colaboración requiere de mecanismos que faciliten la interoperabilidad, para el intercambio de información y de recursos tecnológicos (Luna-Reyes, Gil-Garcia, & Cruz, 2007) entre individuos e instituciones. "Un entorno de colaboración ofrece oportunidades para las personas con conocimientos para compartirlos de manera abierta" (S. M. Allameh et al., 2011, p. 1217), ello requiere de un continuo intercambio de conocimientos, habilidades, ideas y valores (Ho et al., 2014, p. 738).

La **confianza** se refiere a la creencia común entre los miembros de una organización que otros hacen esfuerzos de buena fe para comportarse de acuerdo a un compromiso, actúan con honestidad y no toman ventaja de otros aun cuando se presenta una oportunidad de hacerlo (Loo Geok Pee & Kankanhalli, 2008, p. 442). Los empleados comparan el cuidado y la asistencia que les proveen sus jefes y compañeros con sus expectativas, si éstas son rebasadas la confianza crece (Ho et al., 2014, p. 738). La confianza promueve el intercambio abierto de conocimiento, por lo tanto contribuye al éxito de las iniciativas de GC (Loo Geok Pee & Kankanhalli, 2008; Yuen, 2007). Sin embargo, la confianza disminuye cuando existe preocupación sobre la pérdida de autonomía y el uso indebido de la información por parte de otros individuos o instituciones (Luna-Reyes et al., 2007; Yang & Maxwell, 2011).

La **experiencia** "es la capacidad para realizar bien una tarea, permite al individuo aprender nuevas tareas con mayor rapidez, contribuir a la mejora de la organización y producir mejores resultados" (P. R. Massingham & Massingham, 2014). Además, ayuda a observar el grado de domino de una persona o grupo sobre algún tema en específico. El desarrollo de la experiencia es un proceso continuo y complejo que implica la inversión de recursos organizaciones (Hitt, Biermant, Shimizu, & Kochhar, 2001), proceso que se ve reforzado por el aprendizaje y la adquisición de nuevas habilidades. Así, reconocer el *expertise* o el profesionalismo de las personas eleva su estatus profesional y fomenta el intercambio de conocimiento en la organización (Ho et al., 2014). La transferencia de experiencia entre los individuos de la organización contribuye al éxito de la GC (Chawla & Joshi, 2010).

La **innovación** se refiere a no sólo a la creación de nuevas ideas, sino al uso de éstas para crear nuevos productos o procesos (Murray E Jennex & Smolnik, 2011, p. 112), o mejorar los existentes. Está relacionada con "los métodos que usa una organización para estimular exitosamente la creatividad entre sus empleados y después aprovechar los beneficios de esta creatividad para incrementar la competitividad organizacional" (Ho et al., 2014, p. 739). La innovación "es un tema central en el desarrollo económico y en la mejora del bienestar de la sociedad" (OECD, 2014, p. 17), tiene un impacto positivo en la GC en la AP (Ho et al., 2014), y es un elemento clave para resolver problemas complejos del sector público (Puron-Cid, 2014; United Nations, 2014a).

A través de ella, los empleados usan el conocimiento organizacional existente para generar nuevo conocimiento (Ng, Yip, Din, & Bakar, 2012), que puede ser aplicado por otros miembros de la organización para solucionar problemas (Ho et al., 2014) y "generalmente implica la mejora en los servicios, productos o procesos" (OECD, 2010, p. 12). Por tanto, las organizaciones deberían estimular la generación de nuevas ideas y formas creativas de resolver problemas.

La **comunicación** es un componente relacionado con la interacción humana potenciado por la existencia de redes sociales en la organización (Al-Alawi, Al-Marzooqi, & Mohammed, 2007), es clave para la GC (Canary, 2010), debido a que permite intercambiar y procesar información entre los miembros de la organización (Kuwada, 1998, p. 725). Mediante diversos canales de comunicación, el conocimiento puede ser efectivamente transferido dentro de la organización (Chong Siong Choy et al., 2006; Syed-Ikhsan & Rowland, 2004), e intercambiado entre organizaciones (Akhavan, Jafari, & Fathian, 2006; T. A. Pardo, Cresswell, Zhang, & Thompson, 2001; Savvas & Bassiliades, 2009).

La comunicación apoya actividades como el aprendizaje (Conde, García-peñalvo, Casany, & Forment, 2013; Fersini, Messina, Archetti, & Cislaghi, 2013), el trabajo en equipo (Ragab & Arisha, 2013), el intercambio de conocimientos, la interacción entre las personas (Al-Alawi et al., 2007) y el reporte de actividades institucionales (Greiling & Halachmi, 2013; Park, Ribière, & Jr, 2004; Puron-Cid, 2014).

### Infraestructura

La infraestructura constituye uno de los factores más relevantes para la implementación de iniciativas de GC (M. Allameh et al., 2011; Mas-Machuca & Martínez Costa, 2012; Wiig, 2002), dado que da soporte y apoya las principales actividades en donde se crea, acumula, intercambia, utiliza e internaliza el conocimiento (Chang Lee et al., 2005; Pinho et al., 2012). La infraestructura se refiere "al conjunto de tecnologías relacionadas con la informática que se han desplegado para complementar, permitir o apoyar actividades de la organización" (Clyde W. Holsapple & Luo, 1996, p. 14). No obstante, "la tecnología no debería ser vista como la respuesta absoluta para la GC, ya que ésta es sólo una herramienta (Kuan Yew Wong & Aspinwall, 2005, p. 76)".

Este factor está conformado, además, por aquellos recursos físicos, sistemas de información y procesos organizacionales que están disponibles para los empleados (Ragab & Arisha, 2013) y puede apoyar a los trabajadores de la organización en el cumplimiento de sus tareas como recopilar, analizar y difundir conocimiento relevante entre la organización (Clyde W. Holsapple & Luo, 1996). La infraestructura se integra por componentes interdependientes como la infraestructura tecnológica, la formación, el aprendizaje, los sistemas, los procesos y la evaluación organizacional. Su propósito es permitir a los trabajadores realizar su trabajo de forma más eficiente e incrementar el desempeño organizacional.

La **infraestructura tecnológica** se refiere a todas aquellas herramientas tecnológicas producto de tecnologías de la información y sus capacidades de apoyo a la GC (Loo Geok Pee & Kankanhalli, 2008), como sistemas de comunicación, redes, repositorios de conocimiento, programas de formación a distancia (Wiig, 2002), sistemas de manejo de la información, aplicaciones de inteligencia, software experto, bases de datos, tecnologías específicas, y modelos que se emplean para la GC en la organización (Ragab & Arisha, 2013, p. 878). Los instrumentos de infraestructura tecnológica, pueden ser clasificadas en dos tipos: de comunicación (correos electrónicos, mensajes de videoconferencia, tableros de anuncios electrónicos y conferencias de equipo); y tecnología para la toma de decisiones (de sistemas de soporte de decisiones, sistemas expertos y sistemas de información ejecutivos) (S. M. Allameh et al., 2011, p. 1216).

Este tipo de infraestructura permite establecer redes de colaboración, como las wikis, foros, redes sociales, al tiempo que facilitan el flujo del conocimiento entre los empleados y hacia la organización (Tabrizi, Ebrahimi, & Delpisheh, 2011, p. 695), proporciona sistemas de almacenamiento eficiente y mecanismos de recuperación y transferencia del conocimiento (Loo Geok Pee & Kankanhalli, 2008). Por tanto, la GC no se concibe sin una infraestructura tecnológica que dé soporte estos esfuerzos organizacionales (C. S. Choy, 2006). Por tanto, es necesaria para el éxito de las iniciativas de GC.

Las TIC tienen el potencial de transformar las estructuras de gobierno y mejorar la calidad de los servicios gubernamentales, así "la tecnología ofrece dos oportunidades principales para el gobierno: (1) el aumento de la eficiencia operativa al reducir los costos y aumentar la productividad; y (2) una mejor calidad de los servicios prestados por las agencias gubernamentales" (Gil-García & Pardo, 2005, p. 188).

Asimismo, las TIC juegan un papel crucial en la eliminación de los límites a la comunicación que a menudo inhiben la interacción entre las distintas partes de la organización (S. M. Allameh et al., 2011). Los proyectos de GC tienen más probabilidades de éxito cuando se adopta una infraestructura tecnológica más amplia. Las TIC favorecen la accesibilidad del conocimiento en la organización. Por lo tanto, el apoyo tecnológico es necesario para el desempeño de la GC en la organización (S. M. Allameh et al., 2011).

La **formación y el aprendizaje** implica el desarrollo de conocimientos, habilidades y actitudes (competencias) para la GC (P. R. Massingham &

Massingham, 2014), para ello se deberán identificar las necesidades de formación organizacional e implementar programas de entrenamiento y desarrollo de habilidades (Hafeez & Abdelmeguid, 2003; P. Massingham, 2014). Este componente involucra la formación continua de las personas que trabajan en la organización y el esfuerzo institucional para fomentar el deseo, entre los trabajadores, para mantener actualizados sus conocimientos. A nivel de la organización una actitud de aprendizaje es deseable para implementar cambios en sus iniciativas de GC (Chang Lee et al., 2005; Salleh, Chong, Syed Ahmad, & Syed Ikhsan, 2012).

Los **sistemas** de GC son soluciones basadas en las TIC que apoyan la implementación de la GC en las organizaciones (Ragab & Arisha, 2013), mediante actividades de captura y representación, recuperación, intercambio, reutilización, razonamiento y recuperación del conocimiento en una organización (Ding et al., 2014), y dan soporte al PGC (Benbya, 2010; Chang Lee et al., 2005). Algunos ejemplos de sistemas son: los de gestión de documentos, motores de recuperación de información, bases de datos relacionales y de objetos, sistemas de trabajo en grupo y flujo de trabajo, tecnologías de empuje y agentes, así como herramientas de minería de datos (Kuan Yew Wong & Aspinwall, 2005).

Además son un elemento de apoyo para la planeación y la toma de decisiones en las instituciones (P. Jain, 2009), porque permiten generar marcos institucionales de innovación y procesos de colaboración a través sistemas de gestión integrados (Puron-Cid, 2014; United Nations, 2014b) e interoperables entre diferentes instituciones y/o sectores (Savvas & Bassiliades, 2009; Yang & Maxwell, 2011).

Los **procesos** organizacionales de GC implican el uso del método más eficiente para "transformar" el conocimiento implícito, fragmentado y privado de los individuos o grupos, tanto dentro como fuera de la organización en activos intelectuales valiosos para la organización (Ho et al., 2014, p. 736). La claridad en los procesos internos da secuencia y estructura lógica a las actividades de la organización (Savvas & Bassiliades, 2009). Esto contribuye a su medición y mejora continua, y por tanto a la efectividad organizacional (C. W. Chong & Chong, 2009).

De esta manera, los procesos de GC dan paso a la espiral de creación del conocimiento, propuesta por Nonaka y Takeuchi (1999): socialización, externalización, combinación e internalización (SECI). Por tanto, "exigen la interacción y participación de las personas, la tecnología y la información (Alsadhan et al., 2008, p. 813)". En algunos casos, establecer los procesos requiere estrategias como la "reingeniería", que implica la ruptura de viejas formas tradicionales de hacer las cosas y la búsqueda de formas nuevas e innovadoras (Akhavan et al., 2006).

Finalmente, está demostrado que la **evaluación** es un componente clave para medir el progreso y la efectividad en las iniciativas de GC (C. W. Chong & Chong, 2009; K. Y. Wong et al., 2015). De esa forma, se pueden realizar acciones de mejora basadas en juicios objetivos (Andone, 2009), sobre la contribución de la GC a los objetivos estratégicos de la institución (Girard & McIntyre, 2010).

Desde la perspectiva organizacional, ayuda a observar los resultados obtenidos (logros y beneficios) mediante la GC y conocer el nivel de avance en su implementación, "sin un éxito medible, el entusiasmo y apoyo a la GC es difícil que continúe" (Andone, 2009, p. 25; C. W. Chong & Chong, 2009, p. 146). Además, esta actividad apoya la decisiones de los directivos respecto a la estrategia de GC (Bose & Ranjit, 2004). Desde otro aspecto, la evaluación debe ser vista como una actividad necesaria, relacionada con sistemas para recompensar a las personas por compartir sus conocimientos (Ragab & Arisha, 2013; Winter, 2013, p. 115).

### *Estrategia*

En la literatura académica, es ampliamente reconocido que el factor estratégico impulsa las iniciativas de GC (Akhavan et al., 2006; Mas-Machuca & Martínez Costa, 2012). Integrar la estrategia conlleva definir objetivos claros, los métodos a emplear para lograrlos (Loo Geok Pee & Kankanhalli, 2008) y la forma en que se van a medir. La estrategia provee las bases sobre cómo la organización desplegará sus competencias y recursos para alcanzar los objetivos de la GC (Akhavan et al., 2006, p. 108). En suma, la estrategia da sentido y orienta los esfuerzos organizacionales en materia de GC.

Desde la estrategia se debe formular una estructura coherente con la iniciativa de GC, que permita que la información fluya de manera efectiva, asimismo es importante definir las funciones y responsabilidades en la construcción y en la implementación de la iniciativa de GC (De Angelis, 2013), esto incluye desarrollar una cultura propicia para el intercambio de conocimientos (C. W. Chong & Chong, 2009). Este factor integra componentes interdependientes como la estrategia y la visión, el liderazgo, la normatividad y el presupuesto.

La **estrategia y visión** sobre las iniciativas de GC debe ser específica y alinearse con los objetivos organizacionales, "esto ayuda a aclarar el papel de la GC en el apoyo a la consecución de los objetivos de la organización y crea un impulso más fuerte en los miembros de la organización para apoyar a la GC" (Loo Geok Pee & Kankanhalli, 2008, p. 442). Además, debe crear una visión clara y compartida que inspire a otros a unirse a la exploración de cómo gestionar el conocimiento podría aportar valor a la organización y a su gente (Alsadhan et al., 2008). Una estrategia definida y una visión alcanzable son elementos que contribuyen al éxito de las iniciativas de GC (Akhavan et al., 2006) y crean confianza sobre su propósito (Girard & McIntyre, 2010).

El **liderazgo** en un componente clave para explicar el propósito de la GC, crear confianza entre los miembros de la organización, motivar a las personas a compartir sus conocimiento y a esforzarse para alcanzar las metas y objetivos institucionales (Girard & McIntyre, 2010) es tarea de los líderes. Un liderazgo efectivo favorece la creación de conocimiento (von Krogh, Nonaka, & Rechsteiner, 2012) y motiva su intercambio entre los miembros de la organización e incluso entre organizaciones (United Nations, 2007, pp. 102–114). En la AP, "los líderes deben reconocer que, con ánimo y apoyo, los trabajadores pueden ayudarse unos a otros a hacer su trabajo mejor y servir a los usuarios de manera más efectiva" (Gorry, 2008, p. 110).

Por lo anterior, podemos argumentar que los líderes tienen una alta influencia en los empleados, y contribuyen a articular la visión sobre la GC. Así, ellos pueden fomentar la participación de los funcionarios de las OSP en las actividades relacionadas con la GC, lo que indudablemente manda una señal sobre la importancia que tiene la GC para la organización (Loo Geok Pee & Kankanhalli, 2008, p. 442).

Otro componente a considerar es la **normatividad,** o marco de regulación, como un fundamento que favorece la implementación de las iniciativas de GC en las OP (Puron-Cid, 2014). Se integra por las leyes, decretos, reglamentos y disposiciones diversas que, en ocasiones, pueden ser permisivas o restrictivas con respecto al uso de TIC en la AP. Es importante observar que la mayor parte de las organizaciones gubernamentales se crean y operan en virtud de una regla formal específica o un grupo de reglas. Por ello, al tomar cualquier tipo de decisión, incluidos las relacionadas con proyectos de TIC, los administradores públicos deben tener en cuenta un gran número de leyes y regulaciones restrictivas (Gil-García & Pardo, 2005, p. 192).

Generalmente se deben observar que las políticas favorezcan la implementación de las iniciativas de GC. Asimismo, desde una perspectiva organizacional, las normas apoyan la cooperación, la apertura y el trabajo en equipo en una organización. Estas normas, pueden favorecer la colaboración y las actividades de GC, por ejemplo el intercambio de conocimiento (Loo Geok Pee & Kankanhalli, 2008, p. 442).

El **presupuesto** es reconocido como otro componente que apoya la implementación de iniciativas de GC (Purón-Cid, 2013), es uno de los temas con los que las instituciones gubernamentales deben lidiar anualmente (Gil-García & Pardo, 2005, pp. 192–193) y sobre los cuales los líderes deben poner atención para realizar las gestiones adecuadas (von Krogh et al., 2012), debido a que es un reto común en muchos gobierno nacionales y estatales. Lo que, puede afectar los resultados potenciales del despliegue de las iniciativas de GC en las OP.

Por tal motivo, el presupuesto debe planearse e integrarse como parte de la estrategia de GC en la institución (H.-Y. Chong & Phuah, 2013; OECD, 2003b, p. 96). De igual forma, es importante propiciar e implementar reformas, formatos y procedimientos que incentiven que este tipo de iniciativas continúen (Puron-Cid, 2014), incluso más allá de los cambios de administración.

### 2.3.7. Medición de la Gestión del Conocimiento en las Organizaciones Públicas

Como se ha señalado anteriormente, sin importar los beneficios que tiene la GC para las OSP, aún persiste una falta de conciencia sobre ésta en este sector específico (Chawla & Joshi, 2010). Por esta razón, las OSP representan un contexto único para los practicantes y académicos. Desde una perspectiva académica, la GC en las OSP representan un campo poco explorado (Massaro et al., 2015); por otra parte, desde una perspectiva de los practicantes, un campo para hacer frente a distintos actores (Massaro et al., 2015) y para mejorar la gestión de los conocimientos y para mejorar las relaciones asociaciones con todos las partes interesadas, lo que permite incrementar el desempeño global de sector público (P. Jain, 2009).

Por lo tanto, el estudio de la GC dentro de las OP requiere de una agenda particular (Massaro et al., 2015) y es un desafío abierto para que los investigadores desarrollen y validen marcos o modelo específicos para la GC en las OSP (Dumay et al., 2015)

Para la presente investigación, analizamos a la GC, principalmente porque sin importar si hablamos del sector privado o del público, las iniciativas de GC deben demostrar su valor, y los resultados de la GC necesitan ser evaluadas con el objetivo de medir el éxito en la implementación de este tipo de iniciativas.

En primer lugar, se puede reconocer que el desarrollo de un modelo de medición del desempeño de GC es un reto en muchos aspectos (Kuah et al., 2012, p. 9348). De hecho las iniciativas de GC tienen que demostrar su valor y justificar los recursos y esfuerzos que se le dedican. Sin un éxito mensurable, el entusiasmo y el apoyo a a la GC es difícil que continúe (Andone, 2009, p. 25; C. W. Chong & Chong, 2009, p. 146).

En segundo lugar, ya que debido a su naturaleza compleja, la medición de la GC es uno de los temas menos desarrollados (Bose & Ranjit, 2004, p. 457) o investigados (C. W. Chong & Chong, 2009, p. 142; Garlatti et al., 2014). Por lo tanto es muy importante establecer medidas de rendimiento en las diferentes etapas de implementación de la GC, e incluso desde el principio para que su eficacia puede ser identificada (C. W. Chong & Chong, 2009, p. 143).

Tercero, desde que las iniciativas de GC han sido vistas como decisiones de inversión, tanto su desempeño como sus resultados deben ser medidos y evaluados (Tabrizi, Ebrahimi, & Delpisheh, 2011). De hecho, la medición es uno de los elementos clave que permite el seguimiento del progreso y la efectividad de las iniciativas de GC (C. W. Chong & Chong, 2009; Migdadi, 2009; K. Y. Wong et al., 2015). Desde esta perspectiva es posible implementar acciones de mejora basadas en juicios objetivos (Andone, 2009), asociados con la contribución que la GC tiene en los objetivos estratégicos de la institución (Girard & McIntyre, 2010; Vagnoni & Oppi, 2015).

De acuerdo con Chong & Chong (2009, p. 146), una correcto sistema de medición del desempeño deberá estar establecido y ser adoptado a través de toda la organización y no debería sólo limitarse a medir el conocimiento, la experiencia o el desempeño individual de los empleados. Consecuentemente, la falta de una correcta evaluación del conocimiento puede derivar en ignorancia respecto al valor que el conocimiento tiene, o podría ocurrir una duplicación de esfuerzos (Hu, Wen, & Yan, 2015).

En resumen, la medición de la GC es esencial para asegurar que los objetivos trazados están alineados con la estrategia organizacional (Bose & Ranjit, 2004), o bien para observar si necesitan ser alineados (P. R. Massingham & Massingham, 2014). Ciertamente, las métricas deberían mostrar aquellas áreas en donde se necesitan mejoras en la organización (Goldoni & Oliveira, 2010). En consecuencia la medición de la GC podría demostrar mejoras en el desempeño global de la organización (Bontis, 2001). Sin importar que dentro del campo de la GC aún no se tienen estándares o procedimiento para medir el conocimiento dentro de las organizaciones (Bose & Ranjit, 2004).

## 2.4. Organizaciones Públicas

### 2.4.1. Importancia de la Gestión del Conocimiento en las Organizaciones Públicas

La GC se anuncia crucial para las OSP del siglo XXI, debido a que el gobierno es uno de los mayores consumidores y productores de conocimiento (P. Jain, 2009), es una herramienta que, motivada por la actualización y modernización de las OSP, crea mecanismos de carácter innovador para gestionar y producir conocimiento (Braun & Mueller, 2014) y que, indudablemente, puede apoyar a las actividades esenciales de este tipo de organizaciones (Loo Geok Pee & Kankanhalli, 2008).

Por esta razón, los programas de GC se concentran, principalmente, en manejar y distribuir lo que el gobierno sabe dentro y entre las instituciones públicas, con el propósito de tomar acciones colaborativas (Murray E Jennex & Smolnik, 2011), este paradigma de gestión puede contribuir a la reforma de las OSP, para que los gobiernos actúen de forma más eficiente, transparente, sensible a las necesidades de los ciudadanos, y eficaz en el logro de sus objetivos (De Angelis, 2013). Consecuentemente, la GC es crucial no sólo para el éxito de las instituciones públicas, sino para aprovechar las nuevas tecnologías con la finalidad de fortalecer la gobernabilidad pública (Puron-Cid, 2014) y el desarrollo de la sociedad (Ragab & Arisha, 2013, p. 877).

Este tema ha cobrado tal relevancia que, para el año 2003, muchos países industrializados habían emprendido acciones para atender la demanda emergente de GC en el sector público: una encuesta de investigación de 132 agencias del gobierno central de 20 países llevada a cabo por la Organización para la Cooperación y el Desarrollo Económico (OECD, 2003b), en ese año determinó que la mayoría de las organizaciones del gobierno central en los países miembros de la OCDE había ideado estrategias de GC e incorporó este tema como una de las cinco prioridades futuras de gestión interna (S. Kim & Lee, 2006).

Como se ha señalado, la GC puede ser adoptada como una estrategia para recuperar la confianza de los ciudadanos en el gobierno (United Nations, 2007). Su adopción implica innovación y reformas en las organizaciones. Por lo tanto, la implementación de la GC, no sólo requiere una inversión considerable, sino también cambiar la cultura y la estructura de la organización ejecutora (United Nations, 2007). La GC puede contribuir a generar mejores políticas públicas a través de la integración del conocimiento que se produce en la sociedad en su formulación (Criado & Gil-García, 2013; De Angelis, 2013; Riege & Lindsay, 2006).

Además, el combate a la corrupción, como un tema prioritario por considerarse uno de los principales riesgos que amenazan a la instituciones (Red Latinoamericana por la Transparencia Legislativa, 2011; Transparency International, 2012; World Economic Forum, 2014), debería estar presente en la implementación de nuevas prácticas de GC en la AP (United Nations, 2014a). La contribución de la GC en esta lucha es sustancial (Elkadi, 2013; Tung & Rieck, 2005), y quizá aún con un amplio abanico de posibilidades por explorar. Consecuentemente, es necesaria una mayor conciencia pública de los beneficios directos e indirectos que la GC puede traer en las organizaciones públicas (Massaro et al., 2015; Tung & Rieck, 2005).

Contrariamente a lo que podría suponerse, la GC en las OSP en los últimos años ha ido ganando importancia (Moffett & Walker, 2015); debido a que juega un papel muy importante para hacerlas más efectivas (Wiig, 2002). Evidentemente, la GC en las OSP ayuda a los gobierno a abordar los retos creados por la economía del conocimiento (Moffett & Walker, 2015, p. 68). Las OSP son organizaciones que producen y usan conocimiento de manera intensa (Edge, Karen, & Edge, 2005; P. Jain, 2009; Savvas & Bassiliades, 2009).

Durante los últimos años este tema ha crecido en importancia en el mundo académico (Garlatti et al., 2014; Massaro et al., 2015) y un número cada vez más grande de investigadores han desarrollado trabajos en esta área. Sin embargo, aún hoy en día, existe una falta de entendimiento de la GC en el contexto de las organizaciones del sector público (Loo Geok Pee & Kankanhalli, 2008, p. 439). La GC dentro de las organizaciones del sector público es un campo de investigación insuficientemente explorado entre la comunidad académica, y de acuerdo con Garlattir, Massaro, Dumay, & Garlatti (Massaro et al., 2015), aún existe la necesidad de entender cómo la GC está evolucionando dentro del contexto específico de las organizaciones públicas.

De igual forma, diversos autores han argumentado que es necesario mejorar las prácticas de GC dentro de las OSP para el crecimiento de las economías en los países en vías de desarrollo.(A. K. Jain & Jeppesen, 2013, p. 348). Entonces, no debemos olvidar que los administradores de las OSP son empleados de las entidades, servidores públicos y, al mismo tiempo, son ciudadanos y usuarios de los servicios públicos (Massaro et al., 2015).

Ciertamente, existen algunas diferencias entre la GC en las organizaciones del sector privado y las del sector público, distintos autores han dado muestra de ello (Garlatti et al., 2014; Massaro et al., 2015), principalmente, bajo el argumento de que las metas y objetivos de las OSP son más complejas y difíciles de medir que en las organizaciones privadas (Titi Amayah, 2013, p. 456), así como que los objetivos se refieren a bienes públicos (Choy Chong et al., 2011, p. 498), y que su reporte al exterior está impuesto por los distintos ordenamientos jurídicos (Garlatti et al., 2014) que las regulan.

Además, a las OSP frecuentemente se les considera un sinónimo de ineficiencia y de falta de motivación para innovar (Suwannathat et al., 2015), baja motivación para adoptar nuevos métodos y prácticas de gestión (Serrano Cinca, Mar Molinero, & Bossi Queiroz, 2003), y ausencia de indicadores de mercado que permitan medir su desempeño (Loo Geok Pee & Kankanhalli, 2008, p. 444).

De Angelis (2013), argumenta que las OSP de la actualidad están influenciadas por la necesidad de competencia de nuestro tiempo, así como por los estándares de desempeño, el monitoreo, la medición, la flexibilidad, el énfasis en los resultados, el enfoque en el cliente y el control social (De Angelis, 2013, p. 1). Ciertamente las motivaciones que las OSP tienen para adoptar prácticas de gestión del conocimiento responden, en algunos casos, a cambios políticos, presiones legislativas, así como a las necesidades de los ciudadanos y las empresas (Loo Geok Pee & Kankanhalli, 2008).

Como se ha reiterado, el conocimiento dentro de las OSP tiene el potencial de ampliar e incrementar la efectividad global del conocimiento en la organización (Ann Hazlett, Mcadam, & Beggs, 2008, p. 58), de fortalecer la efectividad y competitividad de los gobiernos en los entornos actuales complejos y cambiantes (Moffett & Walker, 2015, p. 68) y de influenciar y mejorar el proceso de renovación del sector público (Edge et al., 2005, p. 45).

### 2.4.2. Beneficios de la Gestión del Conocimiento en las Organizaciones del Sector Público

La GC en la OSP según Wiig (2000, 2002) genera al menos cuatro áreas de interés:

- Mejora de las decisiones dentro de los servicios públicos;

- Ayuda al público a participar de manera efectiva en la toma de decisiones públicas;
- Construye capacidades sociales (capital intelectual); y
- Desarrolla la competitividad en la fuerza de trabajo basada en el conocimiento.

Los beneficios derivados de la implementación de iniciativas de GC en las instituciones pueden ser observados desde distintas perspectivas: individual, la institucional y desde la sociedad como un todo, principalmente.

A nivel individual, en la literatura, se le atribuyen beneficios como: permitir a las personas hacer mejor su trabajo y en menor tiempo, crear un sentido de confianza, fomentar el desarrollo constante de los empleados (Edvardsson & Durst, 2013) y permitir que las personas sean reconocidas por sus contribuciones a la organización (Tianyong Zhang, 2010).

A nivel institucional, la GC beneficia mediante la reducción de costos en las operaciones, mejora la conveniencia, promueve una mayor eficiencia, efectividad e innovación (Edvardsson & Durst, 2013), fomenta la transparencia, mejora la calidad de los servicios ofrecidos (Kumar, Mukerji, Butt, & Persaud, 2007), ayuda a generar información más precisa, y promueve una mayor responsabilidad y flexibilidad (Weerakkody et al., 2013). Al mismo tiempo, a través de la implementación de prácticas exitosas de GC, se promueve una comunicación más efectiva y permite la comparación y evaluación entre distintas instituciones (Tianyong Zhang, 2010).

Incluso la GC favorece la interoperabilidad del conocimiento entre las instituciones que integran las OSP que se refleja en un incremento de la eficiencia (acelerando tareas recurrentes) y la eficacia (ayudando a tomar decisiones acertadas basadas en el conocimiento) (Savvas & Bassiliades, 2009), como resultado del intercambio de información intra e inter institucional (OECD, 2003b).

Asimismo, la GC es reconocida por optimizar los servicios públicos mediante el uso de TIC (OECD, 2003b; United Nations, 2014a, 2014b). De esta manera, la GC mejora el desempeño global de las OSP (P. Jain, 2009), porque fortalece la rendición de cuentas y la transparencia (United Nations, 2007), reduce la corrupción (o discrecionalidad burocrático-administrativa) e incentiva la participación de los ciudadanos (Elkadi, 2013; Tung & Rieck, 2005). Asimismo, la GC permite eliminar sistemáticamente la duplicación de esfuerzos institucionales e interinstitucionales (P. Jain, 2009) y puede mejorar la percepción ciudadana sobre las OSP al fomentar confianza entre los ciudadanos y sus gobiernos (OECD, 2003b; Srivastava, 2011).

Por otra parte, de manera global, en la sociedad la GC promueve la integración del conocimiento entre diferentes instituciones (T. a. Pardo & Tayi, 2007; United Nations, 2014a), mejora la satisfacción del usuario (Kumar et al., 2007), ayuda a disminuir la burocracia y a prevenir posibles fraudes, mediante el fortalecimiento de la transparencia, y permite tener un mejor control de la información disponible (Tung & Rieck, 2005). Además, promueve la construcción de redes de colaboración entre las instituciones que utilizan el conocimiento (Tianyong Zhang, 2010), así como contribuir a mejorar la calidad de vida de los ciudadanos (Syed-Ikhsan & Rowland, 2004).

Además, la GC en las OSP se relaciona ampliamente con la reducción del costo de los servicios, con su mejoramiento que facilita el avance de una verdadera sociedad del conocimiento (OECD, 2003a, 2004, 2015b; The World Bank, 2011; United Nations, 2014b). Por ello, la GC es vista como una solución efectiva que puede favorecer la modernización de los gobiernos y dar soporte a sus principales funciones administrativas (Mitre-Hernández et al., 2015). La GC en las OSP se ha desarrollado y extendido paralelamente con el desarrollo y uso de las TIC (OECD, 2004; E. Tsui, 2005; United Nations, 2007).

Conjuntamente, en la literatura académica se hace latente que la GC en las OSP podría incrementar la satisfacción de los ciudadanos con los servicios públicos (Serrano Cinca et al., 2003), ayudar a las OP a producir mejor conocimiento (Braun & Mueller, 2014), favorecer la gobernabilidad pública (Puron-Cid, 2014), contribuir al desarrollo de la sociedad (Ragab & Arisha, 2013), una mejor gestión de los recursos naturales (United Nations, 2014b) y, finalmente, ayudar a tomar mejores y más colaborativas decisiones (Murray E Jennex & Smolnik, 2011).

Finalmente la GC en las OSP puede generar una mayor competencia entre instituciones gubernamentales y optimizar el uso del conocimiento que se produce en la esfera pública para tomar mejores decisiones (Wiig, 2002), y para formular políticas públicas más inteligentes (Criado & Gil-García, 2013; Mbhalati, 2014).

.

# CAPÍTULO II

# 3. PROPUESTA METODOLÓGICA

En este capítulo se describe la propuesta metodológica para el desarrollo de la presente tesis. Para lograr el objetivo general y los objetivos particulares descritos anteriormente se ha empleado el (i) enfoque cualitativo, en la realización de las revisiones sistemáticas de la literatura y el (ii) enfoque cuantitativo en el estudio a profundidad realizado entre distintas organizaciones públicas. En la ilustración 3.1, se representan los pasos seguidos para lograr el objetivo de la presente tesis, finalmente derivado de los resultados obtenidos, y a partir de la propuesta metodológica, se han construido algunas propuestas de estrategias y se proponen métricas para la GC en las organizaciones públicas.

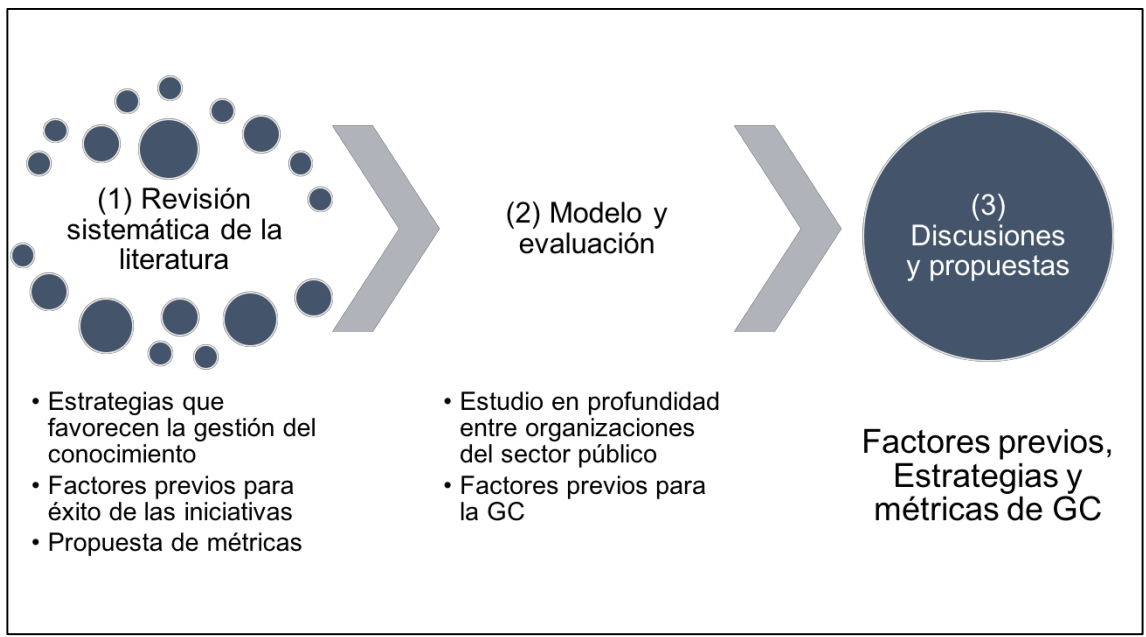

**Ilustración 3-1. Propuesta metodológica**
Elaboración propia (2016).

Esta propuesta metodológica permite obtener información relevante de fuentes primarias y validarla mediante un modelo de evaluación en el contexto objeto de estudio del presente trabajo de investigación. Es por tanto una propuesta académica rigurosa y potencialmente replicable en otros contextos para futuras intervenciones.

# 4. Revisión sistemática de la literatura para analizar los factores previos para la gestión del conocimiento en las organizaciones públicas

Como hemos observado, en el mundo académico existe un extenso corpus literario sobre la GC, esta afirmación se ha visto demostrada por los estudios previos realizados por distintos autores, véase: (Asrar-ul-Haq et al., 2016; Ding et al., 2014; Garlatti et al., 2014; Liao, 2003; Tabrizi, Ebrahimi, & Al-Marwai, 2011).

Existen diversos estudios que analizan factores específicos que favorecen el éxito de la implementación de programas de GC; la evaluación de estos factores es considerada una herramienta estratégica para las OSP porque permite manejar con mayor eficiencia la implementación de los programas de GC.

Sin embargo, factores endebles como, por ejemplo, una cultura de GC débil puede observarse a través de los problemas de confianza, poca colaboración; problemas tecnológicos, como SGC difíciles de usar, que no están interconectados o que no son interoperables entre las distintas instituciones; y finalmente los problemas de estrategia, como los cambios políticos, la falta de continuidad de estas iniciativas entre administraciones y la falta de voluntad política para implementar cambios.

En esta Revisión Sistemática de la Literatura (RSL) observamos los factores que favorecen la implementación de programas de GC en las instituciones. Para ello, se ha realizado una RSL de los FCE para crear programas de GC en las OP. Se han analizado 20 estudios relacionados, resumiendo los beneficios y los atributos de calidad de dichos estudios. Finalmente, derivado de la RSL se han propuesto algunas estrategias para fortalecer a las iniciativas de GC en las instituciones públicas.

### 4.1.1. Objetivos de la Revisión Sistemática de la Literatura

Los objetivos de la presente RSL son:

- Resumir los beneficios y atributos de calidad (comprensibilidad, claridad y credibilidad) de la GC en las OP a través de una RSL.
- Obtener los beneficios, retos, evaluaciones y propuestas de los estudios sobre la GC de una manera sistemática e identificar las estrategias culturales, tecnológicas y estratégicas para fomentar los programas de GC en las OP mexicanas.

El objetivo de la RSL es explorar el estado del arte en la GC en las OSP. Utilizando las directrices formuladas por Kitchenham & Charters (2007) y Petersen et al. (2008), se ha buscado, seleccionado, evaluado y presentado los estudios que se consideran relevantes para responder a las preguntas de investigación, propuestas en la tabla 4.1.

Consecuentemente, las respuestas obtenidas de los estudios seleccionados se han categorizado de acuerdo con los factores críticos de éxito asociados con la GC: cultura, infraestructura y estrategia. Ello, ha permitido descubrir algunas actividades que, dentro del objeto de estudio, son especialmente relevantes, lo que denominamos dominio del conocimiento de los factores de la GC en las OSP. Para dar mayor claridad del estudio propuesto se ha modelado en un proceso, que a su vez está alineado al PGC propuesto por diversos autores (Chang Lee et al., 2005; Ding et al., 2014), como se detalla en la ilustración 4.1

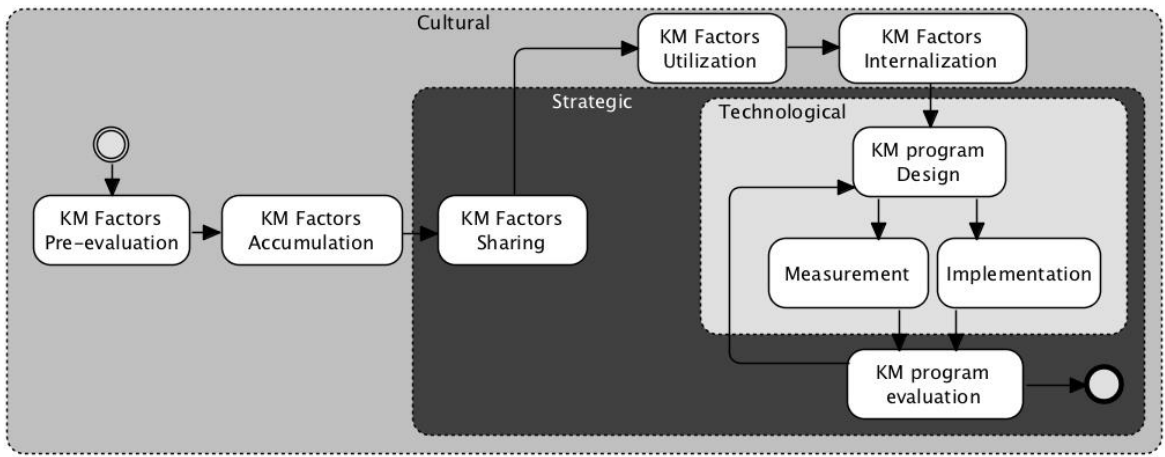

**Ilustración 4-1. Alcance de la revisión sistemática modelado**
Fuente: (Mitre-Hernández et al., 2015)

Dentro del factor cultura, se han agrupado cuatro actividades: (i) La pre-evaluación de los elementos para la gestión del conocimiento; (ii) la acumulación del conocimiento; (iii) la utilización del conocimiento; y (iv) la internalización del conocimiento. Este factor, normalmente está relacionado con la confianza, la colaboración y la apertura (Alsadhan et al., 2008). Una cultura que favorece la GC, valora el conocimiento e impulsa su creación, intercambio y aplicación (Yew Wong, 2005), es un ambiente de colaboración que ofrece oportunidades para que las personas con conocimientos puedan compartirlos de manera abierta y tener programas de GC exitosos (S. M. Allameh et al., 2011, p. 1217).

- ▪ ***Pre-evaluación de elementos para la GC:*** El objetivo de esta actividad es entender mejor los recursos del conocimiento que están disponibles, para identificar que conocimientos son necesarios, por qué lo son y quién podría tener estos conocimientos. Ejemplos de esta actividad, como el mapa de conocimiento y la auditoria de conocimientos, propuestas por Chong & Chong (2009); la primera ayuda a una organización para visualizar las relaciones y procesos que conectan sus recursos de conocimiento, lo que incluye a las personas, la documentación (C. W. Chong & Chong, 2009, p. 147); la auditoría de conocimiento, por su parte, ayuda a las organizaciones a identificar dónde reside el conocimiento dentro de ellas.
- ▪ ***Acumulación de conocimiento:*** Esta actividad está relacionada con la comprensión de la información personalmente adquirida y las búsquedas de los empleados a través de bases de datos corporativas para obtener los conocimientos necesarios para las tareas dadas (Ho et al., 2014). Almacenar el conocimiento de la organización es uno de los elementos más importantes de un SGC (Akhavan et al., 2006, p. 109).
- ▪ ***Utilización***: Esta actividad facilita las tareas mediante la aplicación de conocimientos, es sobre el grado de conocimiento utilizado por la organización (Chang Lee et al., 2005; Ho et al., 2014).
- ▪ ***Internalización***: esta actividad produce nuevos conocimientos desde el aprendizaje y la aplicación de las mejores prácticas (Ho et al., 2014).

Las siguientes dos actividades se asocian con la dimensión estratégica por la alineación fundamental que puede ocurrir entre las estrategias de GC con los objetivos organizacionales básicos (Murray E Jennex & Smolnik, 2011, Chapter 5).

- ▪ **Intercambio de conocimientos:** esta actividad promueve la difusión del conocimiento y contribuye a que el proceso de trabajo astuto y de conocimiento intensivo (Chang Lee et al., 2005; Ho et al., 2014). El intercambio del conocimiento juega un papel importante en la implementación y ejecución del sistema de GC, el intercambio de conocimientos entre los empleados necesita una cultura de GC fuerte, confianza y también la transparencia en toda la organización (Akhavan et al., 2006), el liderazgo, la asignación de tiempos y la confianza fomentan el intercambio de conocimientos (Mas-Machuca & Martínez Costa, 2012), y la interacción social, las recompensas, y el apoyo de la organización tienen un efecto significativo sobre el intercambio de conocimientos (Titi Amayah, 2013). Sin el intercambio de conocimientos es casi imposible que el conocimiento sea transferido a otras personas (Syed-Ikhsan & Rowland, 2004) u otras instituciones.

▪ ***Evaluación del programa de GC***: Se ha demostrado que la medición es un paso importante para la GC (W. Wong, 2013). Esta actividad busca evaluar completamente la implementación del programa de GC con el fin de observar lo que está funcionando, o no, dentro del programa de GC, y por lo tanto, crear un juicio bien versado para ajustar lo que se estime necesario (Andone, 2009). También ayuda para apoyar las decisiones de los directivos sobre la estrategia para mejorar el rendimiento de la GC, y el proceso de GC dentro de la organización (Bose & Ranjit, 2004; W. Wong, 2013).

Estas tres actividades están en la dimensión tecnológica porque se influyen mutuamente a trabajar como conductores de infraestructura que permiten los programas de GC en las OP (Syed-Ikhsan & Rowland, 2004), y también porque hay actividades asociadas comúnmente dentro de esta dimensión (Mas-Machuca & Martínez Costa, 2012)

• ***Diseño del programa de GC***: esta actividad representa la creación de un proyecto de GC, establece sus metas y objetivos para facilitar procesos organizativos. Se puede observar como proponer pautas para la resolución de problemas específicos de vincular el conocimiento y los procesos de organización, un mejor uso y provisión de conocimiento cuando y donde sea necesario. El objetivo de esta actividad es la conceptualización completa de los programas de GC; el diseño abarca todas las barreras y facilitadores que permiten la GC en las organizaciones (S. M. Allameh et al., 2011; Loo Geok Pee & Kankanhalli, 2008).

▪ ***Medición***: La GC no puede demostrar su valor sin un éxito mensurable (Andone, 2009). Distintos autores han observado y propuesto métricas para evaluar la GC (Bose & Ranjit, 2004; Kuah & Wong, 2011; Ragab & Arisha, 2013). La implementación de la GC puede ser vista como una decisión de inversión y, por lo tanto, sus resultados de desempeño deben ser evaluados y medidos de alguna forma (Tabrizi, Ebrahimi, & Delpisheh, 2011). La medición en la GC es esencial a su vez para asegurar que se están alcanzando los objetivos previstos o para saber sí éstos están alineados (P. R. Massingham & Massingham, 2014), así como para realizar un seguimiento de los progresos de la GC en la organización y para determinar sus beneficios y su eficacia (C. W. Chong & Chong, 2009; Migdadi, 2009). Las herramientas de medición del desempeño de la GC deben ser capaces de indicar las áreas de oportunidad para mejorar los programas de GC (Kuah & Wong, 2011).

- ▪ *Implementación:* Esta actividad implica llevar a cabo proyectos de GC diseñados de acuerdo con los patrones establecidos. Para que ocurra el éxito en la implementación de los proyectos de GC en las OSP se tiene que favorecer cierta alineación con los aspectos políticos y también considerar los elementos clave de la organización (Syed-Ikhsan & Rowland, 2004).

### 4.1.2. Protocolo de la Revisión Sistemática de la Literatura

En esta sección se describen las preguntas de investigación (PI) que han sido formuladas de acuerdo con las recomendaciones propuestas por Kitchenham (2007). Como se ha señalado antes, en la tabla 4.1. Protocolo de revisión sistemática, se encuentran estas preguntas de investigación, los atributos de calidad (AC) observados, así como los criterios de inclusión (I) y de exclusión (E) para los estudios consultados.

| Elementos | Descripción |
|---|---|
| **Preguntas de investigación (PI)** | PI1: ¿Cuáles son los beneficios (B1 hasta B5) a las propuestas de medición de la gestión del conocimiento en la administración pública?<br><br>PI2: ¿Ha habido alguna contribución a la entendibilidad, claridad o credibilidad para la organización pública? |
| **Atributos de Calidad (AC)** | AC1: Comprensibilidad<br>AC2: Claridad<br>AC3: Credibilidad |
| **Beneficios (B)** | B1: Alineación de la gestión del conocimiento con los objetivos estratégicos<br>B2: Reducción del índice de corrupción<br>B3: Reducción de los juicios laborales<br>B4: Reducción en tiempos y costos<br>B5: Mejora de la percepción ciudadana |
| **Criterios de inclusión (I)** | I1: El estudio fue llevado a cabo en organizaciones pública o puede ser aplicado a la mismas<br><br>I2: El estudio contribuye a la mejora de algún proceso organizacional |
| **Criterios de exclusión (E)** | E1: Sí el estudio ha sido publicado dos o más veces, el menos maduro es excluido<br><br>E2: Si el estudio no presenta evidencia, se excluye |

**Tabla 4.1. Protocolo de la RSL**

Fuente: (Mitre-Hernández et al., 2015)

### 4.1.3. Proceso de búsqueda

En primer lugar, se ha diseñado un protocolo de revisión sistemática de la literatura para guiar el proceso de búsqueda de la literatura. Consecuentemente, los estudios, que por su contenido, se consideran relevantes se han obtenido desde las bases de datos electrónicas, así como de revistas especializadas, actas de conferencias y demás documentos académicos que contribuyan al objetivo de la RSL.

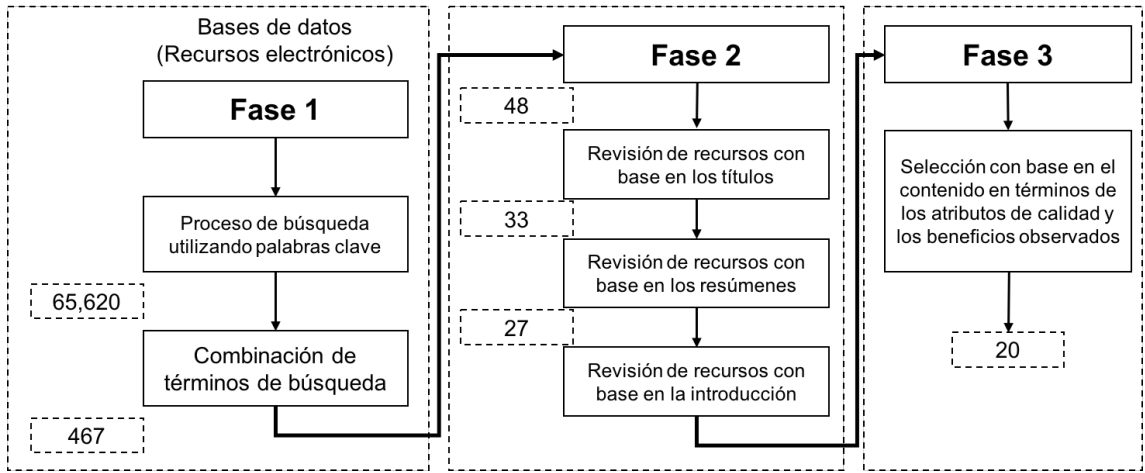

**Ilustración 4-2. Proceso de búsqueda de la RSL**

Fuente: (Mitre-Hernández et al., 2015)

Enseguida, para realizar la búsqueda, en primer lugar, se establecieron los criterios de búsqueda mediante palabras clave: gestión del conocimiento, medición, organización pública, sector público, gobierno electrónico y objetivos estratégicos. Se estableció aceptar estudios en tres idiomas distintos: Inglés, Español y Portugués. El periodo de la RSL abarca desde el año 2005 hasta el año 2014. La búsqueda se llevó a cabo utilizando los comandos lógicos "AND" y "OR" para enlazar dos palabras clave, con el objetivo de obtener el máximo posible de estudios relevantes para dar respuesta a las preguntas de investigación. El proceso de selección de los estudios se realizó en tres fases, como se puede ver en la ilustración 4.2.

La fase 1 se llevó a cabo a través de la búsqueda de información en bases de datos electrónicas en donde se pudo encontrar hasta 65,620 estudios (entre artículos, ponencias, etc.). Posteriormente, refinando los términos de búsqueda se obtuvieron 467 estudios. Consecuentemente, en la fase 2, se realizó un filtrado por título (48), seguido de otro por el resumen o *abstract* (33), y finalmente basados en la introducción se obtuvieron 27 estudios potencialmente relevantes. Finalmente, en la fase 3, de acuerdo con los beneficios y los atributos de calidad esperados, fueron seleccionados los 20 estudios finales de nuestra RSL.

### 4.1.4. Resultados de la RSL

En los siguientes párrafos se resumen los resultados encontrados en la RSL

**Pregunta de investigación 1:** ¿Cuáles son los beneficios (B1 hasta B5) a las propuestas de medición de la GC en la administración pública?

- **B1:** Alineamiento de la GC con los objetivos estratégicos de la organización. La implementación de iniciativas de GC provee información para aplicar sistemas de control desde la gerencia, que le permiten a la organización observar y monitorear el cumplimiento de sus metas (Dalkir, Wiseman, Shulha, & McIntyre, 2007), a través de la medición de sus planes y objetivos estratégicos (P. R. Massingham & Massingham, 2014).
- **B2:** Reducción del índice de corrupción. A través de los sistemas de monitoreo y las herramientas de GC se puede reducir la corrupción e incrementar la transparencia y la rendición de cuentas en la organización (Tung & Rieck, 2005).
- **B3:** Reducción de los juicios laborales. La GC puede reducir los tiempos y costos de los juicios laborales, a través de la redefinición interna de los procesos y la utilización de los recursos de conocimiento de la organización, incidiendo en la eficiencia y en la efectividad de las instituciones públicas (Fersini et al., 2013).
- **B4:** Reducción en tiempos y costos. La GC beneficia a las organizaciones a través de la reducción de tiempos y costos a través de procesos de gestión del conocimiento (Chang Lee et al., 2005) y herramientas tecnológicas (Liberona & Ruiz, 2013); y el desarrollo de trabajadores capaces (Mbhalati, 2014). La GC mejora potencialmente la productividad (Chong Siong Choy et al., 2006).
- B5: Mejora de la percepción ciudadana. La GC mejora la percepción ciudadana como resultado de la mejora de la eficiencia administrativa (Savvas & Bassiliades, 2009) y de la calidad de los servicios públicos (Anantatmula & Kanungo, 2007; Brito et al., 2012; Mbhalati, 2014).

**Pregunta de investigación 2***: ¿Ha habido alguna contribución a la entendibilidad, claridad o credibilidad para la organización pública derivada de la implementación de la iniciativa de GC?

- **AC1: Comprensibilidad.** Los PGC mejoran la entendibilidad en las organizaciones a través de la comprensión de las actividades claves del conocimiento y sus efectos en el desempeño organizacional. Los procesos y sus actividades ayudan a las personas y a las organizaciones a entender mejor el propósito de las iniciativas de GC (Chang Lee et al., 2005; Chong Siong Choy et al., 2006) y su implementación estratégica. Este atributo está relacionado con la claridad (AC2).
- **AC2: Claridad**. La implementación de GC provee una mejor claridad acerca de los roles, tareas y responsabilidades de los funcionarios de la administración pública (Puron-Cid, 2014). Además, la claridad de propósito puede verse incrementada a través de la estandarización y la documentación de políticas de GC, así como procedimiento para asegurar la claridad de roles y procesos (Ragab & Arisha, 2013). También, a través de la medición la organización gana claridad acerca de las necesidades organizaciones para redefinir sus procesos de GC (Bose & Ranjit, 2004).

- **AC3: Credibilidad:** La GC y su evaluación ofrecen una mayor credibilidad acerca de los resultados para las partes interesadas de la organización, y ayuda a la organización a observar el cumplimiento de sus metas (Dalkir et al., 2007). La credibilidad puede verse incrementada a través de una mejor calidad del conocimiento que se produce (Torres-Narváez M, Cruz-Velandia I, & Hernández-Jaramillo J, 2014), esto a su vez le dará mayor credibilidad al conocimiento que se produce (Dalkir et al., 2007).

### 4.1.5. Evaluación de la calidad del estudio

Para evaluar la calidad de esta revisión sistemática de la literatura, se priorizo el nivel de evidencia, tal como se muestra en la tabla 4.3, como un criterio de calidad para la evaluación, similar al utilizado por Ali et al. (2010) and Dybå & Dingsøyr (2008). La calidad de la evaluación del estudio es útil para conocer el grado de precisión de la información obtenida podría decirse que es un criterio de validez y confiabilidad.

En general, se asume que la calidad de los estudios es aceptable. Se obtuvo finalmente una revisión sistemática de la literatura con un factor de calidad de 4.43 dentro de una escala del 1 al 5, como se presenta en la tabla 4.2.

| Estudio | Referencia | Q1 | Q2 | Q3 | Q4 | Q5 | Puntaje total |
|---------|-----------|-----|-----|-----|-----|------|--------------|
| [S1] | (S. Pardo, Coronel, Bertone, & Thomas, 2013) | 1 | 1 | 1 | 1 | 0.50 | 4.5 |
| [S2] | (Dalkir et al., 2007) | 0.8 | 1 | 1 | 1 | 1 | 4.8 |
| [S3] | (Braun & Mueller, 2014) | 1 | 1 | 1 | 1 | 0.5 | 4.5 |
| [S4] | (Fersini et al., 2013) | 0.8 | 1 | 1 | 1 | 0.5 | 4.3 |
| [S5] | (Liberona & Ruiz, 2013) | 0.6 | 1 | 1 | 1 | 0.5 | 4.1 |
| [S6] | (Brito et al., 2012) | 0.8 | 1 | 0.5 | 0.5 | 0 | 2.8 |
| [S7] | (Torres-Narváez M et al., 2014) | 0.8 | 1 | 0.5 | 1 | 0.5 | 3.8 |
| [S8] | (Migdadi, 2009) | 1 | 1 | 1 | 1 | 1 | 5 |

| [S9] | (Alsadhan et al., 2008) | 1 | 1 | 1 | 1 | 1 | 5 |
|---|---|---|---|---|---|---|---|
| [S10] | (Mas-Machuca & Martínez Costa, 2012) | 0.8 | 1 | 1 | 1 | 1 | 4.8 |
| [S11] | (Tabrizi, Ebrahimi, & Delpisheh, 2011) | 0.8 | 1 | 1 | 1 | 1 | 4.8 |
| [S12] | (Loo Geok Pee & Kankanhalli, 2008) | 0.8 | 1 | 0.5 | 1 | 0 | 3.3 |
| [S13] | (Anantatmula & Kanungo, 2007) | 0.8 | 1 | 1 | 1 | 1 | 4.8 |
| [S14] | (Akhavan et al., 2006) | 0.8 | 1 | 1 | 1 | 1 | 4.8 |
| [S15] | (Chong Siong Choy et al., 2006) | 0.8 | 1 | 1 | 1 | 1 | 4.8 |
| [S16] | (Ragab & Arisha, 2013) | 0.6 | 1 | 0.5 | 1 | 0 | 3.1 |
| [S17] | (Ho et al., 2014) | 1 | 1 | 1 | 1 | 1 | 5 |
| [S18] | (Puron-Cid, 2014) | 1 | 1 | 1 | 1 | 1 | 5 |
| [S19] | (Chang Lee et al., 2005) | 1 | 1 | 1 | 1 | 0.5 | 4.5 |
| [S20] | (P. Massingham, 2014) | 0.8 | 1 | 1 | 1 | 1 | 4.8 |
| | **Puntaje promedio** | **0.85** | **1.00** | **0.90** | **0.98** | **0.70** | |
| | **Puntaje total** | **4.43** | | | | | |

**Tabla 4.2. Evaluación de la calidad de la Revisión Sistemática de la Literatura**

Fuente: (Mitre-Hernández et al., 2015)

| Q1: Nivel de evidencia |
|---|
| L0: 0 - Sin evidencia |
| L1: 0.2 - Evidencia obtenida de demostraciones matemáticas |
| L2: 0.4 - Evidencia obtenida de opiniones de expertos u observaciones |
| L3: 0.6 - Evidencia obtenida de experimentos académicos |
| L4: 0.8 - Evidencia obtenida de estudios industriales |
| L5: 1.0 - Evidencia obtenida de prácticas industriales |
| Q2: ¿Son claros los beneficios del estudio realizado? |
| Q3: ¿Son claramente usados los indicadores en el estudio? |
| Q4: ¿Existe claramente mejora a los Atributos de Calidad (AC)? |
| Q5: ¿Están claramente discutidas las limitaciones del estudio? |
| Q2…Q5: 1=Sí; 0.5=Parcialmente; 0=No |

**Tabla 4.3. Preguntas del protocolo de la RSL y nivel de evidencia**
Fuente: (Mitre-Hernández et al., 2015)

### 4.1.6. Estrategias para favorecer los programas de Gestión del Conocimiento en la Organizaciones Públicas

A continuación, en las tablas 4.4, 4.5 y 4.6, con consonancia con la RSL, se presentan algunas estrategias para fomentar los programas de gestión del conocimiento en las instituciones públicas. Estas estrategias están alineadas, como se ha señalado, con los factores críticos de éxito encontrados en la literatura, las dificultades y los posibles beneficios derivados de la implementación de iniciativas de GC en las OSP

.

**Tabla 4.4. Estrategias para el factor Cultura**
Fuente: (Mitre-Hernández et al., 2015)

| Dificultad o problema | Estrategias | Beneficios potenciales |
|---|---|---|
| 1. Débil confianza o colaboración | • (1, 4) Las comunidades de práctica, la construcción de equipos y el liderazgo en los programas de creación del conocimiento [S2], fomentar la difusión de conocimientos entre la organización para evitar la concentración de conocimiento en algunas posiciones [S6] e implementar el uso de credenciales para garantizar la privacidad del conocimiento. | • El intercambio de conocimientos, mejoras en la calidad del Conocimiento.<br><br>• Mejorar el uso y reutilización del conocimiento. |
| 2. Capacidades del personal con e niveles por debajo de lo mínimo satisfactorio | • (3) Promover normas y políticas para la GC [S16], la cooperación y la reciprocidad [S12], talleres y conferencias para compartir conocimientos y crear un ambiente para la GC [S10]. | • Enriquecimiento de los atributos de calidad del conocimiento. |
| 3. Capacidades del personal insatisfactorias con respecto al perfil requerido | • (2) Dar recompensas, reconocimientos o beneficios para las actividades de personal relacionados con el intercambio de conocimientos (alineado con los objetivos estratégicos de la iniciativa de GC) [S6], [S16]. | • Evitar la discrecionalidad en la toma de decisiones y el exceso de burocracia. |
| 3. Duplicidad de funciones y actividades | • (2) Mapeo de capacidades [S20] y estrategias para mejorar las capacidades de personal [S2], [S3], [S18] y desarrollar programas de formación ad-hoc [S14] y fomentar la transferencia de tecnología. | • Mejorar la percepción de los ciudadanos y la credibilidad de los recursos de información con que se cuentan en la organización. |
| 4. Reutilización limitada del conocimiento | • Para tener una contribución adicional en México, los líderes necesitan incentivar más la confianza, generar esquemas de incentivos, favorecer la colaboración y la confianza en sí mismo (autoestima) del personal. | |

**Tabla 4.5. Estrategias para el factor Infraestructura**

Fuente: (Mitre-Hernández et al., 2015)

| Dificultad o problema | Estrategias | Beneficios potenciales |
|---|---|---|
| 1. Los SGC no son adecuados, difíciles de usar, o no están interconectadas ni son sistemas interoperables. | (1, 6) Incrementar la facilidad de uso y acceso de las tecnologías de acceso y sistemas de diseño [S1], sobre la base de la experiencia del usuario [S1]. Evaluación de los SGC con base en atributos de calidad del conocimiento (ACC) [S1], [S18], integración de los sistemas de recuperación de información [S4], implementar ontologías de conocimiento para representaciones del conocimiento [S4], considerar la inclusión de meta-datos, taxonomías, tesauros, documentos en los SGC, así como la gestión de contactos; sistemas de apoyo que favorezcan el aprendizaje, y repositorios de conocimiento [S12], variedad de herramientas para la GC y de apoyo a las actividades diarias [S12], [S16], garantía de conectividad [S20]; integración sistemática de las actividades de gestión [S15]. | • Facilitar el funcionamiento del sistema de GC. <br>• Mejorar el intercambio de conocimientos, <br>• Mejor el proceso de toma de decisiones. <br>• Mejorar las políticas públicas; <br>• Mejorar la eficiencia y la eficacia; <br>• Reducción de costes y disminuciones de esfuerzo. |
| 2. La infraestructura de las TIC es insuficiente | | |
| 3. Las actividades son repetitivas o los procesos sin sentido, y no están vinculadas ni son rastreables | | |
| 4. Falta de procesos estandarizados y canales de comunicación claros | (4) Definir las normas y protocolos para la Interoperabilidad entre las OP [S1], [S2] y establecer relaciones con otras instituciones [S2]. | • Mejorar el funcionamiento interno de los SGC. |
| 5. Falta de medición y control de los activos de | (3, 4, 7) Identificar los conocimientos clave [S5], [S17], [S19], así como los procesos centrales [S18], [S19] y el proceso de | |

| | |
|---|---|
| conocimiento de la organización | documentación [S4], [S16] la normalización, y el proceso de reingeniería organizacional [S14], [S18]. |
| 6. Ausencia de homologación que apoye la aplicación tecnológica en las organizaciones | (5) Medición de la GC, sus actividades y efectos sobre las OP [S10]. Propiciar la medición estratégica para las actividades de gestión [S6], incentivar la mejora continua de la calidad del proceso de GC basado en los estándares de desempeño [S7], [S19]. |
| 7. Muchas actividades básicas deben realizarse manualmente; en consecuencia, los procesos de ejecución son lentos y costosos. | (2) Actualización constante de las plataformas de software como un servicio de apoyo. |

**Tabla 4.6. Estrategias para el factor Estrategia**
Fuente: (Mitre-Hernández et al., 2015)

| Dificultad o problema | Estrategias | Beneficios potenciales |
|---|---|---|
| 1. La visión débil, o la confusión respecto al propósito de la GC | (2, 3, 4, 5, 6) La alta dirección debe dar soporte [S3], [S4], [S8], [S9], [S14], [S18]; mantener el compromiso [S5], así como el apoyo legislativo [S18], las políticas y normas de las TIC [S18], y mejorar conciencia sobre los beneficios de la GC en las OP [S5], y garantizar recursos financieros para la implementación y consolidación de la estrategia de GC [S5]. | • Mejorar los resultados de implementación de las iniciativas de GC.<br><br>• Mejorar los procesos de GC.<br><br>• Fomentar la innovación y mejorar de la calidad de la GC. |
| 2. Los cambios políticos y la no continuidad entre administraciones | | |
| 3. La falta de voluntad política | (1, 3) Establecer la estrategia de GC [S9], [S12], [S14], [S18] y el programa de cambio organizacional [S20], con un enfoque estratégico, y promover la medición de los resultados; asegurar la calidad del contenido y la colaboración [S13], vincular la estrategia de GC a la estrategia global de la institución [S16] y alinear la GC con los recursos de la organización y su presupuesto [S9], y comunicar sus beneficios [S16] entre la organización. Adoptar una visión a largo plazo de la GC [S9], [S12]. | • Garantizar la continuidad de las iniciativas de GC. |
| 4. La falta de liderazgos y apoyo a la GC | | |
| 5. Estatutos de organización que no fomentan la creación de valor, uso y reutilización de activos de conocimiento de la organización | (1) Asignar personal para las iniciativas de GC [S5] y una mejor formación y desarrollo del personal involucrado con la estrategia de GC [S11]. | |

| | |
|---|---|
| 6. La desalineación de la medición en las iniciativas de gestión del conocimiento | (6) La supervisión de los elementos que favorecen la GC [S17] y medir la aplicación del programa de GC [S17] y los resultados y beneficios internos [S17], tener en cuenta las necesidades futuras de las personas, sistemas y procesos [S20] derivados de la GC.

(1) Promover una visión compartida acerca de la intención de la GC.

(6) Establecer objetivos que cuenten con alineamiento político estratégico con las iniciativas de GC.

(1, 2, 3) Informar sobre los beneficios de la GC en maneras que los políticos entienden, reconozcan y aprecian estos en el corto y largo plazo.

(2, 3, 4) Explicar el propósito, el alcance, los roles, en la iniciativa de GC a los interesados en las OP. |

## 4.2.	Propuesta de métricas gestión del conocimiento para organizaciones públicas

### 4.2.1. Introducción

En esta sección se analiza la medición de la GC en las OSP como un elemento central del éxito en las iniciativas de GC. Se analizó a la GC, las OSP y la medición de la GC, además de la relevancia que la GC tiene para las organizaciones de la actualidad. Sin embargo, la medición de la GC aún no ha sido un tema suficientemente explorado dentro de las investigaciones en organizaciones públicas. A partir de la RSL, desarrollamos una propuesta de métricas para la GC en las OP, basada en la RSL.

Es posible, reconocer que no obstante de la profusa y relevante literatura desarrollada en los últimos años en este tema, aún quedan algunas brechas que necesitan cerrarse, relacionadas con la medición de la GC, sus métricas específicas y las herramientas para llevar a cabo la medición de la GC en las OSP, la evidencia empírica de la evaluación de los resultados de desempeño de la GC en las OSP aún es limitada (L.G. Pee & Kankanhalli, 2016, p. 189).

### 4.2.2. Revisión sistemática de la literatura

Con el fin de lograr los objetivos de esta investigación, como ya se ha señalado la SLR se realizó siguiendo las directrices propuestas por Kitchenham & Charter (2007), y los procedimientos desarrollados por Ding et. al. (2014), Ali, Ali Babar, Chen, & Stol (2010), Dybå & Dingsøyr and Garlatti et al.(2014). En primer lugar, establecimos un protocolo de investigación que incluye los criterios de inclusión y exclusión, con el fin de obtener información pertinente de fuentes primarias.

La RSL se circunscribió a un tema académico específico: la medición de la GC en las OSP, con especial atención en observar todas las prácticas llevadas a cabo por académicos y profesionales en el tema, almacenadas en bases de datos académicas. Por esa razón, el primer paso de esta RSL fue la búsqueda en bases de datos académicas, como EBSCO, Scopus y Web of Science; con el fin de obtener estudios primarios (de revistas, libros, conferencias, entre otros), utilizando los operadores "AND" y "OR" para combinar los términos de búsqueda, y así obtener la mayor cantidad de información disponible y relevante.

Acto seguido, de acuerdo con el proceso de selección y evaluación, seleccionamos los estudios primarios para esta RSL. Por último, se evaluó cada estudio de conformidad con el protocolo de búsqueda de la RSL, atendiendo a las Preguntas de Investigación (PI) y los Atributos de Calidad (AC), previamente definidos, véase la tabla 4.8.

| Pasos | Estudios encontrados |
|---|---|
| (1) Estudios obtenidos de las bases de datos usando combinación de los términos de búsqueda establecidos en el protocolo | 3289 |
| (2) Estudios obtenidos después de la depuración utilizando los criterios de inclusión o exclusión | 1119 |
| (3) Estudios seleccionados con base en el título del estudio | 521 |
| (4) Estudios seleccionados con base en el resumen del estudio | 151 |
| (5) Estudios seleccionados con base en la introducción del estudio | 72 |
| (6) Estudios seleccionados con base en el contenido del estudio | 20 |
| (7) Estudios seleccionados de acuerdo con las Preguntas de Investigación (PI) y los Atributos de calidad (AC) | 17 |

**Tabla 4.7. Proceso de búsqueda**
Elaboración propia (2016)

| Elementos | Descripción |
|---|---|
| Criterios de inclusión (I) y exclusión (E); | (I1) Contiene las palabras clave: GC, Organización Pública, Sector Público, Gobierno electrónico (*e-government*), medición, evaluación de resultados, métricas. (I2): Periodo de publicación del año 2005 al 2015 (E1): El documento está en algún otro idioma aparte del Inglés, Español o Portugués (E2): No está disponible el documento completo |

| Preguntas de Investigación (PI) | RQ1: ¿Cuáles son los principales enfoques o planteamientos de medición de GC? RQ2: ¿Cuáles son los principales indicadores, criterios de medición propuestas por la literatura? RQ3: ¿Cuáles son los niveles de medición propuestos o implementados? RQ4: ¿Cuáles serían los beneficios relacionados con la medición de la gestión del conocimiento en las organizaciones del sector público? |
|---|---|
| Atributos de Calidad (AC) | AC1: El estudio se llevó a cabo en organizaciones del sector público AC2: Hay pruebas de validación empírica o evidencia AC3: El método de estudio es cuantitativo o cualitativo |

**Tabla 4.8. Elementos del protocolo de búsqueda**
Fuente: Elaboración propia (2016)

### 4.2.3. Resultados de la RSL: Preguntas de investigación

**PI1:** En cuanto a los principales enfoques para describir, analizar y realizar una medición de la GC en OSP se observó que la perspectiva financiera (Abdel-Maksoud, Elbanna, Mahama, & Pollanen, 2015; Chang Lee et al., 2005; C. W. Chong & Chong, 2009; Choy Chong et al., 2011; Dehghani & Ramsin, 2015; Hu et al., 2015; R. S. Kaplan & Norton, 2000; Robert S Kaplan & David, 2007; Cheng Sheng Lee & Wong, 2015; K. Y. Wong et al., 2015), cuantitativa (Goldoni & Oliveira, 2010; Kuah et al., 2012; Powell & Snellman, 2004) o perspectiva de valor (Hu et al., 2015) se mantiene como la más popular entre la comunidad académica; y que, además, existe una mezcla de perspectivas, entre elementos cuantitativos y cualitativos (Dalkir et al., 2007; Goldoni & Oliveira, 2010; K. Y. Wong et al., 2015) que está ganando atención.

Esto podría ser debido, principalmente, a algunos investigadores y profesionales están tratando de evaluar el conocimiento y el valor de la gestión del conocimiento vinculación distintos enfoques, con el fin de validar sus propuestas y para dar respuesta a la configuración tan única y específica que suponen las organizaciones del sector público.

De hecho, como se ha revisado antes, los fines utilitarios de las OSP difieren de los del sector privado. Mientras que el sector privado se centra más en ser rentable; el sector público, por su parte, se centra en las personas, en la forma de mejorar la calidad de los servicios públicos y el proceso para la toma de decisiones para el progreso y el bienestar general. Por otro lado, la perspectiva no financiera se enfoca en medir el rendimiento de los recursos de conocimiento (Abdel-Maksoud et al., 2015).

**PI2:** A pesar de que, aún en la actualidad, no existen métricas ni criterios estandarizados para la medición del conocimiento o la gestión del conocimiento (K. Y. Wong et al., 2015). Encontramos que los autores seleccionados de la RSL proponen métricas e indicadores que están relacionados con el uso (Ding et al., 2014), la usabilidad del conocimiento (Serrano Cinca et al., 2003) y el uso de sistemas de GC (Goldoni & Oliveira, 2010; Ragab & Arisha, 2013) para producir conocimiento nuevo y diferente (Abdel-Maksoud et al., 2015). Además, algunos estudios observan indicadores relacionados con los productos de la gestión del conocimiento (Hu et al., 2015; Powell & Snellman, 2004) y algunas veces están relacionados con distintos tipos de resultados del PGC (Kuah et al., 2012), o las mejoras organizacionales derivadas de los resultados de la GC (P. R. Massingham & Massingham, 2014; K. Y. Wong et al., 2015) o los resultados financieros asociados o esperados de la iniciativas de GC (Chang Lee et al., 2005; Dalkir et al., 2007), por ejemplo, dentro del *Balance Score Card* (Choy Chong et al., 2011).

De esta manera confirmamos que, sin importar que en la literatura académica la medición es reconocida como un paso crucial, posterior a la implementación de iniciativas de GC (Cheng Sheng Lee & Wong, 2015), aún los límites y clasificaciones de métricas e indicadores de desempeño de la GC en las OSP no sea han hecho específicas ni son claras. En suma, observamos que las métricas de GC están principalmente relacionadas con el uso o la usabilidad del conocimiento y los resultados o productos de la misma, e incluso algunos de ellos están relacionados con la mejora a los factores críticos de éxito de la GC en la organización.

**PI3:** En relación con los niveles de medición del conocimiento y de las iniciativas de GC, hemos encontrado en la literatura académica que principalmente se realiza una evaluación de las etapas de implementación de las iniciativas de GC (Choy Chong et al., 2011), el desempeño del proceso de GC (Chang Lee et al., 2005; Dalkir et al., 2007; Dehghani & Ramsin, 2015; Goldoni & Oliveira, 2010; Kuah et al., 2012; Cheng Sheng Lee & Wong, 2015; K. Y. Wong et al., 2015) y los productos de la GC (Abdel-Maksoud et al., 2015; Bose & Ranjit, 2004; Choy Chong et al., 2011; Dalkir et al., 2007; Goldoni & Oliveira, 2010; Hu et al., 2015; P. R. Massingham & Massingham, 2014; Powell & Snellman, 2004; H. Tsui et al., 2009).

**PI4**: Finalmente, se puede estar de acuerdo en que la GC necesita demostrar su valor. Consecuentemente, en relación con los beneficios potenciales, esperados o percibidos de la medición de la GC, observamos que están enfocados en la mejora del desempeño organizacional (Abdel-Maksoud et al., 2015; Choy Chong et al., 2011; Kuah et al., 2012; Kuah & Wong, 2011; Cheng Sheng Lee & Wong, 2015; Ragab & Arisha, 2013; K. Y. Wong et al., 2015) el incremento en la satisfacción de los clientes o usuarios (Bose & Ranjit, 2004; Serrano Cinca et al., 2003) mejoras en la calidad de los productos o servicios, a demostrar el valor de las iniciativas de GC (Ragab & Arisha, 2013), beneficios organizaciones de largo plazo (Choy Chong et al., 2011), como la competitividad, la eficiencia en las operaciones y la innovación organizacional (Powell & Snellman, 2004), la alineación estratégica (P. R. Massingham & Massingham, 2014), la efectividad, la eficiencia y la mejora de los procesos de GC (Goldoni & Oliveira, 2010; Hu et al., 2015). Además, de algunos otros beneficios como la mejora de la calidad en el diseño de políticas (H. Tsui et al., 2009)

### 4.2.4. Propuesta de Métricas de Gestión del Conocimiento como resultado de la RSL

En esta sección, de acuerdo con la RSL realizada, proponemos algunas métricas para evaluar y medir la GC en las organizaciones públicas. Se han definido algunas métricas con base en la RSL que podrían (i) demostrar el valor de las iniciativas de GC; (ii) ayudar a mejorar sistemática y continuamente el proceso de GC; e incrementar la satisfacción de los usuarios o la percepción de los ciudadanos; y proveer de una ventaja competitiva sustentable en el largo plazo a las OP. En la tabla 4.9, se presenta la propuesta organizada por niveles de análisis (conocimiento, iniciativas de GC, productos de la GC y resultados de la GC.

| Nivel | Métricas |
|---|---|
| Conocimiento | • Comprensibilidad, Claridad, Precisión, Recuperabilidad, Trazabilidad, Consistencia, Exactitud, Integridad, Utilidad, Credibilidad (Ding et al., 2014, p. 555) y uso (Serrano Cinca et al., 2003) del conocimiento.<br>• Uso de los sistemas de gestión del conocimiento (Goldoni & Oliveira, 2010; Ragab & Arisha, 2013)<br>• Número de usuarios frecuentes del sistema de gestión del conocimiento (Kuah et al., 2012)<br>• Cantidad de conocimiento almacenado en sistemas tradicionales (sistemas de llenado manual y computadoras)<br>• Calidad del conocimiento almacenado<br>• Número de unidades de propiedad intelectual (Cheng Sheng Lee & Wong, 2015) |
| Iniciativas de GC y PGC | • Incremento en la calidad de los resultados del proceso de gestión del conocimiento (Chang Lee et al., 2005)<br>• Incremento en la capacidad de adquisición del conocimiento (Cheng Sheng Lee & Wong, 2015), creación, generación, aplicación y utilización del con conocimiento (Cheng Sheng Lee & Wong, 2015)<br>• Incremento en la contribución a los repositorios organizacionales (Goldoni & Oliveira, 2010)<br>• Número de veces que los empleados asisten a sesiones de entrenamiento, seminarios o cursos para adquirir nuevos conocimiento<br>• Número de veces que los empleados<br>• Número de veces los empleados adquirir conocimiento del sus superiores<br>• Número de veces que los empleados contactan con los clientes o proveedores para adquirir nuevo conocimiento<br>• Cantidad de tiempo dedicado a navegar por Internet para adquirir conocimientos<br>• Número de veces que los empleados acceden a repositorios de conocimiento de la empresa para adquirir nuevos conocimientos<br>• Número de veces que los empleados trabajan en equipos para crear nuevos conocimientos |

- Número de veces que los empleados participar en sesiones de reflexión para crear nuevos conocimientos
- Cantidad de tiempo dedicado a codificar y almacenar el conocimiento en repositorios de conocimiento de la organización
- Cantidad de tiempo dedicado a la actualización de repositorios de conocimiento de la organización
- Nivel de disposición de los empleados para contribuir a repositorios de conocimiento de la organización
- Número de veces que los empleados participar en discusiones informales para compartir conocimientos.
- Frecuencia de las sesiones o reuniones para desarrollar conocimiento
- Frecuencia con que los empleados utilizan herramientas tecnológicas para transferir el conocimiento
- Inversión en básica en Tecnologías de la Información y la Comunicación (computadoras, internet, intranet, etc.)
- Inversión en infraestructura de la organización (sala de reuniones, estante de presentación, etc.)
- Frecuencia del mantenimiento de la infraestructura de la organización
- Grado de alineación entre la estrategia de GC y la estrategia de la organización
- Claridad de la estrategia de GC de la organización
- Grado de conciencia y apoyo hacia la estrategia de gestión del conocimiento de los empleados de la organización
- Monto del presupuesto asignado para iniciativas de GC
- Número de empleados que participan en iniciativas de GC
- Cantidad de tiempo asignado para los empleados para llevar a cabo actividades de gestión del conocimiento
- El nivel de esfuerzo puesto en el reclutamiento de empleados
- El nivel de esfuerzo puesto en la retención de empleados

| | |
|---|---|
| **Productos y Resultados de la GC** | • Número de actividades de desarrollo profesional organizados para los empleados (Cheng Sheng Lee & Wong, 2015)<br>• Nuevos conocimientos, ideas y soluciones creadas por empleado;<br>• Nuevos productos, invenciones y servicios generados (Hu et al., 2015; Powell & Snellman, 2004)<br>• Modelos matemáticos, resultados de investigación, artículos científicos (Hu et al., 2015)<br>• Documentos y artículos a que se accede o descargados por empleado;<br>• Documentos y artículos subidos o actualizados por empleado;<br>• Comunidades de práctica activas, grupos de investigación y grupos de intereses especiales;<br>• Comunicaciones por empleado por mes<br>• Problemas resueltos e ideas implementadas por empleado;<br>• Los activos de conocimientos generados por año (Kuah et al., 2012)<br>• Aceleración de los tiempos de aprendizaje<br>• Aumento en el intercambio de experiencias<br>• Disminución de la brecha de capacidades entre las habilidades del personal y los requisitos del trabajo<br>• Aumento de la eficiencia o el tiempo del ciclo de búsqueda para encontrar los conocimientos necesarios para realizar una nueva tarea<br>• Mejora de la percepción del valor de la organización de las partes interesadas<br>• Aumento de la recompensa interna y el reconocimiento a empleados<br>• Mejora de la moral y la productividad del personal (P. R. Massingham & Massingham, 2014)<br>• Mejora de las capacidades de los empleados;<br>• Uso eficiente del presupuesto asignado<br>• Cantidad de productos o servicios prestados<br>• Aumento de la satisfacción de los clientes, o la percepción de los ciudadanos<br>• Recompensas a los empleados (Abdel-Maksoud et al., 2015)<br>• Aumento de la percepción de la comunicación interna (Goldoni & Oliveira, 2010) |

- Ingresos generados por la propiedad intelectual
- Cantidad de recompensas dada a los empleados que crean nuevos conocimientos, ideas y soluciones
- Número de veces que los empleados aplican propuestas útiles o ponen en práctica sus ideas de mejora
- Número de veces que los empleados aplican el conocimiento para resolver problemas
- Número de nuevos productos o servicios lanzados (Cheng Sheng Lee & Wong, 2015)

**Tabla 4.9. Métricas de GC para las OP**
Fuente: Elaboración propia (2016)

# 5. Modelo de evaluación de los factores previos para la gestión del conocimiento en las organizaciones públicas

Esta sección tiene como objetivo definir un modelo de evaluación de factores culturales, estratégicos y de infraestructura para obtener un diagnóstico previo a la implementación de iniciativas de GC en las OP. Para posteriormente, ejecutar y discutir la implementación del modelo de evaluación (previamente mencionado) para conocer la contribución (influencia) de los constructos de cultura, infraestructura y estrategia sobre la GC.

En ese sentido, en primer término se analiza la composición de la Administración Pública en un contexto determinado, en este caso en Costa Rica; posteriormente se analizan los factores propuestos por la literatura; y finalmente se propone un modelo de medición que utiliza un instrumento de evaluación empleado entre distintas organizaciones públicas costarricenses. Finalmente se analizan los resultados y se ofrecen algunas conclusiones preliminares.

### *Composición y descripción de la Administración Pública Costarricense*

La Administración Pública Costarricense, desde el punto de vista legal, está constituida por el Estado y los demás entes públicos, cada uno con personalidad jurídica y capacidad de derecho público y privado (Art. 1, Ley General de Administración Pública, No. 6227) (Asamblea Legislativa de la República de Costa Rica, 1978). Así, actuará sometida al ordenamiento jurídico y sólo podrá realizar aquellos actos o prestar aquellos servicios públicos que autorice dicho ordenamiento, según la escala jerárquica de las fuentes del derecho.

Desde el punto de vista organizacional, está conformada por los tres Poderes del Estado: Ejecutivo, Legislativo y Judicial, en un enfoque de pesos y contrapesos; adicionalmente, se considera al Tribunal Supremo de Elecciones, como un cuarto poder, dadas sus prerrogativas y el respeto a la democracia que impera en el país. Dicha conformación está respaldada por la Constitución Política de la República, de la cual se desprende que los órganos constitucionales superiores de la Administración del Estado serán: el Presidente de la República, los Ministros, el Poder Ejecutivo y el Consejo de Gobierno. El Poder Ejecutivo lo conforman el Presidente de la República y el Ministro del Ramo.

Una de sus principales características es que cuenta con 329 instituciones en lo que se denomina sector público costarricense, que incluye Poderes de la República, Órganos del Poder Legislativo, Organismo Electoral, Ministerios,

Órganos adscritos a Ministerios, Órganos adscritos a la Presidencia, Instituciones autónomas, Órganos adscritos a Instituciones autónomas, Instituciones semiautónomas, Empresas públicas estatales, Empresas públicas no estatales, Entes públicos no estatales, Entes Administradores de Fondos Públicos, Municipalidades y Consejos Municipales de Distrito (MIDEPLAN, 2014).

En el marco administrativo, cada institución del Estado tiene mandatos específicos dictados por el marco jurídico, las políticas públicas y la estrategia nacional que se denominada Plan Nacional de Desarrollo, en las cuales se incluyen las etapas de planificación, presupuestación, ejecución y evaluación con instrumentos como el plan estratégico y operativo de cada organización. Esto genera una serie de datos e información que se constituyen en insumos para la mejora continua, procesada, sistematizada y ensamblada en documentos como informes de resultados y evaluaciones periódicas que aplican instituciones como el Ministerio de Planificación Nacional y Política Económica (MIDEPLAN), el Ministerio de Hacienda (MH) y la Contraloría General de la República (CGR).

Finalmente, la Administración Pública Costarricense cuenta con un andamiaje legal y organizacional sólido, que requiere una serie de esfuerzos interinstitucionales para que los datos y la información se generen y transmitan de forma adecuada. De tal manera que, fomentar el uso y aplicación de la misma para incentivar la estabilidad socio-económica, mejorar los servicios públicos y aumentar la competitividad en el actual contexto global es una tarea en desarrollo.

### *Contexto nacional*

En los resultados del informe sobre la encuesta de las Naciones Unidas de gobierno electrónico (*e-government*) 2014 (United Nations, 2014b), Costa Rica se ubicaba en el lugar 54 de entre 193 países, 23 posiciones por debajo con respecto al año 2012. Esta encuesta mide el Índice de Desarrollo del Gobierno Electrónico (*E-Government Development Index*, EGDI), integrado por la (i) oferta de servicios en línea (*online services*), (ii) la conectividad en telecomunicaciones y (iii) la capacidad humana de las naciones.

Además, para el caso de Costa Rica, la Secretaría Técnica de Gobierno Digital (STGD) ha formulado un Programa Gobierno Digital (PGD), y creó la Comisión Intersectorial de Gobierno Digital, un órgano de coordinación y definición de la política pública y la ejecución de proyectos; lo que ha sido esencial en el avance del país hacia la sociedad de la información y el conocimiento (UCR, 2013). También se han presentado diversas iniciativas para incorporar el tema de las

TIC en las políticas públicas (UCR, 2013, pp. 59–62), como un instrumento que apoye el combate a la corrupción, una mayor agilidad de la función pública y el ahorro de recursos públicos (UCR, 2013).

En este sentido, Rabinovitch (2009) indica que los cambios significativos no ocurren necesariamente por las mejoras tecnológicas, sino por las mejoras en el comportamiento, es decir, si bien es cierto que las TIC y el gobierno electrónico facilitan el acceso a información, ejecución de trámites y la transparencia, también es cierto que "el proceso de GC no es un fin en sí mismo, pero sí es un medio para mejorar la satisfacción ciudadana" (Rabinovitch, 2009, p. 3).

Así, el modelo de sociedad evoluciona y requiere organizaciones abiertas, flexibles, interconectadas, orientadas al funcionamiento en red y con capacidad de reacción inmediata (Martínez, Lara-Navarra, y Pilar, 2006). Consecuentemente, las OSP no pueden permanecer aisladas y ajenas a las transformaciones que las TIC producen en todo el orbe, sino que deben integrarse por medio del tratamiento de la GC para aprovechar toda su potencialidad, con el fin de mejorar su posición de servicio a los ciudadanos y a la sociedad.

En el caso de Costa Rica la GC ha tomado mayor énfasis a partir del año 2000, por ejemplo, con investigaciones que inician a sistematizar alguna presencia de los elementos de la GC en las OP. Así, se pueden identificar investigaciones como "Gestión del conocimiento en el primer nivel de atención de salud, en Heredia (Costa Rica)" de Villanueva (2002), en donde se explica que el estudio se origina con el reconocimiento de que el sistema de salud pública puede estar funcionando ajeno al momento histórico de reconfiguraciones aceleradas que se vive en el mundo, producto del avance en las comunicaciones que están influyendo de manera decisiva en los cambios de escenarios y paradigmas de cualquier proceso organizacional.

Esto hace suponer un estadio de madurez en la cual las organizaciones públicas inician la identificación del proceso básico de recopilar datos, sistematizar resultados, generar información y gestionar conocimiento (tácito y explícito), con una implementación incipiente, ya que, "hacer accesible la información y el conocimiento es clave para la productividad y el bienestar de una sociedad, además de la innovación", conclusión de la conferencia "Gestión del conocimiento en el siglo XXI" del Dr. Eugenio Trejos (2010), organizada por la Escuela de Bibliotecología y Ciencias de la Información de la Universidad de Costa Rica.

Adicionalmente, el Dr. Trejos (2010) manifestó que "la cantidad de datos e información que se produce en Internet es enorme, por eso su sistematización y clasificación representa todo un reto para profesionales del área de bibliotecología e informática […] Los datos permiten generar información y ésta a su vez, conocimiento". Por ello, indican Riquelme, Cravero y Saavedra (2008), que el conocimiento es una mezcla fluida de la experiencia, valores, información contextual y visión experta que proporciona un marco teórico para evaluar e incorporar nuevas experiencias e información.

Dado que desde la función pública se busca el desarrollo sostenible, económico y social, las sociedades que aprovechan mejor el conocimiento, son las que probablemente, producto de la actividad económica, van a producir mayor bienestar (Trejos, 2010). Así, el incremento de la productividad en las organizaciones se basará más en el conocimiento que en la información (Villanueva, 2002). La Dra. Lizette Brenes, Vicerrectora de Investigación de la Universidad Estatal a Distancia (UNED), indicó en su charla denominada "Las redes como estrategia clave para las sociedades del conocimiento" (2012), que:

> "[…] en la era cristiana, la humanidad tardó 1750 años en duplicar el conocimiento; durante el siglo pasado, el conocimiento se duplicó cada 50 años y en este siglo se duplica cada cinco años. Las proyecciones estiman que para el año 2020, este se duplicará cada 73 días, un volumen de conocimiento que hace imposible que la forma de organización tradicional pueda tan solo asimilar".

Esta situación conduce a las condiciones en las cuales puede operar la GC en la AP costarricense: existe un marco legal que promueve el acceso a la información, y las organizaciones están obligadas por Ley a incluir y brindar información a los usuarios que la soliciten, por ejemplo:

Ley del Sistema Nacional de Archivos (No. 7202 de 24 de octubre de 1990):

> Artículo 10. "Se garantiza el libre acceso a todos los documentos que produzcan o custodien las instituciones a las que se refiere el artículo 2o. de esta Ley. Cuando se trate de documentos declarados secreto de Estado, o de acceso restringido, perderán esa condición después de treinta años de haber sido producidos, y podrá facilitarse para investigaciones de carácter científico - cultural, debidamente comprobados, siempre que no se irrespeten otros derechos constitucionales."

Ley del Sistema de Estadística Nacional (No. 7839 de 15 de octubre del 1998):

> Artículo 4. "Las dependencias y entidades que conforman el SEN recopilarán, manejarán y compartirán datos con fines estadísticos, conforme a los principios de confidencialidad estadística, transparencia, especialidad y proporcionalidad".

Ley de Protección al ciudadano del exceso de requisitos y trámites administrativas, (No. 8220 de 4 de marzo del 2002 y su reforma No. 8990 de 29 de septiembre del 2011):

> Artículo 5. "Todo funcionario, entidad u órgano público estará obligado a proveerle al administrado información sobre los trámites y requisitos que se realicen en la respectiva unidad administrativa o dependencia. Para estos efectos, no podrá exigirle la presencia física al administrado, salvo en los casos en que la ley expresamente lo requiera".

Los elementos anteriores, contextualizan brevemente la situación de la GC en la AP costarricense con base en literatura científica y memorias de charlas que apuntan hacia la importancia de la GC para el sector público en Costa Rica. Por lo anterior señalado, podemos argumentar que para las OP costarricenses, la GC es un tema estratégico, ya que, como se dijo en apartados anteriores, el Estado es el mayor generador de datos e información, y por ello, se indican a continuación un conjunto de factores que le brindan tal relevancia:

Los países de la OCDE han agendado a la GC como parte de su estrategia país (OCDE, 2003), y Costa Rica se encuentra en proceso de adhesión a este organismo.

El Estado por medio de las políticas públicas direcciona el crecimiento económico, y dado que el conocimiento es un factor de producción, éste puede propiciar un aumento en la competitividad, dicho de otra forma "lo interesante del conocimiento, a diferencia de otros factores de producción […] Puede que se vuelva viejo, pero como tal no se gasta. No se pierde si lo presto a otros, más bien, el potencial de que crezca es mayor si lo comparto" (Trejos, 2010).

Costa Rica cuenta con una plataforma de servicios en línea de diversas instituciones públicas, y esto provoca que "no sólo hay que compartir la información, sino que también significa que hay que hacerla más accesible y conectar conocimientos unos con otros" (Trejos, 2010).

La AP Costarricense trabaja basada en Sectores (Salud, Trabajo, Educación, etc.), los cuales están integrados por una serie de instituciones que tienen relaciones temáticas, funcionales y estructurales con responsabilidades conjuntas en el logro de resultados macro; así bajo el enfoque de redes:

> "[…] una red que no está conectada con otras redes no funciona bien. La red debe estar conectada con la mayor cantidad de redes posibles y debe estar siempre en crecimiento. Es necesario facilitar la identificación de las personas adecuadas en el momento oportuno para que sus conocimientos puedan ser complementarios con otros. Si una red no facilita eso, tiene una gran debilidad" (Brenes, 2012).

Es importante señalar que en las OP costarricenses se han realizado importantes esfuerzos por introducir las TIC en sus procesos. No obstante, este impulso no ha sido el mismo en todas las instituciones y, por lo tanto, su impacto no ha tenido el efecto que se esperaría en el país (UCR, 2013, p. 163). Sin embargo, la transición hacia una nueva forma de gestión de la AP es un desafío abierto que requiere el diálogo y los esfuerzos de los actores que interactúan e intervienen en la formulación de las políticas públicas, además de un compromiso y voluntad política (Mbhalati, 2014) que orqueste las reformas necesarias para materializar la modernización de las OP costarricenses. La gestión de estas acciones y procesos resulta imprescindible para que los servicios públicos funcionen de forma inteligente (Riquelme et al., 2008, p. 49) e integrada.

### 5.1.1. Factores críticos de éxito para la Gestión del Conocimiento en las Organizaciones Públicas

Los FCE para la GC son "elementos que facilitan la adquisición, creación, intercambio y transferencia del conocimiento dentro de y entre las organizaciones" (Pinho et al., 2012, p. 217), y que deben ser abordados con el fin de asegurar la implementación exitosa de las iniciativas de GC (S. M. Allameh et al., 2011; Alsadhan et al., 2008). Generalmente, se atienden aquellos factores del ambiente interno de la organización que pueden controlarse (C. S. Choy, 2006).

Con base en una extensa revisión de la literatura académica, para la presente sección se han analizan los factores: (i) cultura, (ii) infraestructura y (iii) estrategia, como elementos que acompañan, y hacen posible, la implementación de la GC en las OP. Estos factores se analizan debido a que han sido profusamente relacionados con el éxito de las iniciativas de GC (S. M. Allameh et al., 2011; Alsadhan et al., 2008; C.W. Holsapple y Joshi, 2001; Mas-Machuca y Martínez Costa, 2012; Pee y Kankanhalli, 2008; Kuan Yew Wong y Aspinwall, 2005; Kuan Yew Wong, 2005).

### 5.1.2. Metodología

Esta es una investigación cuantitativa de corte descriptivo - correlacional, que tiene por objetivo observar la influencia de los factores cultura, infraestructura y estrategia, y sus componentes en la GC, para determinar cuáles de éstos son los que tienen una mayor relevancia y cuáles deberían ser atendidos previo a la implementación de iniciativas de GC en las instituciones de las OP costarricenses. Además, se busca conocer cuáles son los beneficios, relacionados con la GC, que son más percibidos por los funcionarios.

Para ello, se ha definido el alcance del estudio como da muestra la ilustración 5.1. Esta investigación además busca proponer recomendaciones prácticas, alineadas con los factores críticos de éxito de la GC, a los tomadores de decisiones en las instituciones públicas, para mejorar la implementación de este tipo de iniciativas, y contribuir al desarrollo teórico en la materia.

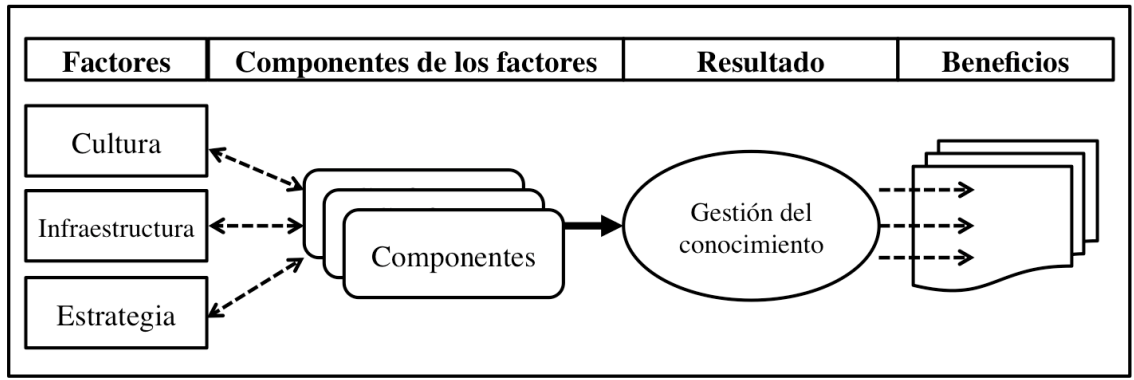

**Ilustración 5-1. Alcance del estudio propuesto**
Fuente: (Pérez López-Portillo, Romero Hidalgo, & Mora Martínez, 2016)

### 5.1.3. Instrumento de evaluación

Para el presente estudio se utiliza un instrumento de evaluación diseñado con base en una extensa revisión de la literatura académica sobre los factores críticos de éxito en las iniciativas de GC, sus componentes y los beneficios observados, de conformidad con el alcance del estudio presentado en la ilustración 5.1. Para ello, se han integrado las variables y se han construido los ítems del instrumento de evaluación, considerando las siguientes fases:

(1) Se realizó una revisión sistemática de la literatura, utilizando bases de datos académicas, para observar los factores críticos de éxito y sus componentes en las iniciativas de GC, así como los beneficios asociados. El criterio de inclusión de la literatura fue su representatividad interdisciplinar, su validez y contribución teórica y empírica al debate sobre la GC;

(2) Con base en los factores observados se diseñaron los ítems del cuestionario y se integraron con respecto a los componentes de cada factor, asimismo se formularon los ítems sobre el proceso de GC y sobre los beneficios asociados;

(3) Se integró una versión preliminar del cuestionario que fue consultada, nutrida y revisada con académicos expertos en el tema vía correo electrónico. Derivado de la retroalimentación se estableció una nueva versión con ajustes semánticos, principalmente;

(4) Posteriormente; se aplicó un pre-test a un pequeño grupo de funcionarios de las OSP para evaluar el nivel de entendimiento de los ítems.

(5) Finalmente, se hicieron algunos ajustes menores en los ítems y se obtuvo una versión final del cuestionario.

(6) La versión final del cuestionario se presenta en el Apéndice I, está integrada por 54 ítems en escala Likert de 5 puntos, en donde 1 representa "totalmente en desacuerdo", hasta 5 que representa "totalmente de acuerdo". El cuestionario se divide en 5 secciones: (1) datos generales; (2) evaluación del factor cultural; (3) evaluación del factor infraestructura; (4) evaluación del factor estrategia; y (5) beneficios percibidos y atributos de calidad del conocimiento.

(7) Enseguida, con el apoyo de las bases de contactos del Centro de Investigación y Capacitación en Administración Pública (CICAP) de la Universidad de Costa Rica, el cuestionario fue distribuido de forma virtual a través de correos electrónicos entre distintas instituciones de la AP costarricense. La recolección de los datos se realizó durante un periodo de cuatro semanas durante el mes de mayo del año 2015.

Es importante señalar que el cuestionario incluía una nota sobre la confidencialidad en el tratamiento de la información, por tal motivo en este estudio solamente se presentan datos generales sobre los participantes y las instituciones.

### 5.1.4. Participantes

Los participantes fueron seleccionados de manera intencional, utilizando bases de datos de contacto. Se ha tomado en cuenta el criterio de inclusión que fuesen trabajadores de las OSP costarricense y el criterio de exclusión que sean cuestionarios contestados en su totalidad. De esta manera, se han obtenido un total de 267 cuestionarios válidos, respondidos por funcionarios de un total de 80 distintas instituciones de la AP costarricense. En la tabla 5.1, se muestra la descripción de los participantes en este estudio.

| Tamaño de la organización | | | | |
|---|---|---|---|---|
| Menos de 50 (5.99%) | Más de 50 y hasta 100 (10.86%) | Más de 101 y hasta 200 (9.74%) | Más de 201 (73.41%) | |
| **Formación académica de los participantes** | | | | |
| Bachillerato universitario (6.82%) | Licenciatura (57.95%) | Maestría (33.33%) | Doctorado (1.89%) | |
| **Puesto del funcionario en la organización** | | | | |
| Nivel 1 (32.33%) | Nivel 2 (42.11%) | Nivel 3 (18.8%) | Nivel 4 (4.14%) | Otros (2.63%) |
| **Antigüedad del funcionario en la organización** | | | | |
| 0 a 10 años (37.02%) | 11 a 20 años (25.57%) | 21 a 30 años (28.24%) | 31 o más (9.16%) | |

**Tabla 5.1. Descripción de las características de los participantes en el estudio**

Fuente: (Pérez López-Portillo et al., 2016)

Dada las múltiples respuestas sobre los puestos asignados a los funcionarios, con el objetivo de una mejor categorización y presentación de los resultados, los puestos han sido clasificados de la siguiente manera:

- **Nivel 1** (operativos): puestos de técnico, secretaria, de apoyo, asistente, ejecutivo, encargado, oficial, profesional 1 y 1B, administrativo, analista;
- **Nivel 2** (mandos intermedios): jefe, auditor, administrador, asesor, fiscalizador, líder, planificador, supervisor, responsable de área, profesional 3 y 4;
- **Nivel 3** (mandos medios): coordinador, contralor, director administrativo o de área, profesional 4 o superior;
- **Nivel 4**: director general, subdirector y alta dirección; y
- **Otros**: docente, proveedor (sin embargo, no han sido considerados como válidos, dado el criterio de inclusión mencionado).

### 5.1.5. Modelo de investigación

Para la construcción del modelo de investigación, ilustración 5.2, se han tomado como referencia algunos estudios previos que operan variables latentes sobre alguna variable convergente (Chou, Wang, & Tang, 2015; Y. J. Kim, Lee, Koo, & Nam, 2013), con la intención de conocer, mediante un modelo de prueba de hipótesis la contribución o peso específico que tiene una variable sobre otra, con la intención de conocer su contribución. Para el presente estudio, en congruencia con el objetivo general y los particulares de este trabajo de investigación, se han propuesto las siguientes hipótesis, a manera de supuestos de investigación:

- H1. El factor cultura y sus elementos contribuyen significativamente a la gestión del conocimiento;
- H2. El factor es infraestructura y sus elementos contribuyen significativamente a la gestión del conocimiento;
- H3. El factor estrategia y sus elementos contribuyen significativamente a la gestión del conocimiento.

Consecuentemente, en la sección análisis de datos y resultados se explican los hallazgos más relevantes a partir del estudio propuesto.

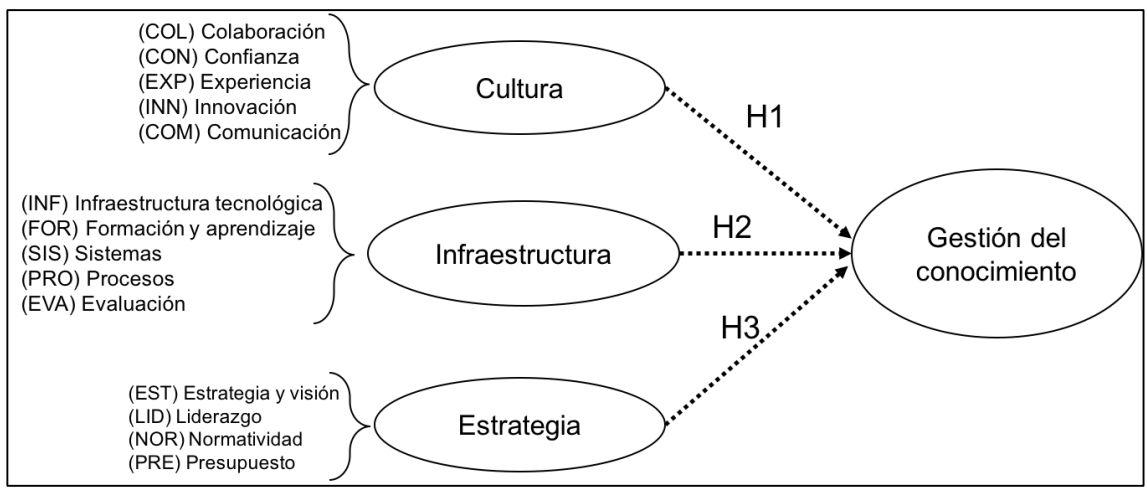

**Ilustración 5-2. Modelo de investigación propuesto**
Fuente: Elaboración propia (2016)

### 5.1.6. Análisis de los datos

Para realizar el análisis de los datos obtenidos se ha utilizado el software de análisis estadístico SmartPLS (versión 3) (Ringle, Wende, y Becker, 2015). Con apoyo del software citado, se emplea el análisis mediante mínimos cuadrados parciales (PLS, por sus siglas en inglés), bajo la técnica propuesta por Wold (1982), que permite observar de manera predictiva los coeficientes, cuya ruta representa una posible unión causal que relaciona directamente los valores encontrados entre los factores críticos de éxito y la GC. Este algoritmo de mínimos cuadrados parciales es esencialmente una secuencia de regresiones estadísticas en términos de peso de los vectores. Los pesos de los vectores obtenidos en la convergencia satisfacen las ecuaciones de punto fijo.

Para el presente estudio, este método permite observar relaciones del tipo causa-efecto entre los elementos y sus componentes, y finalmente, su contribución a la GC. Este esquema de ponderación provee el valor más alto de $R^2$ para las variables endógenas latentes y es, generalmente, aplicable para todo tipo de especificaciones y estimaciones de modelados del camino (o patrón) basados en Mínimos Cuadrados Parciales. Para analizar los datos del presente estudio se utilizaron 1,000 iteraciones, el valor máximo permitido por el software.

### 5.1.7. Resultados

Sobre los participantes se puede observar, en la tabla 5.7, que la mayoría de los participantes proviene de organizaciones de más de 201 trabajadores (73.41%). La formación académica de los participantes en su mayoría es de Licenciatura (57.95%) y Maestría (33.33%). La mayoría de los puestos de los respondientes es de nivel 2 (42.11%), de acuerdo con la clasificación propuesta. Finalmente, el grupo de antigüedad de los funcionarios de mayor porcentaje es de 0 a 10 años (37.02%), con una media de 16.4 años. También, de la muestra se aduce que los funcionarios de las OSP tienen en su mayor proporción más de 11 años de antigüedad.

Sobre la fiabilidad compuesta del estudio, ésta se ha determinado a través del α de Chronbach estimada para cada uno de los factores: cultura (*0.843*), infraestructura (*0.925*) y estrategia (*0.904*), lo cual para un estudio de esta naturaleza es un valor aceptable (Nonaka & von Krogh, 2009). Asimismo, el criterio Fornell-Larcker (1981) para cada uno de los constructos excede 0.8, que otorga cierta validez al estudio realizado (Henseler, Ringle, & Sarstedt, 2014).

Consecuentemente, en la ilustración 5.3, podemos observar el peso, o contribución, que tiene cada componente con respecto al factor crítico de éxito que integra en las iniciativas de GC. En este sentido, los valores negativos de los componentes se han marcado en letras itálicas para representar aquellos elementos que resultaron más débiles y que, por tanto, tienen mayores áreas de oportunidad en la implementación de iniciativas de GC en las OSP. Además, el modelo propuesto permite observar, la contribución de cada factor a la GC.

En términos generales se observa como la Cultura (0.370) es el factor que mayor contribución tiene sobre la GC, seguido de la Estrategia (0.183) y, finalmente, la Infraestructura (0.158). Los números sobre las fechas representan los valores residuales de contribución a la Gestión del Conocimiento ofrecidos por el modelo predictivo.

Particularmente, respecto al factor cultural se puede observar que los componentes relacionados con comunicación, innovación y colaboración, son los más débiles. Respecto al factor infraestructura, se observa que los componentes relacionados con *infraestructura tecnológica, sistemas de información, formación y aprendizaje y evaluaciones* son los más débiles. Para el caso del factor estrategia sus componentes más débiles son *presupuesto y liderazgo.* Derivado de este análisis, en la siguiente sección, se proponen algunas recomendaciones para la implementación de iniciativas de GC en las instituciones públicas, considerando los resultados observados y, sobre todo, las características de los funcionarios participantes; como elementos de una radiografía del estado de la GC en las OSP costarricense.

Por otra parte, para analizar los beneficios asociados con la GC, percibidos por los funcionarios de las OSP, se observan los valores en media de cada uno de ellos. El α de Chronbach para esta sección del estudio resultó de 0.92. Como se muestra en la tabla 5.2, señalados con letras en itálicas, los beneficios que principalmente se perciben son: (B8) El impulso de la GC al logro de los objetivos estratégicos de la organización; (B6) la GC mejora la toma de decisiones; y (B2) que la GC ayuda a generar mejores políticas públicas. En contraparte, los beneficios que son menos asociados con la GC son: (B1) La GC ayuda a reducir la corrupción; (B4) La GC Fomenta la transparencia en la organización; y (B3) La GC permite a los funcionarios trabajar de manera más eficiente.

| Beneficios asociados con la GC | Media (Desviación estándar) |
|---|---|
| *(B1) Ayuda a reducir la corrupción* | 3.67 (1.29) |
| *(B2) Ayuda a generar mejores políticas públicas* | 4.19 (0.95) |
| (B3) Permite trabajar de manera más eficiente | 4.02 (1.06) |
| (B4) Fomenta la transparencia en la organización | 3.94 (1.13) |
| (B5) Mejora la percepción ciudadana y la confianza en la AP | 4.12 (1.08) |
| *(B6) Mejora la toma de decisiones* | 4.26 (0.95) |
| (B7) Favorece la profesionalización de los funcionarios | 4.20 (1.04) |
| *(B8) Impulsa el logro de los objetivos estratégicos* | 4.31 (0.92) |

**Tabla 5.2. Resultados de los beneficios percibidos**
Fuente: (Pérez López-Portillo et al., 2016)

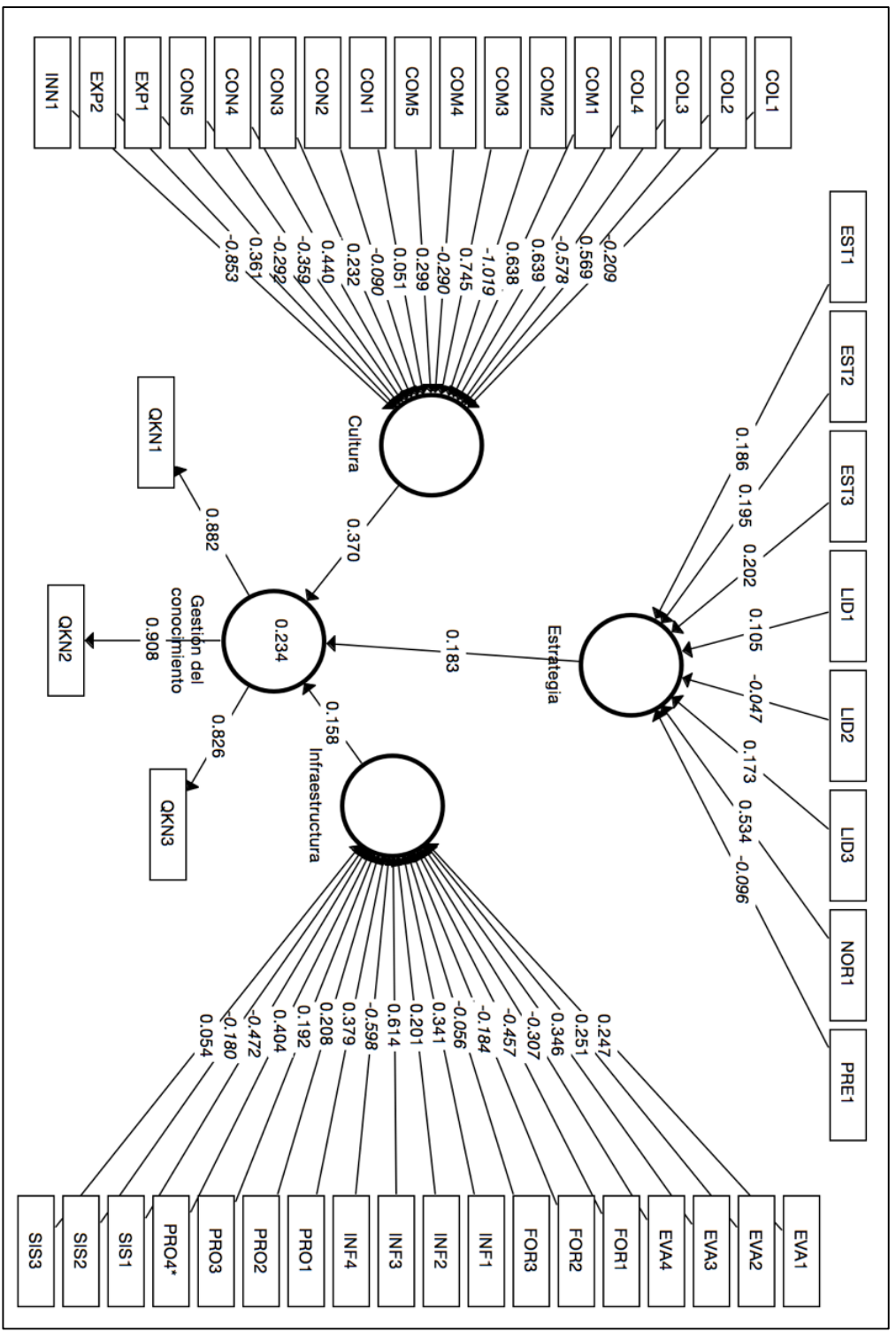

**Ilustración 5-3. Modelo de predicción de factores previos para la Gestión del Conocimiento**

Fuente: (Pérez López-Portillo et al., 2016)

# 6. DISCUSIONES

## 6.1.    RSL para analizar los factores previos para la Gestión del Conocimiento en las Organizaciones Públicas

Con respecto a las discusiones a raíz de la RSL llevada a cabo para conocer los factores previos que favorecen la GC en las OSP, así como para proponer estrategias que impulsen a la GC en este tipo de organizaciones, podemos referir lo siguiente:

Consideramos que la reutilización del conocimiento está altamente relacionada con la dimensión cultural, asociada a la confianza, la transparencia, la honestidad, la colaboración, la profesionalidad, la flexibilidad, el compromiso, el aprendizaje, la innovación y la especialización. El uso de la base de conocimiento existente es una práctica elemental para una buena administración y la reutilización del conocimiento (Savvas & Bassiliades, 2009). Como la experiencia empírica ha demostrado, la GC puede minimizar la duplicación de esfuerzos y promover una mayor consistencia en los servicios (M.-D. P. Lee, 2008). Una cultura organizacional que apoye el intercambio, el uso y la reutilización de conocimientos ha sido identificada como un facilitador para la GC (C. S. Choy, 2006). Por lo tanto, un entorno en donde las personas no se sienten obligados a compartir el conocimiento, sino que tienen un deseo constante de aprender juntos y complementar sus aprendizajes, es deseable.

La dimensión tecnológica consiste en la medición, los procesos de negocio y la infraestructura tecnológica. Por su parte, la infraestructura tecnológica proporciona una base de control para evaluar a el éxito de las iniciativas de GC, con el fin de detectar oportunidades para reorientar la estrategia organizacional de GC (Akhavan et al., 2006). Por otro lado, las actividades de medición de la GC dan una especie de control estratégico para que los administradores puedan mejorar la toma de decisiones y para regular el uso de los activos de conocimiento. Por tal motivo, es importante tener un plan de evaluación de la GC (Migdadi, 2009). A través de esto, es posible detectar qué tipo de habilidades técnicas y las necesidades de infraestructura son necesarias para iniciativas de GC.

En dimensión estratégica, la política pública, vista como la legislación y los marcos normativos, representan un facilitador del desarrollo de iniciativas de GC y de marcos amplios que fomenten actividades de GC como el intercambio y la evaluación de programas. Por ello, se necesita un liderazgo a nivel ejecutivo, representado como el apoyo de los responsables políticos, con el fin de desarrollar políticas relacionadas con las TIC que favorezcan la GC. Esto debe ser legitimado mediante el vínculo entre la GC y los objetivos de la organización, además de un acompañamiento y un complemento de buena voluntad política (Mbhalati, 2014). Los problemas de falta de elementos jurídicos, estatutos o marcos normativos que apoyen las iniciativas de GC, pueden ser causa de ignorancia acerca de la importancia y los beneficios de la GC (Gil-García & Pardo, 2005). Por estas razones, los factores institucionales y de organización como las normas, reglamentos, recursos, planes, normas, procesos, nivel de centralización y jurisdicciones son tan relevantes para una mejor adopción de iniciativas de GC.

## 6.2. Propuesta de métricas gestión del conocimiento para organizaciones públicas

Como podemos ver, las métricas de GC están relacionadas y entrelazados entre sí. Desde que la GC ha ido creciendo en importancia, muchas organizaciones la han promovido dentro de sus sus actividades y proceso esenciales, independientemente de sus diferencias o fines. Las OSP han adoptado este nuevo enfoque de gestión; en primer lugar como un imperativo para promover la innovación, y como un esfuerzo para proporcionar mejores servicios a los ciudadanos. En nuestra RSL, se observó que todavía hay una brecha en el desarrollo de la medición de la GC en este campo.

También se encontró que la medición de la GC en las OP no se analiza solamente desde una perspectiva del valor que tiene en el mercado, sino desde la calidad y cantidad del conocimiento, el proceso e GC, las iniciativas de GC y desde los resultados de la GC, así como desde sus beneficios esperados o demostrados. La medición de la GC parece ser observadas desde diferentes perspectivas (de valor, de no-valor, basada en sus resultados y en sus productos); desde diversos enfoques (cualitativos y cuantitativos, o financieros y no-financieros); y desde distintos niveles (conocimientos, iniciativas de GC, resultados o productos de GC; y la contribución al desempeño de la organización)

## 6.3. Estudio en profundidad para conocer los factores previos para la Gestión del Conocimiento en las Organizaciones Públicas

Como se puede observar en los resultados, hay componentes que contribuyen con menor fuerza al factor, y dado que los factores tienen una contribución hacia la GC, se han formulado algunas recomendaciones dirigidas a los tomadores de decisiones o líderes encargados de implementar iniciativas de GC en las OSP. Estas recomendaciones se han formulado buscando contrarrestar las barreras; se han alineado con los factores y sus respectivos componentes; y considerando las características particulares de los funcionarios de la AP costarricense. Se ha decidido enunciar las recomendaciones para el factor cultura, por ser el factor que ha resultado tener una mayor contribución en la GC, seguido de los factores estrategia e infraestructura.

### (1) Cultura

De manera general se observa la importancia de promover una cultura que favorezca la GC, y sus actividades relacionadas, como la transferencia y el intercambio de conocimientos dentro de la organización. Para ello, es necesario mejorar los canales de comunicación, promover espacios de intercambio de experiencias y monitorear el logro de los objetivos organizacionales.

Es crucial realizar acciones para aumentar el nivel de confianza y colaboración entre las personas de la organización y con otras instituciones, para ello se pueden establecer acuerdos formales de colaboración entre las OSP con reglas claras y comprensibles. Igualmente, respecto a este factor, es importante estimular la generación de nuevas ideas, y favorecer la creatividad sobre formas innovadoras de resolver los problemas de la institución. Así como favorecer el intercambio de experiencias entre las personas de la organización y crear las condiciones necesarias para salvaguardar las experiencias positivas y resguardar los activos valiosos de conocimiento. Además, los espacios colaborativos, de divulgación y transferencia de información propician las condiciones de generación de lecciones aprendidas para el mejoramiento continuo de los procesos basados en la motivación de las personas de la organización, que visualizan su contribución al logro de objetivos.

### (2) Infraestructura

Sobre la infraestructura se advierte necesario integrar sistemas e instrumentos de compatibilidad entre las OSP. También, se debe mejorar el diseño de los sistemas de información actuales, hacerlos más útiles y que realmente respondan a las necesidades de las instituciones, considerando a su vez, la voz de los funcionarios que los utilizaran en su diseño e implementación.

Conjuntamente, se debe favorecer el desarrollo de capacidades entre los funcionarios de las OP para la GC, se ha observado que la poca preparación de las OSP ante el cambio tecnológico que conlleva la GC es una de las principales barrera para su implementación. Así, las estrategias de capacitación, en este orden de ideas, deben estar alineadas las características particulares de los funcionarios. Es decir, desarrollar sistemas de capacitación que acompañen a las iniciativas de GC que tomen en cuenta que los funcionarios tienen un grado académico de licenciatura y maestría, principalmente, así como una antigüedad en media de 16 años y un nivel 2, de acuerdo con la clasificación de este estudio. Además, es importante mejorar los sistemas de evaluación del desempeño de los funcionarios e impulsar el aprendizaje permanente y el desarrollo de nuevas competencias para la GC entre los funcionarios

Resulta importante indicar que existe un alto grado de profesionalización de los funcionarios de las OSP costarricense, lo cual, es una base para mejorar la competitividad de las organizaciones públicas. Por ello, se requiere que las OSP integren de manera articulada y sostenible en el tiempo procesos de GC soportados con TIC que permitan una alta contribución de los funcionarios públicos en la producción de nuevo conocimiento. La oportunidad para aprovechar el nivel de profesionalización de los funcionarios de las OSP costarricense, que está acompañado de la alta experiencia, se materializa para la OP toda vez que se realicen inversiones para mejorar la infraestructura y obtener un mejor aprovechamiento de la GC.

### (3) Estrategia

Para el factor estrategia además de superar las barreras legislativas y los marcos presupuestarios, es necesario un liderazgo que dé mayor acompañamiento a las iniciativas de GC en las OSP, adicionado a la voluntad política de los funcionarios de más alto nivel. Como se ha señalado, la conciencia pública sobre los beneficios potenciales de la GC es necesaria para tener una visión más clara sobre el rumbo que deben llevar estar iniciativas.

Por tanto, es fundamental garantizar apoyo presupuestal a las iniciativas de GC, evaluar continua y sistemáticamente la implementación de estas iniciativas en las instituciones y reconocer sus resultados, así como monitorear la implementación de cambios en las instituciones. También es necesario difundir entre todos los niveles de la OP la visión proyectada sobre la GC y desarrollar un mayor conocimiento, entendimiento y adopción de la estrategia, planes, metas y objetivos de la organización entre sus funcionarios.

De esta forma, la operación de las políticas públicas, planes, programas y proyectos puede ser mejorada toda vez que las OSP tome decisiones orientadas hacia el fortalecimiento de la GC como factor para el logro de resultados de desarrollo y la generación de impactos en la sociedad.

Asimismo, la lucha contra la corrupción y la transparencia deben ser variables que se fortalezcan a partir de la creación de espacios de colaboración y transferencia del conocimiento, aprovechando iniciativas, como por ejemplo: (i) la Red Interinstitucional de Transparencia, promovida por la Defensoría de los Habitantes de la República (DHR) de Costa Rica que busca facilitar a las y los habitantes el acceso a la información relacionada con la administración de los recursos públicos, a través de su publicación en Internet; y (ii) la iniciativa de Gobierno Abierto Costa Rica que persigue mejorar los niveles de transparencia y acceso a la información, facilitar la participación ciudadana y propiciar la generación de espacios de trabajo colaborativo en las instituciones públicas, lo anterior mediante la innovación utilizando las TIC y herramientas novedosas que atiendan la brecha digital.

# 7. CONCLUSIONES, TRABAJO FUTURO Y LIMITACIONES

## 7.1. Conclusiones

Las OP continuarán experimentando cambios en la forma en que prestan sus servicios, en función de las realidades sociales, políticas y económicas del momento (Guthrie & Dumay, 2015). La GC tiene el potencial de ampliar y aumentar la eficacia de la base general de conocimientos de la organización (Ann Hazlett et al., 2008, p. 58); la medición necesita de un plan que identifique los activos de conocimiento específicos en las organizaciones públicas (Dalkir et al., 2007, p. 1449), con el objetivo de proveer de un marco de evaluación valido medir el éxito de las iniciativas de gestión del conocimiento y los resultados del proceso de GC. Además, las OP utilizan cada vez más las tecnologías de información para colaborar entre sí, lo que implica una mayor necesidad de desarrollar una fuerte capacidad de compartir, aplicar y crear conocimiento (L.G. Pee & Kankanhalli, 2016, pp. 188–189).

La GC es un tema que atrae cada vez mayor relevancia en el ámbito público. Sin duda, los beneficios potenciales que puede traer la GC a las instituciones y a la sociedad en general, se observan a la luz de la inminente transformación que requieren las sociedades del siglo XXI. Por tal motivo, estudiar los factores que contribuyen a un mayor éxito en las iniciativas de GC en las instituciones públicas es pertinente. El presente estudio ha logrado su objetivo, conocer los factores, sus componentes asociados y los beneficios percibidos con respecto a la GC en las OSP. Del análisis de los resultados se han generado algunas recomendaciones para potencializar el éxito de las iniciativas de GC en la esfera pública. Los acelerados cambios en la sociedad y el avance en las TIC nos conminan a adoptar estrategias de gestión más inteligentes y efectivas que se transformen en beneficio de la sociedad en su conjunto.

A través de esta tesis se reivindica, para futuras investigaciones, la necesidad de poner atención en el desarrollo de métodos más sofisticados y exhaustivos de medición del conocimiento y de la GC. Atendiendo, sobre todo, a la naturaleza compleja del conocimiento, que integren elementos de sistemas de inteligencia artificial, técnicas de optimización (Hu et al., 2015), minería y análisis de grandes datos, y que atiendan a la configuración específica que tienen las organizaciones públicas (Garlatti et al., 2014; Massaro et al., 2015); organizaciones que producen conocimiento con gran intensidad (Manfreda, Buh, & Indihar Štemberger, 2015).

A las que los sistemas de medición les permitirá incrementar su desempeño y mejorarlo continuamente (Fiona Buick, Deborah Blackman, Michael Edward O'Donnell, & Damian West, 2015), para tomar mejores decisiones (Abdel-Maksoud et al., 2015).

Finalmente, en este trabajo de investigación de definió un modelo de evaluación de los FCE a partir de una revisión sistemática que permite identificar cuáles son los elementos más relevantes a considerar previo, durante y posterior a la implementación de iniciativas de GC en las OSP. A partir de los resultados obtenidos se puede ver el pesos específico que tienen algunos factores y elementos sobre otros y valorar la estrategia particular que se sigue en cada OSP previo al esfuerzo institucional por implementar prácticas de GC. De esta manera la alineación estratégica de los factores, considerar los elementos que favorecen a la GC y las métricas propuestas pueden apoyar este tipo de iniciativas en las OP facilitando el camino y potencializando los resultados.

A nivel personal este trabajo de investigación representó grandes retos, por una parte, porque se circunscribe dentro de un proyecto en curso financiado por el Consejo Nacional de Ciencia y Tecnología, con el apoyo del Centro de Investigación en Matemáticas A.C., y por la Universidad de Guanajuato. Por otra parte, ha constituido una grata y rica experiencia de aprendizaje personal y de crecimiento profesional y académico, me ha ayudado a perfilar un proyecto de investigación que estoy cierto continuaré.

## 7.2. Líneas futuras de investigación

Hacia el futuro, se requieren estudios más exhaustivos, por ventura longitudinales, que permitan analizar el objeto de estudio (la GC) en distintos contextos, atendiendo a múltiples configuraciones socio-organizacionales. Además, se deberá ampliar el espectro de FCE, los elementos que integran a estos factores y se podrían hacer algunas validaciones adicionales contrastando a la GC con elementos que emergen como la minería de datos (Berka, Rauch, & Zighed, 2009), la inteligencia de negocios (Ong, 2016), las ciudades inteligentes (Smart Cities) (Gil-Garcia, Helbig, & Ojo, 2014) y el *Big Data* (OECD, 2015a; Reimsbach-kounatze, 2015)**.**

## 7.3. Limitaciones del estudio

Una de las principales limitaciones del presente estudio ha sido el tiempo para recolectar los datos. Creemos que se puede hacer un estudio longitudinal observando el comportamiento de los factores en un determinado tiempo, así como un estudio comparativo entre distintitos países. Asimismo, después de considerar los factores en la implementación de las iniciativas de GC se podría hacer una evaluación más exhaustiva de los factores que han sido clave y, de esta forma, tener alguna validación empírica posterior la implementación de la GC.

Otra limitación para el desarrollo del presente estudio podría referirse a la delimitación geográfica determinada dentro de la cual se realizó la presente investigación (Corbetta, 2007; Hernández Sampieri, Fernández-Collado, & Baptista Lucio, 2010; Pacheco & Cruz, 2011), lo que pudiera no permitir su comparabilidad en el tiempo y el espacio. La muestra con la que contó el estudio en profundidad pudiera además ser vista como una limitación, que se ha tratado de solventar volviendo cada vez más exhaustiva la RSL.

Finalmente, las limitaciones encontradas en esta RSL son similares entre los estudios seleccionados, lo que sugiere que aún la evidencia o validación empírica de las métricas propuestas hasta hoy es insuficiente; y que además el proceso para definir métricas de cada iniciativa de GC es complejo, e incluso más para las iniciativas de gestión del conocimiento en las OSP. Ciertamente, se puede estar de acuerdo con Massaro et al., (2015), cuando ellos argumentan que la GC en las OP parece estar fragmentada o desarticulada.

# 8. BIBLIOGRAFÍA

# 9. APÉNDICE A: Estudios primarios de la RSL

# 10. APÉNDICE B: Cuestionario aplicado, ítems, valores de media y desviación estándar

| Factor | Componente | Descripción | Media (Desviación estándar) |
|---|---|---|---|
| Cultura | COL | 1. Los miembros de la organización están satisfechos con el grado de cooperación mutua | 3.48 (0.93) |
| | | 2. Las personas de la organización están dispuestas para ayudarse y apoyarse entre ellas | 3.49 (1.00) |
| | | 3. En esta organización existen las condiciones necesarias para que se genere confianza y colaboración | 3.38 (1.12) |
| | CON | 4. Los miembros de la organización estamos dispuestos a cooperar entre distintas unidades y/o departamentos | 3.71 (0.95) |
| | | 1. Los miembros de la organización podemos confiar en las actitudes, comportamientos y habilidades de otros miembros | 3.25 (0.98) |
| | | 2. Los miembros de la organización confiamos en que nuestros esfuerzos ayudan a lograr los objetivos de la organización | 2.92 (0.96) |
| | | 3. En esta organización las decisiones de sus integrantes se toman con base en los objetivos de la organización | 3.28 (1.06) |
| | | 4. Siento confianza para compartir con mis compañeros los conocimientos que poseo para lograr los objetivos de mi área | 3.93 (1.08) |
| | | 5. Las relaciones entre los miembros de esta organización están basadas en la confianza mutua | 3.19 (1.03) |

| | | |
|---|---|---|
| EXP | 1. Compartir mi experiencia con otras personas puede ayudar a resolver problemas y a generar nuevas oportunidades de mejora | 4.37 (0.81) |
| | 2. Se toma en cuenta la experiencia de las personas para mejorar los servicios que se ofrecen | 3.25 (1.13) |
| INN | 1. Se estimulan la generación de nuevas ideas y las formas creativas para resolver los problemas | 2.98 (1.18) |
| COM | 1. Se cuenta con herramientas y medios efectivos para la comunicación entre sus miembros | 3.46 (1.19) |
| | 2. Los equipos de trabajo de la organización mantienen reuniones o sesiones para intercambiar experiencias y/o avances en sus actividades | 3.22 (1.13) |
| | 3. Los líderes (jefes) facilitan los espacios de discusión y realimentación del cumplimiento de objetivos y metas | 2.89 (1.22) |
| | 4. La organización cuenta con un espacio para divulgar los conocimientos adquiridos por sus miembros | 2.96 (1.21) |
| | 5. La organización ofrece a sus beneficiarios y a la comunidad en general las experiencias sistematizadas de sus resultados por medio de boletines electrónicos y/o boletines impresos. | 3.22 (1.29) |

| Infraestructura | | | |
|---|---|---|---|
| INF | | 1. La infraestructura tecnológica (sistemas, hardware, software, etc.) de esta organización es la apropiada para ofrecer un excelente servicio y/o producto | 3.43 (1.22) |
| | | 2. Los sistemas y las tecnologías con que cuenta la organización me permiten realizar mi trabajo de mejor manera | 3.64 (1.16) |
| | | 3. Los sistemas y tecnologías son efectivos para facilitar compartir la información y el conocimiento entre sus miembros | 3.59 (1.10) |
| | | 4. Los sistemas de esta organización están integrados y/o son compatibles entre distintos departamentos e instituciones | 3.26 (1.23) |
| FOR | | 1. Las personas de esta organización tienen los conocimientos, las habilidades y las aptitudes necesarias para el desarrollo de sus funciones | 3.57 (0.91) |
| | | 2. En esta organización se apoya el aprendizaje permanente y el desarrollo de nuevas competencias y habilidades en las personas | 3.29 (1.21) |
| | | 3. En esta organización se incentiva el uso y deseo de aprender a utilizar nuevas tecnologías | 3.26 (1.15) |
| SIS | | 1. En esta organización los sistemas funcionan correctamente, generan información útil y cumplen con su propósito | 3.36 (1.07) |
| | | 2. Los sistemas con que cuenta la organización me resultan fáciles de usar | 3.82 (0.92) |
| | | 3. Los sistemas actuales de esta organización proveen información actualizada y precisa cuando la necesito | 3.41 (1.05) |
| PRO | 1. En esta organización las actividades, los procesos y los procedimientos son realizados bajo estándares de calidad y/o marcos de regulación | | 3.35 (1.18) |
| | 2. El conocimiento en la organización está correctamente explicitado (en manuales, procedimientos, en la normatividad, políticas, etc.) | | 3.35 (1.14) |
| | 3. Existen criterios comunes (sobre formatos, actividades, tareas) entre personas que realizan las mismas funciones | | 3.28 (1.04) |
| | 4. Con frecuencia utilizo formatos que no me han sido proporcionados por la organización para realizar más eficientemente mi trabajo | | 3.09 (1.26) |
| EVA | 1. Existen procedimientos establecidos evaluar el conocimiento y la información que se produce en la organización | | 3.01 (1.10) |
| | 2. En esta organización los sistemas de evaluación permiten mejorar constantemente los servicios y/o productos que se ofrecen | | 2.82 (1.15) |
| | 3. En la organización existen sistemas de evaluación que ofrecen recompensas para incentivan la mejora continua en nuestras actividades | | 2.11 (1.10) |
| | 4. Esta organización cuenta con sistemas y tecnologías que contribuyen a tener evaluaciones objetivas y puntuales sobre el desempeño de las personas | | 2.42 (1.19) |

| Calidad | Estrategia | | |
|---------|------------|---|---|
| EST | | 1. Los miembros de esta organización conocen los objetivos, planes y/o políticas de la misma | 3.43 (1.05) |
| | | 2. Existe un firme compromiso de los líderes (jefes) de esta organización con la implementación de nuevos sistemas y tecnologías | 3.20 (1.11) |
| | | 3. Los miembros de esta organización conocen sus roles y responsabilidades para el cumplimiento de sus deberes | 3.67 (0.91) |
| LID | | 1. En esta organización existe una visión institucional clara sobre la gestión del conocimiento y sus beneficios | 3.03 (1.11) |
| | | 2. Los esfuerzos para la implementación de sistemas de gestión del conocimiento han sido efectivos hasta el momento | 2.94 (1.08) |
| | | 3. Con frecuencia se implementan cambios para mejorar los servicios y/o productos que se ofrecen | 3.17 (1.04) |
| NOR | | 1. Las políticas y regulaciones actuales favorecen la implementación de gestión del conocimiento | 3.26 (1.08) |
| PRE | | 1. Existe apoyo presupuestal para la implementación de nuevos sistemas y tecnologías | 3.22 (1.13) |
| QKM | | 1. En esta organización el conocimiento y la información que se producen son reutilizados constantemente para producir nuevo conocimiento | 3.17 (0.97) |
| | | 2. El conocimiento y la información que produce la organización son utilizados racionalmente para tomar decisiones | 3.36 (0.91) |

| | | Beneficios | |
|---|---|---|---|
| | KMB | 3. El conocimiento y la información que produce esta organización son útiles para otras entidades de la administración pública | 3.62 (1.12) |
| | | 1. La gestión del conocimiento en la administración pública ayuda a reducir la corrupción | 3.67 (1.29) |
| | | 2. La gestión del conocimiento ayuda a generar mejores políticas públicas | 4.19 (0.96) |
| | | 3. La gestión del conocimiento permite a las personas de esta organización realizar su trabajo de una forma más eficiente y en menor tiempo | 4.02 (1.06) |
| | | 4. La gestión del conocimiento fomenta la transparencia en la organización | 3.94 (1.14) |
| | | 5. La gestión del conocimiento mejora la percepción ciudadana y la confianza en la organización | 4.13 (1.08) |
| | | 6. La gestión del conocimiento ayuda a tomar mejores decisiones a la organización | 4.26 (0.96) |
| | | 7. La gestión del conocimiento favorece la profesionalización de los funcionarios | 4.20 (1.04) |
| | | 8. La gestión del conocimiento impulsa el logro de los objetivos organizacionales | 4.32 (0.93) |